\documentclass[preprint,1p]{elsarticle}
\journal{Journal of Logical and Algebraic Methods in Programming}

\usepackage{times}
\usepackage{amsmath}
\usepackage{amssymb}
\usepackage{graphicx}
\usepackage{multirow}
\usepackage[usenames,dvipsnames,svgnames,table]{xcolor}
\usepackage{subcaption}
\usepackage{mathpartir}
\usepackage{etoolbox}
\usepackage{xifthen}
\usepackage{mathtools}
\usepackage{pgf-umlcd}
\renewcommand{\umlfillcolor}{white}
\usepackage[hidelinks]{hyperref}
\usepackage{todonotes}
\usepackage{enumitem}
\setlist{itemsep=3pt}
\setlist[itemize]{leftmargin=2em}
\setlist[enumerate]{leftmargin=1.5em}
\usepackage{booktabs}
\usepackage{natbib}

\newdefinition{rmk}{Remark}
\newtheorem{definition}{Definition}
\newtheorem{example}{Example}
\newtheorem{remark}{Remark}
\newtheorem{proposition}{Proposition}
\newproof{proof}{Proof}
\newproof{proofpart}{}

\newcommand{\optdefault}[3][]{\ifthenelse{\isempty{#2}}{#3}{#1{#2}}}
\newcommand\figdec[1]{(#1)}
\newcommand\rfig[2][]{Fig.~\ref{#2}\optdefault[\figdec]{#1}{}}
\newcommand\rfigb[2][]{Figure~\ref{#2}\optdefault[\figdec]{#1}{}}
\newcommand{\secref}[1]{Section~\ref{#1}}
\newcommand{\defref}[1]{Definition~\ref{#1}}
\newcommand{\exref}[1]{Example~\ref{#1}}
\newcommand\bipmin{7cm}
\newcommand\bipAtomicExample{B}

\newcommand{\twopartdef}[4]
{
  \left\{
    \begin{array}{ll}
      #1 & #2 \\
      #3 & #4
    \end{array}
  \right.
}
\newcommand{\npartsdef}[1]
{
  \left\{
    \begin{array}{ll}
		#1
    \end{array}
  \right.
}

\newcommand\rtable[2][]{Table~\ref{#2}\optdefault[\figdec]{#1}{}}
\newcommand\rcode[2][]{Listing~\ref{#2}\optdefault[\figdec]{#1}{}}
\newcommand\code[1]{\ensuremath{\mathtt{#1}}}

\newcommand{\getcode}[5][]{%
	\lstinputlisting[float=t, #5, caption={#3}, label=lst:#1]{src/#2}%
}

\makeatletter
\newcounter{getexamplesfig}
\newcommand{\getexamples}[5][]{ %
	\begin{figure}[t!] %
		\centering%
		\newcommand\maketikz@var{#4} %
		\newcommand\maketikz@split{#5} %
		\setcounter{getexamplesfig}{0}  %
		\maketikz{#2} %
		\renewcommand\maketikz@var\relax %
		\renewcommand\maketikz@split\relax %
		\caption{#3} %
		\label{fig:#1} %
	\end{figure} %
}
\newcommand\maketikz[1]{%
    \forcsvlist{\getlist@item}{#1}
}
\newcommand\getlist@item[1]{
	\scalebox{\maketikz@var}{\input{tikz/#1}}
	\stepcounter{getexamplesfig}
	\ifnum\value{getexamplesfig}=\maketikz@split ~\\\setcounter{getexamplesfig}{0}\fi
}

\newcommand{\casepath}[1]{case/#1}
\newcommand{\sideaspect}[6][]{%
	\begin{figure}[t!] %
	  \centering%
	  \framebox{\scalebox{1}{\lstinputlisting[style=bip, frame=r, language=aopbip, linewidth=#3\textwidth, #1]{\casepath{#2.abip}}}%
      \hspace{#4}\scalebox{1}{\lstinputlisting[style=txt, linebackgroundwidth=5.2cm, #6]{\casepath{#2-output.txt}}}%
    }%
		\caption{#5}%
		\label{lst:aspect-#2}%
	\end{figure}%
}

\renewcommand{\sideaspect}[6][]{%
\begin{figure}[t!] %
	\centering%
	\begin{subfigure}[b]{0.5\textwidth} %
		\scalebox{.9}{\lstinputlisting[style=bip, frame=none, language=aopbip, linewidth=.4\textwidth, #1]{\casepath{faults.abip}}}%
    \end{subfigure}	%
	\begin{subfigure}[b]{0.35\textwidth}%
		\scalebox{1}{\lstinputlisting[style=txt, linebackgroundwidth=5.2cm, #6]{\casepath{#2-output.txt}}}%
  \end{subfigure}	%
	\caption{#5}%
	\label{lst:aspect-#2}%
\end{figure}%
}%
\newcommand{\aopbip}{\texttt{AOP-BIP}}

\newcommand{\events}{{\cal E}}
\newcommand\tmap{\ensuremath{\mathrm{m}}}
\newcommand{\btrue}{\code{true}}
\newcommand{\bfalse}{\code{false}}
\newcommand{\vglobal}{\mathrm{g}}

\newcommand{\setof}[1]{\ensuremath{\left \{ #1 \right \}}}
\newcommand{\tuple}[1]{\ensuremath{\left \langle #1 \right \rangle }}

\newcommand{\atoms}{\ensuremath{\mathcal{B}}}
\newcommand{\codes}{\ensuremath{\mathcal{F}}}
\newcommand{\composite}{\ensuremath{\mathcal{C}}}
\newcommand{\atomic}{\ensuremath{\mathit{B}}}
\newcommand{\subgamma}{\ensuremath{\mathcal{I}}}

\newcommand{\varsof}[1]{\ensuremath{\mathit{#1.vars}}}
\newcommand{\portsof}[1]{\ensuremath{\mathit{#1.ports}}}
\newcommand{\locsof}[1]{\ensuremath{\mathit{#1.locs}}}
\newcommand{\transof}[1]{\ensuremath{\mathit{#1.trans}}}

\newcommand{\valuations}{\ensuremath{\textbf{X}}}

\newcommand{\func}{\mathit{func}}
\newcommand{\port}{\mathit{port}}

\newcommand{\guard}{\mathit{guard}}
\newcommand{\source}{\mathit{src}}
\newcommand{\dest}{\mathit{dest}}
\newcommand{\aspectlocal}{\mathit{LA}}

\newcommand{\aspectglobal}{\mathit{GA}}

\newcommand{\weaveall}{\mathrm{weaveAll}}
\newcommand{\weaveserial}{\mathrm{weaveSerial}}

\newcommand{\fweaveg}{\mathrm{weave_g}}
\newcommand{\fweavel}{\mathrm{weave_\ell}}

\DeclareMathOperator{\varguard}{readguard}
\DeclareMathOperator{\varread}{readvar}
\DeclareMathOperator{\varwrite}{writevar}

\newcommand{\pcloc}{\mathrm{atLocation}}
\newcommand{\pcguard}{\mathrm{readVarGuard}}
\newcommand{\pcfunc}{\mathrm{readVarFunc}}
\newcommand{\pcwrite}{\mathrm{write}}
\newcommand{\pcportenabled}{\mathrm{portEnabled}}
\newcommand{\pcportexec}{\mathrm{portExecute}}

\newcommand{\pcmatch}{\mathrm{select_{\ell}}}
\DeclareMathOperator{\pcedit}{\ensuremath{\mathrm{edit}}}
\newcommand{\pcmatchg}{\mathrm{select_{g}}}

\newcommand{\EP}{\mathit{EP}}
\newcommand{\WPreviousBefore}{\mathit{PB}}
\newcommand{\WPreviousAfter}{\mathit{PE}}
\newcommand{\WCreate}{\mathit{RUN}}
\newcommand{\WCurrentBefore}{\mathit{CB}}
\newcommand{\WCurrentAfter}{\mathit{CE}}
\newcommand{\WPrevious}{P}

\DeclareMathOperator{\WGetP}{prev}
\DeclareMathOperator{\WOrigin}{origin}
\DeclareMathOperator{\WDest}{dest}
\DeclareMathOperator{\WSiblings}{siblings}

\DeclareMathOperator{\weave}{\triangleleft_{\ell}}

\DeclareMathOperator{\weaveg}{\triangleleft_{\mathit g}}

\DeclareMathOperator{\weaveframe}{wframe}
\DeclareMathOperator{\weavereset}{wReset}

\newcommand{\aopvar}{\ensuremath{b_\mathrm{aop}}}
\newcommand{\aopport}{\ensuremath{ip}}
\newcommand{\aopset}{\ensuremath{F_\mathrm{set}}}
\newcommand{\aopclear}{\ensuremath{F_\mathrm{clear}}}
\newcommand{\resetlocs}{\ensuremath{L_{\mathrm{R}}}}

\newcommand{\mkAddGuard}{\mathrm{mkGuard}}
\newcommand{\trmap}{\mathrm{map}}
\newcommand{\trproject}{\mathrm{map}}

\newcommand{\erem}{\mathrm{rem}}
\newcommand{\fvar}[1]{\mathit{#1}}

\newcommand{\evalLocal}[2]{#1 \vDash #2}
\newcommand{\stateproject}{\upharpoonleft}

\usepackage{tikz}
\usetikzlibrary{shapes,fadings,shadows,trees,matrix,intersections,patterns,decorations.markings,decorations.pathreplacing,arrows,automata,calc,decorations,fit,positioning,backgrounds}
\newcommand{\colorselect}{magenta}
\newcommand{\colorhigh}{blue}
\newcommand{\colorcreate}{green!50!black}

\newcommand\tmark[4]{
	\node[circle,shift={(#4)}, draw, fill=yellow!50, minimum size=0pt, inner sep=0cm,text width=0.4cm] (#1) at($(#2)$) {\scalebox{0.7}{#3}};
}

\tikzset{
	bip/.style      = {auto,node distance=1cm,line width=2pt,>=to,thick,align=center},
	component/.style= {auto,node distance=1.5cm,thick,
						align=center,inner sep=0cm,minimum size=0pt, transform shape,
						 ->,>=stealth',shorten >=1pt},
	location/.style = {state,fill=white!89!black,draw=white!50!black,minimum size=1cm,inner sep=0cm},
	var/.style	    = {draw,thick,minimum width=0.2cm,minimum height=0.7cm},
	portvar/.style  = {dashed, draw=black!40},
	export/.style   = {state,fill=black,minimum size=1.5mm,inner sep=0mm},
	ilabel/.style   = {node distance=-0.1cm, font=\boldmath},
	clabel/.style	= {yshift=1cm, minimum height=1cm}
}

\tikzset{
	normal/.style	= {->, draw=black, fill=black},
	fade/.style		= {->, color=black!20},
	selected/.style	= {line width=0.5mm, draw=\colorselect},
	high/.style		= {line width=0.4mm, color=\colorhigh, draw=\colorhigh},
	created/.style	= {dotted, line width=0.4mm, color=\colorcreate!30!black,
						draw=\colorcreate},
	previous/.style	= {loosely dashed, ->, line width=0.3mm, draw=\colorhigh},
	hidden/.style	= {draw=black!0,fill=black!0, color=black!0},
	sample/.style	= {text opacity=0},
	sel/.style={fill=black!10},
}

\tikzset{
	atomic/.style	= {minimum size=6.3cm},
	a-big/.style	= {minimum size=7.5cm},
	composite/.style= {minimum size=2.5cm},
	explode/.style	= {node distance= 3.5cm}
}

\newcommand{\component}[5][]{
  \node[rectangle,draw] (#2) #4 [#3] {
    \\ \\
    \begin{tikzpicture}[component, #1]
      \tikzset{
      place/.style={location},
      prime/.style={place,fill=white!95!black,draw=white!70!black}
    }
      #5

    \end{tikzpicture} 
  }
}

\newcommand{\componentb}[6][]{
  \node[rectangle,draw] (#2) #4 [#3] {
    \\ \\
    \begin{tikzpicture}[component, #1]
      \tikzset{
      place/.style={location},
      prime/.style={place,fill=white!95!black,draw=white!70!black}
    }
      #5

    \end{tikzpicture} 
  };
  \node at ($(#2.south west)!.5!(#2.south east)$) [label=below:#6]    {};
}

\pgfmathtruncatemacro\distance{1}
\newcommand{\hsel}{|[sel]|}

\newcommand{\chigh}[1]{\textcolor{\colorhigh}{#1}}
\newcommand{\cselect}[1]{\textcolor{\colorselect}{#1}}

\newcommand{\fwrap}[4][\colorselect]{\ffunc{\textcolor{#1}{#2} #3 \textcolor{#1}{#4}}}
\newcommand{\ffunc}[1]{\ensuremath{\langle #1 \rangle}}
\newcommand{\ifunc}[2]{\lbrack#1\rbrack\\\scalebox{0.9}{#2}}
\newcommand{\tilabel}[2]{
	\footnotesize{\lbrack\texttt{#1}\rbrack} \\ \footnotesize{#2}
}

\newcommand{\tconnect}[9][]{
	\path[->,#2] (#3) edge [         ] node (#1)[label={[label distance=#5]#6:\footnotesize{\lbrack\texttt{#7}\rbrack} \\ \footnotesize{#8} \\ \footnotesize{\texttt{#9}}}]   {} (#4);
}

\newcommand{\bipcomposite}{2.5cm}

\newcommand{\tconnector}[7][]{
	\draw[-,#1] (#2) -| ([shift={(#4)}]$(#2)!0.5!(#3)$) node[#5](#7){#6} |- (#3); 
}

\newcommand{\tconnectorgap}[7][]{
	\draw[-,#1] (#2) 
			 -- ([shift={(#4)}]$(C_p2.east)$)node {}
			 -- ([shift={({#4}[0],0)}]$(#2)!0.5!(#3)$)node [#5](#7) {#6}
			 -- ([shift={(#4)}]$(#3)$)node {}
			 -- (#3);
}

\usepackage{lstlinebgrd} 

\newcommand\singlespacing{}
\lstdefinelanguage{aopbip}
{
	morekeywords={Aspect, with, do, readVarGuard, readVarFunc, write, portEnabled, portExecute, atLocation, data, ports, writePortVars, readPortVars},
	sensitive=\btrue,
	morecomment=[l]{//},
	morecomment=[s]{/*}{*/},
}
\lstdefinelanguage{bip}
{
	language=c,
	morekeywords={model,on, data, connector, atomic, compound, port, do, from, to, component, initial, end, type, from, data. define, up, down, export, place, int, bool, header, module},
	sensitive=\btrue,
	morecomment=[l]{//},
	morecomment=[s]{/*}{*/},
}
\lstdefinelanguage{aspectj}
{
	language=java,
	morekeywords={aspect, call, ||, pointcut, before, after, around, returning},
	sensitive=\btrue,
	morecomment=[l]{//},
	morecomment=[s]{/*}{*/},
}
\lstdefinelanguage{bnfgrammar}
{
  moredelim=[s][\color{green!50!black}\ttfamily]{'}{'},
  alsoletter={:,|,;,+,?,*},
  morekeywords={:,|,;,+,?,*},
  morecomment=[l]{//}, 
	morecomment=[s]{/*}{*/},
  emph={
        IDENTIFIER, CODE
  } 
}
\lstdefinestyle{grammar}{
  language=bnfgrammar,
  frame=lrtb,
  showstringspaces=false,
  basicstyle=\scriptsize\ttfamily\singlespacing,
  emphstyle={\color{green!30!black}\ttfamily},
  keywordstyle=\bfseries\color{blue!80!black}
}
\lstdefinestyle{bip}{
  language=bip,
  breaklines=false,
  breakautoindent=false,
  frame=lrtb,
  showstringspaces=false,
  basicstyle=\scriptsize\ttfamily\singlespacing,
  keywordstyle=\bfseries\color{blue!80!black},
  commentstyle=\itshape\color{green!50!black},
  identifierstyle=\color{black},
  stringstyle=\color{red}
}
\lstdefinestyle{javaex}{
  language=Java,
  breaklines=false,
  breakautoindent=false,
  frame=lrtb,
  tabsize=2,
  keepspaces,
  xleftmargin=none,
  showstringspaces=false,
  basicstyle=\tiny\ttfamily\singlespacing,
  keywordstyle=\bfseries\color{blue!80!black},
  commentstyle=\itshape\color{green!50!black},
  identifierstyle=\color{black},
  stringstyle=\color{red}
}
\lstdefinestyle{aspectj}{
  language=aspectj,
  breaklines=false,
  breakautoindent=false,
  frame=lrtb,
  tabsize=2,
  showstringspaces=false,
  basicstyle=\tiny\ttfamily\singlespacing,
  keywordstyle=\bfseries\color{blue!80!black},
  commentstyle=\itshape\color{green!50!black},
  identifierstyle=\color{black},
  stringstyle=\color{red}
}
\lstdefinestyle{javasrc}{
  language=Java,
  breaklines=false,
  breakautoindent=false,
  frame=lrtb,
  tabsize=2,
  xleftmargin=none,
  showstringspaces=false,
  basicstyle=\tiny\ttfamily\singlespacing,
  keywordstyle=\bfseries\color{blue!80!black},
  commentstyle=\itshape\color{green!50!black},
  identifierstyle=\color{black},
  stringstyle=\color{red}
}
\lstdefinestyle{txt}{
  breaklines=false,
  breakautoindent=false,
  frame=n,
  showstringspaces=false,
  basicstyle=\scriptsize\ttfamily\singlespacing,
}
\newcommand\hl[1]{
	\btline{#1}{yellow!60}
}
\newcommand{\btline}[2]{
	\ifnum\value{lstnumber}=#1 \color{#2} \fi
}

\usepackage[ruled]{algorithm2e}

\SetAlFnt{\small}
\SetAlCapFnt{\small}
\SetAlCapNameFnt{\small}
\SetAlCapHSkip{0pt}
\IncMargin{-\parindent}
\begin{document}

\title{Modularizing Behavioral and Architectural Crosscutting Concerns in Formal Component-Based Systems\\{\normalsize Application to the Behavior Interaction Priority Framework}}

\author[fr]{Antoine El-Hokayem}
\ead{antoine.el-hokayem@univ-grenoble-alpes.fr}
\author[fr]{Yli\`es Falcone\corref{cor1}}
\ead{ylies.falcone@univ-grenoble-alpes.fr}
\author[aub]{Mohamad Jaber}
\ead{mj54@aub.edu.lb}

\address[fr]{Univ. Grenoble Alpes, CNRS, Inria, Grenoble INP, LIG, 38000 Grenoble, France}
\address[aub]{American University of Beirut, Beirut, Lebanon}
\cortext[cor1]{Corresponding author}

\begin{abstract}
We define a method to modularize crosscutting concerns in Component-Based Systems (CBSs) expressed using the Behavior Interaction Priority (BIP) framework. 
Our method is inspired from the Aspect Oriented Programming (AOP) paradigm which was initially conceived to support the separation of concerns during the development of monolithic systems.
BIP has a formal operational semantics and makes a clear separation between architecture and behavior to allow for compositional and incremental design and analysis of systems.
We distinguish local from global aspects.
Local aspects model concerns at the component level and are used to refine the behavior of components.
Global aspects model concerns at the architecture level, and hence refine communications (synchronization and data transfer) between components.
We formalize local and global aspects as well as their composition and integration into a BIP system through rigorous transformation primitives.
We present \aopbip{}, a tool for Aspect-Oriented Programming of BIP systems, demonstrate its use to modularize logging, security, and fault tolerance in a network protocol, and discuss its possible use in runtime verification of CBSs.
\end{abstract}

\maketitle

\section{Introduction} \label{sec:introduction}
A component-based approach~\cite{BasuBBCJNS11,Darwin, Szyperski} consists in building complex systems by composing components (building blocks).
Such approach confers numerous advantages (e.g., productivity, incremental construction, compositionality) that facilitate dealing with the complexity in the construction phase.
Component-based design is based on the separation between interaction and computation.
The isolation of interaction mechanisms allows for a global treatment and analysis on interactions between components even if local computations on components are not visible (i.e., when components are ``black boxes'').

A typical system consists of its main logic along with tangled code that implements multiple other functionalities. Such functionalities are often seen as secondary to the system.
For example, logging is not particularly related to the main logic of most systems, yet it is often scattered throughout multiple locations in the code.
Logging and the main code are separate domains and represent different \emph{concerns}.
A concern is defined in~\cite{czarnecki97} as a ``\emph{domain used as a decomposition criterion for a system or another domain with that concern}''.
Domains include logging, persistence, and system policies such as security.
Concerns are often found in different parts of a system, or in some cases multiple concerns overlap one region.
These are called \emph{crosscutting concerns}.
Aspect-Oriented Programming (AOP)~\cite{KiczalesLMMLLI97,AspectJ} aims at \emph{modularizing} crosscutting concerns by identifying a clear role for each of them in the system, implementing each concern in a separate module, and loosely coupling each module to only a limited number of other modules.
Essentially, AOP defines mechanisms to determine the locations of the concerns in the system execution by introducing \emph{joinpoints} and \emph{pointcuts}.
Then, it determines what to do at these locations by introducing \emph{advices}.
Finally, it provides a mechanism to coordinate all the advices happening at a location by introducing a process called \emph{weaving}.
%
%
\paragraph{Motivations and challenges}
In Component Based Systems (CBSs), crosscutting concerns arise at the levels of components~\cite{Duclos02,Lieberherr03} (building blocks) and architectures (communications).
Integrating crosscutting concerns in CBSs allows users to reason about crosscutting concerns in separation, and favors correct-by-construction design.
Defining an AOP paradigm for CBSs poses multiple challenges.
While the execution of a sequential program can be seen as a sequence of instructions, the semantics of a CBS is generally more complex and relies on a notion of architecture imposing several constraints on their execution.
As a consequence of these constraints, AOP matching, instrumentation and modifications need to be extended to account for the architecture (i.e., data transfer and rendez-vous between components).
Multiple approaches~\cite{PessemierSDC08,Fractal,Duclos02,Lieberherr03,SAFRAN} have sought to apply AOP for CBSs.
However they have not formalized the AOP concepts in the context of CBSs, nor have they defined formal transformations and semantics that allow us to reason about the transformed systems rigorously.
%
%
\paragraph{Approach overview}
We formalize AOP for CBSs.
We rely on a general abstraction of CBS executions as traces, AOP is then concerned with matching segments of the trace and modifying them.
We identify two views for CBSs: local and global.
The local view is concerned with the behavior of a component, the component is seen as a white box and its internals are inspected.
The global view is concerned with the architecture of the system, i.e., the interaction between components and their interfaces, the components are seen as black boxes.
We formalize our approach by extending an existing formalism of CBSs: the Behavior Interaction Priority (BIP) framework~\cite{BasuBBCJNS11,FACS2014}.
The BIP component-based framework uses formal operational semantics.
BIP consists of: (1) Behavior which is handled by atomic components; (2) Interaction which describes the collaboration between the atomic components; (3) Priority determines which interaction to execute out of many.
Multiple formalizations for CBS exist such as BRIC~\cite{BRIC}, Pi-Calculus~\cite{PiCalc}, and Fractal~\cite{Fractal}.
The choice of BIP is mainly due to compatibility.
BIP models explicitly the two views~\cite{BIPAbstract}, behavior models the local view while interaction and priority model the global architectural view.
Furthermore, BIP can also be viewed as an architecture description language (ADL)~\cite{Darwin}, and is used for systems modeling~\cite{BasuBBCJNS11}.
In particular, an AADL model (which is a superset of ADL~\cite{AADLDEF}) can be transformed into a BIP model~\cite{AADL}.
In addition, BIP supports a full set of tools~\cite{BIPTools} for manipulating the BIP model, source-to-source transformations, model transformations, code generation, and compositional verification~\cite{DFinder}.
We implement our approach as a model-to-model transformation tool.
We transform an existing BIP model using an AOP description to a new BIP model that implements it.
For each view, we define pointcuts to target transitions and interactions as joinpoints, and modify them by appending additional computation before and after their execution.
Additionally, we allow advices to change the state of atomic components.
As such, we are able to implement logging, authentication, congestion control and fault tolerance to a simple network protocol (\secref{sec:evaluation:case}).
In \secref{sec:evaluation:rv}, we show the application to runtime verification~\cite{FalconeJNBB15,KiviluomaKM06}.
Our approach can also be used for various crosscutting concerns in CBSs such as testing, runtime enforcement~\cite{FalconeJ17} and monitor synthesis~\cite{JavaMOP}.
However, we note that our approach is intended to cover a basic set of advices and not all.
For example, we do not allow for a change in priorities (and thus the scheduling) between the interactions of components.
We also do not allow for the disabling of interactions.
These advices can be implemented by defining more transformations following our methodology.
%
%
\paragraph{Paper Structure}
We begin by presenting the concepts of the BIP framework and AOP in \secref{sec:bip} and \secref{sec:aop}, respectively.
In \secref{sec:aopcbs}, we formalize the identification and description of concerns in the context of BIP.
In general, concerns are expressed by determining their locations in the system, and their behavior at the given locations.
Based on the formalization of concerns, we determine the \emph{rules} that govern the integration of these concerns in a BIP system.
Therefore, given an initial BIP system, and a description of concerns, we transform it so as to include the desired concerns.
We distinguish and define two types of aspects: \emph{Global} and \emph{Local}, in Sections~\ref{sec:global} and~\ref{sec:local}, respectively.
Global aspects are used to model crosscutting concerns at the architecture level and are thus useful to refine communications (synchronization and data transfer) between components.
Local aspects are used to model crosscutting concerns within components and are thus particularly useful to refine the behaviors of components.
Moreover, in \secref{sec:containers}, we define the notion of aspect container which serves as a construct for grouping aspects.
We discuss weaving strategies of aspects and their integration into a BIP system.
Moreover, we define a high-level language for writing local, global aspects, and aspect containers.
Our framework is fully implemented in \aopbip{} and tested on a network protocol refined to add several crosscutting concerns: logging, authentication, congestion control and fail-safe (\secref{sec:evaluation}).
Furthermore, we discuss monitoring CBSs with our approach, since runtime verification is a crosscutting concern.
Finally, we present related work in \secref{sec:related}, then draw conclusions and present future work in \secref{sec:conclusion}.

This paper revises and significantly extends a paper that appeared in the proceedings of the 14th international conference on Software Engineering and Formal Methods (SEFM 2016)~\cite{AOPSEFM}.
The additional contributions can be summarized as follows.
First, we elaborate the general overview of our approach, and explain the applicability of our approach to other formalizations of CBSs (\secref{sec:aopcbs}).
Second, we present the full description of local aspects (\secref{sec:local}) designed to target the local view.
Third, we define the strategies for weaving aspects (\secref{sec:containers}), by introducing containers and weaving procedures.
Fourth, we extend the experimental work to include a case study on using our approach for runtime verification of CBSs (\secref{sec:evaluation:rv}).
Finally, we improve the presentation and readability by adding more details and examples, elaborating on the views (\secref{sec:aop_cbs:overview}), and improving on the notation.
%
%

\section{Behavior Interaction Priority (BIP)} \label{sec:bip}
Behavior Interaction Priority (BIP)~\cite{BasuBBCJNS11,FACS2014} allows to define systems as sets of atomic components with prioritized interactions.
We present components, interactions, priorities, and their composition.
First, we provide definitions for the construction of BIP systems, and then illustrate them with Example~\ref{ex:bip:system}.

We begin by describing the \emph{update function}.
An update function abstracts code execution, which may modify the state of the system by reading and writing to variables.
\begin{definition}[Update function]
\label{def:tfunc}
An update function $\fvar{F}$ over a set of variables $X$ is a sequence of assignments $x_1 := \mathit{expr}_{X_1} \cdots x_n := \mathit{expr}_{X_n} $ such that $ \forall i \in [1,n]: x_i \in X$, and $\mathit{expr}_{X_i}$ is an expression using variables in set $X_i$, for $ i \in [1,n] $.
\end{definition}
The set of all sequences of update functions is denoted by $\codes$.
Furthermore, for $F = x_1 := \mathit{expr}_{X_1} \cdots x_n := \mathit{expr}_{X_n}$, we use $\varread(F)$ and $\varwrite(F)$ to denote variable the \emph{read}  and \emph{modified} variables, i.e., $\cup_{i \in [1,n]} X_i$ and $\cup_{i \in [1,n]} \{ x_i \} $, respectively.
Such sets can be obtained using a simple and standard syntactic analysis of the update function.
Moreover, for two update functions $F_1$ and $F_2$, we note $F_1 \cdot F_2$, update function formed by the concatenation of $F_1$ and $F_2$ (noted $F_1 \ F_2$ to lighten the notation at places).

An atomic component is the basic computation unit.
It is defined by its interface (i.e., a set of ports) and behavior defined as a Labeled Transition System (LTS) extended with data.
Transitions are labeled with update functions, guards, and ports.
Ports define communication and synchronization points for components.
A port can be associated with some variables (of the component), to exchange data with other components.
Ports are said to be exported by the component as they define its interface.

\begin{definition}[Port]
\label{def:port}
A port $\tuple{p, X_p}$ is defined by an identifier $p$ and a set of attached local variables $\fvar{X_p}$ (denoted by $\varsof{p}$).
\end{definition}

\begin{definition}[Atomic component]
\label{def:atomic}
An atomic component is a tuple $\tuple{P,L,T,X}$ s.t.:
\begin{itemize}
  \item $\fvar{P}$ is a set of ports;
  \item $\fvar{L}$ is a set of control locations;
  \item $\fvar{X}$ is a set of variables such that $\bigcup_{p \in P} \varsof{p} \subseteq X$;
  \item $T = L \times P \times \mathbb{B}[X] \times \codes  \times L$ is the set of transitions, where $\mathbb{B}[X]$ (resp. $\codes$) is the set of boolean predicates (resp. update functions) over $X$.
\end{itemize}
\end{definition}
In a transition $\tau = \tuple{\ell, p_\tau, g_\tau, F_\tau, \ell'} \in T$, (1) $\ell$ is the source location; (2) $\fvar{\ell'}$ is the destination location; (3) $\fvar{p_\tau}$ is the port exported by the component; (4) $\fvar{g_\tau}$ is the guard (a boolean predicate), a boolean function over $\fvar{X}$; (5) $\fvar{F_\tau}$ is an update function over $\fvar{X}$.
For a component $\fvar{B} = \tuple{P, L, T, X}$, we denote $\fvar{P}$, $\fvar{L}$, $\fvar{T}$, $\fvar{X}$, by \locsof{B}, \portsof{B}, \transof{B}, \varsof{B}, respectively.
Additionally, we denote by $\atoms$ the set of all atomic components.
Furthermore, for a transition $\tau = \tuple{\ell, p, g, F, \ell'}$, we denote $\fvar{\ell}, \fvar{p}, \fvar{g}, \fvar{F}, \fvar{\ell'}$ by $\fvar{\tau.\source}$, $\fvar{\tau.\port}$, $\fvar{\tau.\guard}$, $\fvar{\tau.\func}$, $\fvar{\tau.\dest}$, respectively.
We denote by $\varguard(g_\tau)$ the set of variables appearing in the expression defining $g_\tau$.

The semantics of an atomic component $\fvar{B}$ is defined as an LTS.
A state of the LTS consists of a location $\ell$ and valuation $\fvar{v}$ of the variables of $\fvar{B}$ where a valuation is a function from the variables of the component to a set of values.
First, we define how an update function modifies a valuation.
For an update function $F$ and a valuation $v$, executing $F$ on $v$ yields a new valuation $v'$, noted $v' = F(v)$, such that $v'$ is obtained in the usual way by the successive applications of the assignments in $F$ taken in order and where the right-hand side expressions are evaluated with the latest constructed temporary valuation.
Moreover, for two valuations $v$ and $v'$, $v'/v$ denotes the valuation where values in $v'$ have priority over those in $v$.

A transition $\tuple{\ell, p[X_p], g_\tau, F_\tau, \ell'}$ is possible to a new state $\tuple{\ell', v'}$ iff $B$ has a transition $\tau = \tuple{\ell,p[X_p], g_\tau, F_\tau, \ell'} \in \transof{B}$ such that: (1) the guard before receiving the new valuation $v_p$ of the port variables holds, i.e., $g_\tau(v) = \btrue$, and (2) the application of the update function $F_\tau(v_p/v)$ yields $v'$.
A transition is labeled with port along with valuation of its variables $\fvar{v_p}$, which is possibly received from other components.

\begin{definition}[Semantics of an atomic component]
\label{def:semantics-atom}
	The semantics of an atomic component $B$ is the LTS $S_B = \tuple{\locsof{B} \times \valuations, \portsof{B} \times \valuations, \rightarrow}$, where: \\
$\rightarrow = \{\tuple{\tuple{\ell,v}, p(v_p),\tuple{\ell',v'}} \mid$ $
\exists \tau = \tuple{\ell, p[X_p], g_\tau, F_\tau, \ell'} \in \transof{T}: g_\tau(v) \land v' = F_\tau(v_p/v)\}$;
and, $\valuations$ denotes the set of possible valuations of the variables in $\locsof{B}$.
\end{definition}
Furthermore, we say that a port $p$ is enabled in a state $\tuple{\ell, v}$, if there exists at least one transition $\tau$ from $\ell$ labeled by $p$ and its guard $g_\tau(v)$ holds.
Interactions serve as the glue that coordinates (i.e., synchronization and data transfer)  the components through their ports.
We consider $\atoms = \setof{B_1, \hdots, B_n}$ a set of atomic components where the semantics of $\fvar{B_i}$ is $\fvar{S_{Bi}} = \tuple{Q_{Bi}, P_{Bi}, \rightarrow}$, $i \in [1,n]$, and a set of interactions $\gamma$.
An interaction consists of one or more ports of different atomic components, a guard on the variables
of its ports, an update function that realizes data transfer between the ports.

\begin{definition}[Interaction]
\label{def:interaction}
An interaction $\fvar{a} \in \gamma$ is a tuple $\tuple{P_a,F_a,G_a}$ s.t.:
\begin{itemize}
	\item $\fvar{P_a} \subseteq \bigcup_{B \in \atoms} (\portsof{B})$  is a set of ports including at most one port per atomic component, i.e., $\forall B \in \atoms : |\portsof{B} \cap P_a| \leq 1$.
	\item $\fvar{F_a}$ is an update function executed with the interaction such that
	\[
	\left( \varread(\fvar{F_a}) \cup \varwrite(\fvar{F_a}) \right) \subseteq \bigcup_{p_i \in P_a} (\varsof{p_i}).
	\]
	\item $\fvar{g_a}$ is a boolean expression, the guard of the interaction.
\end{itemize}
\end{definition}
For an interaction $\fvar{a}$, we denote $\fvar{P_a}, \fvar{g_a}, \fvar{F_a}$, as $\portsof{a}$, $\fvar{a.\guard}$, $\fvar{a.\func}$, respectively.

A composite component is defined by composing atomic components using glue consisting of interactions and priorities.
\begin{definition}[Semantics of a composite component]
\label{semantics-composite}
The semantics of the composite component built with $\atoms$ and $\gamma$ (noted $\gamma(\atoms)$) is the LTS $\tuple{Q, \gamma, \rightarrow}$ where $Q = Q_{B_1} \times Q_{B_2} \times \hdots \times Q_{B_n}$, and
$\rightarrow$ is the least set of transitions satisfying the following rule:
\begin{align*}%
	I \subseteq [1,n] \hspace{18pt} \fvar{a} = (\setof{p_i}_{i \in I}, g_a, F_a) \in \gamma \hspace{18pt} g_a(\setof{v_{p_i}}_{i\in I}) \\
	\inferrule
		{\forall i \in I: q_i \xrightarrow{p_i(v_i)}_i q_i' \land v_i = g_a^i(\setof{v_{p_i}}_{i \in I}) \hspace{18pt} \forall i \not\in I: q_i = q_i' }
		{\tuple{q_1, \hdots, q_n} \xrightarrow{a} \tuple{q_1', \hdots, q_n'}}
\end{align*}
where $\fvar{v_{p_i}}$ is the valuation of the variables attached to port $\fvar{p_i}$ and $\fvar{F_a^i}$ is the partial update function derived from $\fvar{F_a}$ restricted to the variables of $\fvar{p_i}$.
\end{definition}
An interaction $\fvar{a}$ is enabled iff its guard $\fvar{g_a}$ holds and all of its ports are enabled.
An enabled interaction is selected from all interactions, based on the current states of the atomic components.
The BIP engine selects one of the enabled interactions and executes its update function $\fvar{F_a}$, which may modify its port variables.
Then, the involved atomic components (with indices in set $I$) execute their corresponding transitions given the new valuations $\fvar{v_i}$ received by the selected ports.
In the following, we consider a composite component $\composite = \gamma(\atoms)$ with behavior $\tuple{Q, \gamma, \rightarrow}$.



Multiple interactions can be enabled in a configuration.
Priorities are used to filter the enabled interactions and reduce non-determinism.
\begin{definition}[Priority]
A priority model $\pi$ over $\composite$ is a strict partial order on the set of interactions $\gamma$.
We abbreviate $\tuple{a, a'} \in \pi$ by $a \prec_\pi a'$.
Adding $\pi$ to $\composite$ results in a new composite component $\composite' = \pi(\composite)$ which semantics is the LTS $\tuple{Q, \gamma, \rightarrow_\pi}$ where $\rightarrow_\pi$ is the least set of transitions satisfying the following rule:
\[
  \inferrule{q \xrightarrow{a} q' \quad \quad \neg(\exists a' \in \gamma, \exists q'' \in Q : a \prec_\pi a' \land q \xrightarrow{a'} q'')}{q \xrightarrow{a}_\pi q'}
\]
\end{definition}
Whenever according to $\pi$ an interaction $\fvar{a} \in \gamma$ is selected, there does not exist an enabled interaction in $\gamma$ which has higher priority than $\fvar{a}$.

A composite component, obtained by the composition of a set of atomic components, can be composed with other components (composite or atomic) in a hierarchical and incremental fashion using the same operational semantics.
Our method can be applied to a hierarchical system using two different approaches: (1) by flattening the system into one composite component, and (2) using scoping rules to target individual composite components.
Without loss of generality, in the first approach, we flatten a hierarchical composite component to obtain a non-hierarchical one (i.e., consisting only of atomic components and simple interactions) using the method presented in~\cite{sourcetosource}.
The method ensures the existence of a mapping between the hierarchical to the non-hierarchical component: all transformations applied to the flattened component can be mapped to the original system.
The non-hierarchical composite component resulting from the flattening is subsequently referred to as the \emph{BIP model}.
The BIP model is a single closed composite component, it has no ports that are exposed to an external entity.
The second approach uses syntax directives to determine the scope of the transformations.
A composite component is a combination of other composite components or atomic components, both behave similarly at the interactions level (i.e., they have a similar interface through ports).
As such, it is possible to simply restrict the scope to a given composite component, and treat it as a global system, using scoping syntax to determine the targeted interactions at a given level in the hierarchy.
In this approach, a composite component is seen as a grey box, where the interactions and interfaces of components that form it are available, and one can iterate over the sub-components as necessary.
A BIP system is the instantiation of a BIP model, it defines the initial locations and variable initialization of atomic components.
\begin{definition}[BIP system]
\label{def:bipsystem}
A BIP system is a tuple $\tuple{\composite, q_0}$, where $q_0 = \tuple{\mathit{Init}, v}$ is the initial state with $\mathit{Init} \in \locsof{B_1} \times \hdots \times \locsof{B_n}$ being the tuple of initial locations of atomic components, and $v \in \valuations^{\mathit{\mathit{Init}}}$ is the tuple formed by the initial valuations of all variables in atomic components $X^{\mathit{Init}} \subseteq \bigcup_{B \in \atoms} (\varsof{B})$.
\end{definition}

	\begin{figure}[t!] %
		\centering %
		\scalebox{0.7}{\begin{tikzpicture}[bip]
\tikzset{
  myloop/.style={loop, looseness=2, min distance=10mm},
  casestudy/.style={minimum height=7cm},
	v1/.style  = {},
	v2/.style  = {},
}
   \componentb{C1}{atomic, minimum height=4.5cm}{}{
      \node[place,double] (C1_IDL) {$\mathrm{IDL_1}$};
      \node[place, below=of C1_IDL] (C1_REP) {SND};
      \tconnect[C1_recv]{normal,bend left}{C1_IDL}{C1_REP}{0.2cm}{right}{$\btrue$}{$send_1$}{\ffunc} 
      \tconnect[C1_ack] {normal,bend left}{C1_REP}{C1_IDL}{0.2cm}{left }{$\btrue$}{$recv_1$}{$\ffunc{p_1 := \mathrm{gen}()}$} 
  }{Ping};
	\node[export,v1,yshift=-.9cm] (C1_send) at (C1.north east) [label=left:$send_1$] {};
	\node[export,v1,below=0.6cm of C1_send] (C1_recv) [label=left:$recv_1$] {};
  \node[var,v1] (C1_p) at ( $(C1.south west) + (0.6,0.6)$) []  {$x_1$};

  \componentb{C}{atomic, minimum height=4.5cm, right=4.0cm of C1}{}{
      \node[place,double] (C_IDL) {$\mathrm{IDL_2}$};
      \node[place, below=of C_IDL] (C_REP) {REP};
      \tconnect[C_recv]{normal,bend left}{C_IDL}{C_REP}{0.2cm}{right}{$\btrue$}{$recv_2$}{$\ffunc{p_2 := \mathrm{ack}(p_2)}$} 
      \tconnect[C_ack] {normal,bend left}{C_REP}{C_IDL}{0.2cm}{left }{$\btrue$}{$send_2$}{\ffunc} 
  }{Pong};
	\node[export,v1,yshift=-.9cm] (C_recv) at (C.north west) [label=right:$recv_2$] {};
	\node[export,v1,below=0.6cm of C_recv] (C_ack) [label=right:$send_2$] {};
  \node[var,v1] (C_p) at ( $(C.south east) + (-0.6,0.6)$) []  {$x_2$};

  \draw[-] (C1_send) 
            -- ($(C1_send.east)!.5!(C_recv.west)$) node [above] {\ifunc{$\btrue$}{$\ffunc{recv_2.p_2 := send_1.p_1}$}} 
            -- (C_recv);
  \draw[-] (C1_recv) 
            -- ($(C1_recv.east)!.5!(C_ack.west)$) node [below] {\ifunc{$\btrue$}{$\ffunc{recv_1.p_1 := send_2.p_2}$}} 
            -- (C_ack);

\end{tikzpicture}} %
		\caption{Two communicating agents} %
		\label{fig:bipsys} %
	\end{figure}
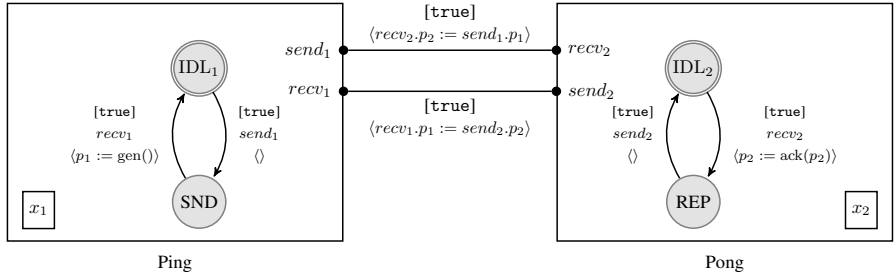 %

\begin{example}[BIP system]\label{ex:bip:system}
\rfigb{fig:bipsys} depicts a BIP system composed of two atomic components: \code{Ping} and \code{Pong}.
The \code{Ping} component has one variable $\fvar{x_1}$ initialized to a random number and two locations $\code{IDL_1}$ and \code{SND}, and two ports $\code{send_1}$ and $\code{recv_1}$.
We associate the variable $\fvar{x_1}$ with both $\code{send_1}$ and $\code{recv_1}$.
Initially, the \code{Ping} and \code{Pong} components are at the $\code{IDL_1}$ and $\code{IDL_2}$ locations, respectively.
From location $\code{IDL_1}$, in component \code{Ping}, port $\code{send_1}$ is enabled, since the guard of the transition from $\code{IDL_1}$ to \code{SND} holds.
Similarly the transition from $\code{IDL_2}$ to \code{REP} in \code{Pong} is possible, and $\code{recv_2}$ is enabled.
The interaction that has both ports $\code{send_1}$ and $\code{recv_2}$ enabled, and its guard holding, is now enabled.
Since no other interaction is enabled, it executes.
Its update function executes the data transfer using the ports $\code{send_1}$ and $\code{recv_2}$ and their associated variables $\fvar{x_1}$ and $\fvar{x_2}$.
Then, the update function of each transition executes, generating the acknowledgment packet in \code{Pong}.
\code{Ping} will move to location \code{SND} and \code{Pong} to \code{REP}.
Similarly, on the next step, the acknowledgement is sent back to \code{Ping}  and it generates a new number.
The two interactions ensure synchronization between the two components.
\end{example}

We abstract the execution of a BIP system as a BIP trace.
The trace of a BIP system consists of the state changes and interactions taken.
\begin{definition}[BIP trace]
\label{def:trace-global}
A BIP trace $\rho = (q_0 \cdot a_0 \cdot q_1 \cdot a_1 \cdots q_{i-1} \cdot a_{i-1} \cdot q_i)$ is an alternating sequence of states of $S$ and interactions in $\gamma$; where $q_k \xrightarrow{a_k} q_{k+1} \in {\rightarrow}$, for $k \in {[0, i-1]}$.
\end{definition}
The trace of a BIP system consists of the state changes and interactions executed.
\begin{example}[BIP Trace] \label{ex:bip:trace}
	Recall the system presented in \exref{ex:bip:system} consisting in the two components \code{Ping} and \code{Pong}.
	The first change in the state results in the following trace:
	\[
  \tuple{\tuple{\code{IDL_1}, \tuple{1}}, \tuple{\code{IDL_2}, \tuple{0}}}
  \cdot \setof{\code{send_1}, \code{recv_2}}
  \cdot \tuple{\tuple{\code{SND}, \tuple{1}}, \tuple{\code{REP}, \tuple{2}}}
	\]
	The above trace shows the alternations of locations and valuations of the components, and the interaction that is only represented by its ports for clarity.
	Initially \code{Ping} is in $\code{IDL_1}$ and \code{Pong} is in $\code{IDL_2}$, with the variables $\fvar{x_1}$ and $\fvar{x_2}$ being initialized to $1$ and $0$ respectively.
	We use a shorthand of the interaction for clarity referring only the ports, in which case it is the interaction that consists of the set $\setof{\code{send_1}, \code{recv_2}}$.
	Then, we show the resulting state after executing the interaction.
	\code{Ping} (resp. \code{Pong}) moves to the location \code{SND} (resp. \code{REP}), and the values for $\fvar{x_1}$ and $\fvar{x_2}$ are respectively $1$ and $2$.
	The value of $p_2$ has been first set to $1$ through the interaction, then to $2$ as the result of applying $\mathrm{ack}(1)$ in the transition.
\end{example}

\section{Aspect-Oriented Programming (AOP)} \label{sec:aop}
%
The implementation of crosscutting concerns mentioned in the introduction leads to two typical problems: \emph{scattering} and \emph{tangling}~\cite{Lieberherr03}.
\begin{itemize}
	\item \emph{Tangling} happens when concerns overlap in one region of the program.
Consequently, enforcing one concern may affect others.
	\item \emph{Scattering} is the dual situation of tangling.
It happens when one concern is spread across different regions of the program.
Scattering concerns go against encapsulation. Developers have to manually keep track of the location of a specific concern in multiple areas of the system.
\end{itemize}
\begin{figure}[t]
	\centering
	\scalebox{0.65}{\newcommand{\concern}[1]{\texttt{\color{blue}{#1}}}
\begin{tikzpicture}
	\begin{class}[text width=6cm]{Account}{0,0}
		\operation{\concern{----} + getOwner() : User}
		\operation{\concern{---S} + getBalance() : Dollars}
		\operation{\concern{PL-S} + deposit(amt : Dollars)}
		\operation{\concern{PL-S} + withdraw(amt : Dollars)}
	\end{class}
	\begin{class}[text width=10cm]{AccountController}{10,0}
		\operation{\concern{PL-S} + wire(from : Account, to : Account, amt : Dollars) : bool}
		\operation{\concern{PL-S} + close(acct : Account) : bool}
		\operation{\concern{PL-S} + open(user : User, balance : Dollars) : Account}
		\operation{\concern{-LC-} + list(user : User) : Account[ ]}
	\end{class}
	\begin{class}[text width=10cm]{UserController}{10,-3.5}
		\operation{\concern{PL-S} + create(data : UserData) : User}
		\operation{\concern{--C-} + get(code : Integer) : User}
		\operation{\concern{--C-} + find(name : String) : User[ ]}
		\operation{\concern{PL-S} + block(user : User) : bool}
	\end{class}
	\begin{class}[text width=6cm]{User}{0, -3.5}
		\operation{\concern{--CS} + getInfo() : UserData}
		\operation{\concern{----} + getCode() : Integer}
		\operation{\concern{-L--} + getLastActivity() : Date}
	\end{class}
\end{tikzpicture}}\\
	{\footnotesize\emph{P: Persistence  L: Logging  C: Caching  S: Security Policy}}
	\caption{Multiple concerns in a simple system}
	\label{fig:crosscuts}
\end{figure}
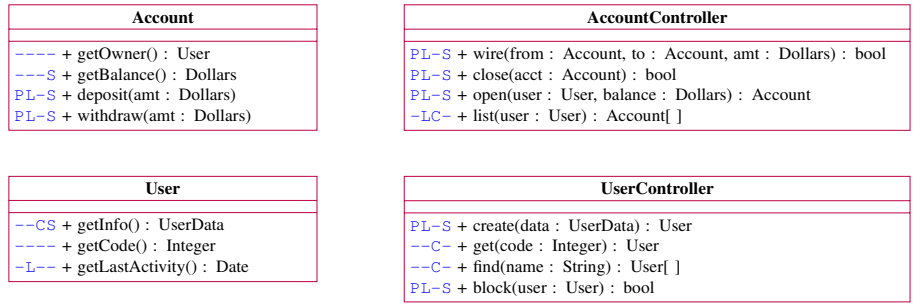
In the following example, we illustrate the above two problems on an example.
\begin{example}[Crosscutting concerns]\label{ex:aop:cross}
Figure \ref{fig:crosscuts} illustrates four different: logging, caching, persistence and security policy.
A class diagram describing the classes' main methods is presented, we omitted to describe their relationships for clarity.
The class diagram methods are prefixed with the four concerns as flags.
If a method has a concern then some code related to the logic of that concern is included in the method.
For example, the method \code{Account.withdraw} has three \emph{tangled} concerns: persistence, logging, and policy.
Thus, the method \code{withdraw} has to include code for persistence, logging, and logic.
This code enforces the policy in addition to its own main logic.
The policy concern is scattered across all four classes, hence maintaining it requires one to modify all four classes when a change is needed.
\end{example}
The purpose of Aspect-Oriented Programming (AOP) is to localize crosscutting concerns in an aspect.
An aspect is defined in~\cite{Kiczales01} as ``\emph{a well-modularized implementation of a crosscutting concern}''. These concerns are separated from the main program logic and contained in separate logical units.
One example of separation of concerns is achieved by AspectJ~\cite{Kiczales01}, which is an aspect-oriented extension to the Java programming language.
A \emph{joinpoint} is a well-defined point in program execution where a concern needs to be handled.
It acts as a reference point to coordinate the behavior of multiple concerns.
A \emph{pointcut} refers to a set of \emph{joinpoints} and execution context information.
Basic pointcuts can be composed and identified so as to increase re-usability.
Pointcuts are the syntactic elements used to select joinpoints.
A pointcut specifies a function signature, a variable name, and a module that needs to be matched.
Furthermore, pointcuts are able to specify dynamic execution constraints, such as a function being invoked while inside another function (e.g., \code{cflow} pointcut in AspectJ).
A pointcut regulates scattering by describing the joinpoints needed to implement the concern.
An \emph{advice} defines the additional behavior to be executed at each specific joinpoint selected by a pointcut.
An \emph{aspect} serves as the modular unit that encapsulate advices, pointcuts, and additional behavior.
Furthermore, aspects may introduce their own variables, methods, and fields.
This is referred to as inter-type declarations.
The term inter-type designates the fact that these extra objects and code are accessible in different types (based on the matching joinpoints).
The main task of an AOP language implementation is to coordinate the execution of the non-aspect code with the aspect code.
This coordination has to ensure a correct execution at the joinpoint of both primary and secondary concerns.
This process is called \emph{weaving} and can be done at compile-time, load-time, or run-time.

\hspace*{1em}
\scalebox{.9}{
\begin{lstlisting}[style=bip, label={lst:aspect},caption={Logging concern}, numbers=left, frame=L, language=java, linewidth=.4\textwidth]
public aspect Logging {
	private static Logger logger = Logger.getLogger(Logging.class.getName());
	pointcut log() :
	call(void Account.deposit(Dollars))
		|| call(void Account.withdraw(Dollars))
		|| call(* AccountController.*(*))
		|| call(Date User.getLastActivity())
		|| call(User UserController.create(UserData))
		|| call(bool UserController.block(User)

	after() returning(Object res) : log() {
		logger.info(thisJoinPoint.getSignature().toShortString()
			+ Arrays.toString(thisJoinPoint.getArgs())
			+ " -> " + res);
	}
}
\end{lstlisting}}

\begin{example}[Logging concern]
Listing \ref{lst:aspect} implements parts of the logging concern shown in Example~\ref{ex:aop:cross} using AspectJ.
In the case of logging, the inter-type variable is a \code{Logger} object (Line 2).
The pointcut expression (lines 4-9) specifies the various method invocations to be intercepted and names the pointcut \code{log}.
The advice implements the logging code.
It consists of code necessary to (i) capture the arguments of the method invoked  using the \code{getArgs()} method on the \code{thisJoinPoint} object (ii) capture the return value of the method invoked, and (iii) pass it to the logger.
The advice is set to trigger after the pointcut (line 11), in which case, it means after a method returns.
Effectively, for any of the methods defined in the pointcut, the logger will log the name of the method called, its arguments and return value.
\end{example}

\section{Overview and Preliminaries of Aspect-Oriented Programming for CBSs}
\label{sec:aopcbs}
Defining execution points for crosscutting concerns in a program implicitly relies on the programming paradigm.
In the case of AspectJ, the pointcuts are described using object-oriented terminology.
For example, a pointcut is defined by specifying object types, their methods (messages) with their arguments and return values, and the fields accesses of a given object.
As such the crosscutting concerns are described using object-oriented terminology.
To illustrate this point, consider the fact that it is not possible to intercept changes to local variables inside methods using AspectJ because of the encapsulation and data-hiding concepts that are specific to object-oriented design.
Similarly, when designing AOP pointcuts for CBSs, we restrict the pointcuts to what is relevant to component-based design.
%
\subsection{Component-Based AOP}
\label{sec:aopcbs:general}
%
Developing component-based systems is a process of progressively repeating the following two stages.
The first stage consists in building components that follow a certain interface.
The interface defines the behavior the component must implement.
In this stage, a component is a white-box, its internals are exposed to the component itself.
The second stage, components are composed using their interfaces to form a system (e.g. an architecture).
In this stage, a component is a black-box, its internals are not visible to the system, only its interface is.
As such for CBSs in general, it is important to handle crosscutting concerns that may arise in both stages.
Therefore, one can distinguish two views of CBSs, \emph{the local view and the global view}.
The local view is concerned with the component design itself (i.e. the first stage), and the global view is concerned with the interaction of components (i.e. the second stage).

The local view and global view are independent, in the sense that components must only implement interfaces but they can do so separately.
This separation effectively characterizes the joinpoints that one can reason about.
In the local view, we consider the state of the component itself and actions that modify the state.
In the global view, we examine the interaction between components by examining the passing of messages across the interfaces and synchronization.

Our notion of advice includes additional computation that executes before and after a given joinpoint.
In the case of local advices, we include the ability to change the location of the component.
The advice can require storage of additional state information.
As such, intertype variables are required for both the local and global views.
They need to be accessible for the advice and each of the levels separately.
Once woven into the system, advices can modify the system behavior.
While it is possible to modify the system arbitrarily, we define our notion of correctness.
In the scope of this paper, we consider the correctness of advice \emph{application}, i.e., we verify the advice has been placed correctly at the matched joinpoint.
%
\subsection{Overview of AOP for BIP}
\label{sec:aop_cbs:overview}
%
The BIP framework is endowed with formal semantics that describes both views.
Atomic components (\defref{def:atomic}) describe the internals of a given component and its implementation,
while the BIP model (\defref{def:bipsystem}) formalizes the composition of atomic components and their priorities.
We recall the distinction between three various notions.
First, the BIP model consists of the elements needed to represent a given system.
It is used to form the LTS that represents the behavior of the system.
Second, the BIP system is a runtime instantiation of the model.
And lastly, the BIP trace  contains the elements of the model that are executed by the LTS during runtime.
While the global BIP trace (\defref{def:trace-global}) contains all information, we consider restrictions so that we separate the internals of components from the interactions between components.
Using the information from each view, we define two types of joinpoints: global and local.
\paragraph{Global joinpoints}
In the global view, concerns are at the level of interactions.
Therefore, we focus on the interface of components.
This view evaluates concerns that crosscut interactions (i.e., ports, synchronization and data transfer).
The components export only their ports, on which interactions are defined.
Generally, each component computes its enabled ports.
The interactions that have all their ports enabled and their guard evaluating to $\btrue$, are said to be enabled.
The system executes the enabled interaction with the highest priority.
Therefore, at the interaction level, the following operations exist: interaction enablement and interaction execution.
We do not consider interaction enablement, since to inspect and instrument around it requires knowledge of the internals of components.
Therefore, it is not compatible with the global view.
Whenever an interaction executes, three kinds of global joinpoints can be identified: (1) synchronization between different atomic components (2) one or more atomic components sending data (3) one or more atomic components receiving data.
These three joinpoints are captured by the interaction: the ports define synchronization, and variables read or written define data transfer.
From a BIP trace (\defref{def:trace-global}), one can extract the sequence of executed interactions called the global trace.
\begin{definition}[Global Trace]
\label{def:events-global}
The global trace extracted from the BIP trace $\rho = (q_0 \cdot a_0 \cdot q_1 \cdots a_{i-1} \cdot q_{i-1} \cdot a_{i-1} \cdot q_{i} )$, noted  $T(\rho)$, is $(a_0 \cdot a_1 \cdots a_{i-1} )$.
\end{definition}
A global joinpoint is an interaction execution moving the system from a state to another state.
We suppress the state information from the BIP trace, as the atomic components are blackboxes for the global view.
For the rest of the paper, we consider $\events$ to be the set of all possible interaction executions in $\tuple{ \composite , q_0 }$ with $\rightarrow$, and fix an arbitrary BIP trace $\rho$.
We use two sets $\gamma$ and $\events$ to distinguish between the \emph{syntax} representing the interactions, and the actually \emph{executed} interaction, respectively.%
%
%
\paragraph{Local joinpoints}
In the local view, we focus on atomic components seen as white boxes and seek to refine their behavior.
We study the state of an atomic component to locate possible points where crosscutting concerns apply.
The behavior of the atomic component consists of an LTS that changes states when a transition is taken.
Therefore, we consider concerns are at the level of locations, states, transitions, guards and computations on transitions.
In order to study the local view, we must define first how an atomic component contributes to the global execution of the BIP system.
Since in this view we see components as white boxes, we have knowledge of the full BIP system and can extract a local execution trace for a given atomic component.
We fix an atomic component $B_k$ with semantics defined by the LTS $S_{B_k} = \tuple{Q_{B_k}, P_{B_k}, \rightarrow_{B_k}}$.
In order to define events local to a given atomic component, we first define the projection of the global state on a local component to extract its own state.
\begin{definition}[State projection]
The projection of a global state $q = \tuple{q_0, \hdots q_n}$ on a local component $B_k$ is defined as $q \stateproject_{B_k} = q_k$.
\end{definition}
An event local to a component is defined as a triple $\tuple{\tuple{l,v}, \tau, \tuple{l',v'}}$.
This event denotes that an atomic component has moved from the state with location $l$ and valuations $v$ to another state with location $l'$ and valuations $v'$ using a transition $\tau$.
Given a composite component $\composite$ composed of atomic components in a set $\atoms = \setof{ B_1, \ldots, B_n }$, and its BIP trace $\rho = (q_0 \cdot a_0 \cdot q_1 \cdot a_1 \cdots q_{i-1} \cdot a_{i-1} \cdot q_i)$, we say that $E_i = \tuple{q_i, a_i, q_{i+1}}$ is a global event.
A global event represents a global move in the BIP system from a state to another, after executing an interaction.
\begin{definition}[Event projection]
\label{def:trmap}
We project a global event $E_i$ to a local event $e_i$ in the atomic component $B_k$ by using $\trmap(E_i, B_k)$ where:
\[
	\trmap(E_i, B_k) = \left\{
    \begin{array}{ll}
      \tuple{\tuple{l_i,v_i}, \tau, \tuple{l_{i+1},v_{i+1}}} & \text{if } \tuple{l_i,v_i} = q_i \stateproject B_k \\
    &\land \tuple{l_{i+1},v_{i+1}} = q_{i+1} \stateproject B_k \land p = \tau.\port \\
		& \land\ p \in \portsof{a_i} \land  \tau.\guard(v_i) \\
					   &  \land \, v_p = F_a^p(\{v_{p_j}\}_{p_j \in a.ports})\\
					   &  \land \,  v_{i+1} = \mathit{\tau.\func}(v_p/v_i)	\\
      \epsilon & \text{otherwise}
    \end{array}
  \right.
\]
\end{definition}
The $\trmap$ function searches within the interaction $a_i$ executing globally for any ports in the atomic component that are both enabled and included in $a_i$.
If no ports are found, then other components are involved in $a$ and thus the global event does not concern the local component (and $\trmap$ returns $\epsilon$).
Otherwise, the function $\trmap$ projects both the global state before and after the interaction to the local component and set $\tau$ to be the interaction that enabled the port for the interaction to execute, and takes the local component from $\tuple{l_i,v_i}$ to $\tuple{l_{i+1},v_{i+1}}$.
For a local event $e_i = \tuple{\tuple{l_i,v_i}, \tau, \tuple{l_{i+1},v_{i+1}}}$  we denote $\tuple{l_i,v_i}, \tuple{l_{i+1},v_{i+1}}, l_i, l_{i+1}, v_i, v_{i+1}, \tau$ by $\mathrm{e.q}$, $\mathrm{e.q'}$,  $\mathrm{e.l}$,  $\mathrm{e.l'}$,  $\mathrm{e.v}$,  $\mathrm{e.v'}$, and $\mathrm{e.\tau}$, respectively.
We then extend function $\trproject$ to a sequence of global events: $\trproject(E_0 \cdots E_i, B_k) = \trmap(E_0, B_k) \cdots \trmap(E_i, B_k)$,  where $\epsilon$ is interpreted as the neutral element of concatenation (i.e., $E \cdot \epsilon = \epsilon \cdot E = E$).
In the sequel, we denote  $T_k$ the local trace of an atomic component $B_k$.
\begin{example}[Traces and views] \label{ex:traces}
	Table~\ref{table:views} shows traces associated with the views.
	We use the ports as shorthands to interactions and transitions for clarity.
	The BIP trace from Example~\ref{ex:bip:trace} is presented in the first row. We recall the two atomic components $B_0 = \code{Ping}$ and $B_1 = \code{Pong}$.
	In the second row, the global trace associated with the global view is extracted from the BIP trace, it only contains the interactions executed, in this case $\setof{\code{send_1},\code{recv_2}}$.
	From the BIP trace, we extract the following global event:
	\[
	E_0 = \tuple{
		\tuple{\tuple{\code{IDL_1}, \tuple{1}}, \tuple{\code{IDL_2}, \tuple{0}}},
		\setof{\code{send_1}, \code{recv_2}},
		\tuple{\tuple{\code{SND}, \tuple{1}}, \tuple{\code{REP}, \tuple{2}}}
	}
	\]
	The third row shows the local trace for the $\code{Ping}$ component ($B_0$).
	It is obtained as follows:
$
	\trmap(E_0, B_0) = \tuple{\code{IDL_1}, \tuple{1}} \cdot
	\code{send_1} \cdot
	\tuple{\code{SND}, \tuple{1}}
$.
	From the BIP trace, we extract only elements relevant to $B_0$.
	We know that the interaction $\setof{\code{send_1}, \code{recv_2}}$ is executed, which is associated with the transition with port $\code{send_1}$ in $B_0$.
\end{example}
\begin{table}[t]
	\caption{Traces According to Views}
	\label{table:views}
	\begin{tabular}{|r|l|}
		\hline BIP Trace ($\rho$) &
		 $\tuple{\tuple{\code{IDL_1}, \tuple{1}}, \tuple{\code{IDL_2}, \tuple{0}}}
		  \cdot \setof{\code{send_1}, \code{recv_2}}
		  \cdot \tuple{\tuple{\code{SND}, \tuple{1}}, \tuple{\code{REP}, \tuple{2}}}$ \\
		\hline Global Trace ($T$) &
		$\setof{\code{send_1}, \code{recv_2}}$\\
		\hline Local Trace ($T_0$) &
		$\tuple{\code{IDL_1}, \tuple{1}} \cdot
		\code{send_1} \cdot
		\tuple{\code{SND}, \tuple{1}}$\\
		\hline
	\end{tabular}
\end{table}
To handle the concerns arising at both the local and global views, we need to formally identify and select the joinpoints described for each view.
We leverage the operational semantics of a BIP model to associate joinpoints, pointcuts, and advices with the original model.
For the rest of the paper, we fix an arbitrary BIP-system $\tuple{\composite, q_0}$ where $\composite = \pi(\gamma(\atoms))$ is the BIP model with semantics $S = \tuple{Q, \gamma, \rightarrow}$.
In \secref{sec:global} and \ref{sec:local} we present the AOP concepts associated with the global and local views, respectively.
For each view, we define an homogeneous notion of execution points in order to define where concerns can arise (i.e., joinpoints), and then the syntax to select these points (i.e., pointcuts).
We recall that joinpoints represent points in the \emph{execution} of a BIP system.
Given a pointcut expression $\mathrm{pc}$ for a given view, we use the assertion ``$\mathrm{e} \vDash \mathrm{pc}$'' to indicate that an element $\mathrm{e}$ of a (local or global) trace matches pointcut $\mathrm{pc}$.
We define the legal actions at these execution points (i.e., advices) and how these actions are integrated into an existing model (i.e., weaving).
We allow advices to store additional state information through the creation of additional variables both at the local and global view.
These additional variables constitute the inter-type variables.
Furthermore for each view, we define the function $\mathrm{select}$, which returns the elements of the \emph{model} (syntax) that are required to instrument so that in the resulting system, whenever the assertion ($\mathrm{e} \vDash \mathrm{pc}$) holds during runtime, we execute the advice.
%
\subsection{Applicability to other CBS Frameworks}
\label{sec:aop_cbs:app}
%
While we use the BIP semantics for implementation, we note that our approach applies to other CBS frameworks.
In general, the program execution is abstracted as an execution trace on which AOP defines its matches and changes.
Given the two defined views, we require a general trace of the system that provides information about the interactions between components.
Then, knowing the internals of the system, we can project the global trace to a local trace, to determine the execution of the local component.
Verifying the correctness of AOP for CBS is done by analyzing the traces and ensuring that the joinpoints match the correct trace fragments, and that advices modify the traces accordingly and minimally (i.e., only when the joinpoints match).
While our approach relies on the BIP semantics for the instrumentation of aspects, one could adapt it to any formal framework that defines the semantics of the behavior of a component, and the semantics of the interactions between components.
%

%

%
\section{Global Aspects} \label{sec:global}
Following the general requirements of the AOP approach for CBSs described in \secref{sec:aop_cbs:overview}, we now address the concerns arising in the global view, namely the view where components are black boxes and only the interactions are visible.
For this section, we consider the composite component $\composite = \pi(\gamma(\atoms))$ as the BIP model with a BIP trace $\rho$.
%
\subsection{Global Pointcuts}
\begin{figure}[tbp]%
	\centering
	\begin{subfigure}[b]{0.49\textwidth}%
	\centering%
		\scalebox{0.67}{\begin{tikzpicture}
\node[rectangle, draw, auto, align=center](pframe){
\begin{tikzpicture}[bip]
	 
    \componentb{A}{composite}{}{}{$A$}
    \componentb{B}{composite,above=of A}{}{}{$B$}
   	\componentb{C}{composite,above=of B}{}{}{$C$}
   	\componentb{D}{composite,right= \bipcomposite of B}{}{}{$D$}

	\node[export] (A_p1) at (A.west) [label=right:$pa_{1}$] {};

    \node[export] (B_p1) at (B.west) [label=right:$pb_{1}$] {};
    \node[export] (B_p2) at (B.east) [label=left: $pb_{2}$]  {};

    \node[export] (C_p1) at (C.west) [label=right:$pc_{1}$] {};
    \node[export] (C_p2) at (C.east) [label=left: $pc_{2}$] {};
	
	\node[export] (D_p1) at (D.west)  [label=right:$pd_1$] {};
	\node[export] (D_p2) at (D.north) [label=below:$pd_2$] {};

	\node[var] (A_x) at ( $(A.east) + (-0.4,0.0)$) []  {$x_a$};
	\node[var] (B_x) at ( $(B.center) + (0.0,0.0)$) []  {$x_b$};
	\node[var] (D_x) at ( $(D.east) + (-0.4,0.0)$) []  {$x_d$};


	\tconnector[selected]{A_p1.west}{B_p1.west}{-0.5cm,0cm}{left}{}{}
	\tconnector[selected]{B_p1.west}{C_p1.west}{-0.5cm,-1.5cm}{left}{\tilabel{$g_0$}{$\ffunc{x_b := x_a}$}}{i0}

	\tconnector[selected]{B_p2.east}{D_p1.west}{0,0}{yshift=-0.7cm,below}{\tilabel{$g_1$}{$\ffunc{x_d := x_d + x_b}$}}{i1}

	\tconnector{C_p2.east}{D_p2.north}{1.85cm,0}{right}{\tilabel{$g_2$}{$f_2$}}{i2}
	\tconnectorgap[selected]{C_p2.east}{B_p2.east}{0.3,0.3}{right}{\tilabel{$g_3$}{$\ffunc{x_b := 0}$}}{i3}

	\node[ilabel,above =of i0]  {$a_0$};
	\node[ilabel,above =of i1]  {$a_1$};
	\node[ilabel,right=of i2]  {$a_2$};
	\node[ilabel,right=of i3]  {$a_3$};

\end{tikzpicture}
};
\end{tikzpicture}}
		\caption{Matching} %
		\label{fig:global-aspect-match}%
	\end{subfigure}%
	\begin{subfigure}[b]{0.49\textwidth}%
	\centering%
		\scalebox{0.67}{\begin{tikzpicture}
\node[rectangle, draw, auto, align=center](pframe){
\begin{tikzpicture}[bip]
	 
    \componentb{A}{composite}{}{}{$A$}
    \componentb{B}{composite,above=of A}{}{}{$B$}
   	\componentb{C}{composite,above=of B}{}{}{$C$}
   	\componentb{D}{composite,right= \bipcomposite of B}{}{}{$D$}
   	\componentb{AOP}{composite, created,below=of D}{}{}{$B_{\setof{v_0,v_1}}$}

	\node[export] (A_p1) at (A.west) [label=right:$pa_{1}$] {};

    \node[export] (B_p1) at (B.west) [label=right:$pb_{1}$] {};
    \node[export] (B_p2) at (B.east) [label=left: $pb_{2}$]  {};

    \node[export] (C_p1) at (C.west) [label=right:$pc_{1}$] {};
    \node[export] (C_p2) at (C.east) [label=left: $pc_{2}$] {};
	
	\node[export] (D_p1) at (D.west)  [label=right:$pd_1$] {};
	\node[export] (D_p2) at (D.north) [label=below:$pd_2$] {};

	\node[export,created, yshift=0cm] (AOP_p) at (AOP.west) [label=below left:$p_{\setof{v_0,v_1}}$] {};

	\node[var] (A_x) at ( $(A.east) + (-0.4,0.0)$) []  {$x_a$};
	\node[var] (B_x) at ( $(B.center) + (0.0,0.0)$) []  {$x_b$};
	\node[var] (D_x) at ( $(D.east) + (-0.4,0.0)$) []  {$x_d$};

	\node[var, created] (v0) at ( $(AOP.south west) + (0.8,0.5)$) []  {$v_0$};
	\node[var, created] (v1)  [node distance=0.2cm,right=of v0]  {$v_1$};


	\tconnector[]{A_p1.west}{B_p1.west}{-0.5cm,0cm}{left}{}{}
	\tconnector[]{B_p1.west}{C_p1.west}{-0.5cm,-1.5cm}{left}{\tilabel{$g_0$}{$f_0$}}{i0}

	\tconnector[selected]{B_p2.east}{D_p1.west}{0,0}{yshift=-0.7cm,below}{\tilabel{$g_1$}{$\fwrap[black]{F_b}{f_1}{F_a}$}}{i1}
	\tconnector[selected]{AOP_p.west}{D_p1.west}{-0.3cm,0cm}{left}{}{}

	\tconnector{C_p2.east}{D_p2.north}{1.85cm,0}{right}{\tilabel{$g_2$}{$f_2$}}{i2}
	\tconnectorgap{C_p2.east}{B_p2.east}{0.3,0.3}{right}{\tilabel{$g_3$}{$f_3$}}{i3}


	\node[ilabel,above =of i0]  {$a_0$};
	\node[ilabel,above =of i1]  {$a_1$};
	\node[ilabel,right=of i2]  {$a_2$};
	\node[ilabel,right=of i3]  {$a_3$};

\end{tikzpicture}
};
\end{tikzpicture}}
		\caption{Weaving} %
		\label{fig:global-aspect-weave} %
	\end{subfigure}%
	\caption{Matching and Weaving a Global Aspect} %
	\label{fig:global-aspect} %
\end{figure}
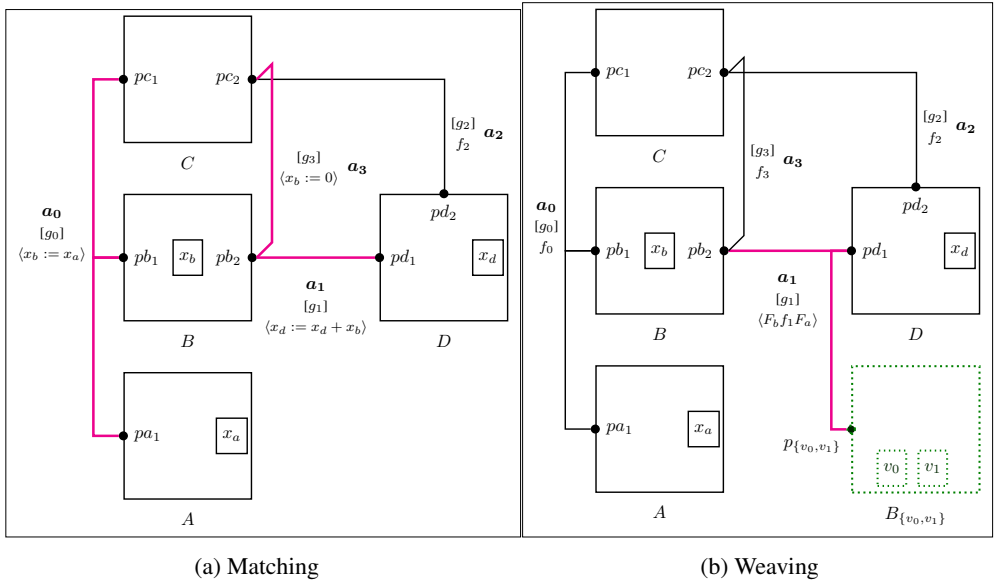

%
Since we consider only the interaction execution joinpoint, we consider criteria for matching interactions and relate them to global joinpoints.
To select a set of interactions, we use constraints over their associated ports (and their variables) and the involved data transfer.
For this, a global pointcut expression has three parts: the ports themselves, a set of read variables, and  a set of write variables.
Note that the port variables should be involved in the computation function of the interaction.
In \secref{sec:global:advices}, we use the read and written variables to define the context information passed to the advice.
\begin{definition}[Global pointcut]
\label{def:gpc}
	A global pointcut is a 3-tuple $\tuple{p, v_r, v_w}$ s.t.:
	\begin{itemize}
		\item $p \subseteq \bigcup_{B\in\atoms} (\portsof{B})$ is a set of ports,
		\item $v_r \subseteq \bigcup_{p_i \in p} (\varsof{p_i})$ is the set of \emph{read} variables, and
		\item $v_w \subseteq \bigcup_{p_i \in p} (\varsof{p_i})$ is the set of \emph{modified} variables.
	\end{itemize}
\end{definition}
We recall from \defref{def:events-global} that the global trace is a sequence of executed interactions $(a_0 \cdot a_1 \cdots a_{i-1})$.
Using the structure of an interaction (syntax only), we can match synchronization and data transfer.
Synchronization is matched using ports and data transfer by syntactic analysis of variables read and written.
\begin{definition}[Matching an executed interaction with a global pointcut]
\label{def:gpc-match}
An \\ executed interaction $a$ is a joinpoint selected by a global pointcut $\tuple{p, v_r, v_w}$ iff $a \vDash \tuple{p, v_r, v_w}$, where: \\ $
	a \vDash \tuple{p, v_r, v_w} \mbox{ iff } p \subseteq \portsof{a}
					 \land v_r \subseteq \varread(\mathit{a.\func})
					 \land v_w \subseteq \varwrite(\mathit{a.\func}).
$
\end{definition}
By construction of the global trace, the executed interaction is the same interaction found in the model.
As such, one can immediately match and modify the executed interactions by modifying the model directly.
Matching a global pointcut consists in selecting a subset of the interactions of a composite component.
\begin{proposition} \label{prop:global-match}
	$\forall a \in \events: a \vDash \mathit{gpc} \mbox{ iff } a \in \pcmatchg(\composite, \mathit{gpc})$, where: $\pcmatchg(\composite, \tuple{p, v_r, v_w}) =$ $\{a' \in \gamma \mid p \subseteq \portsof{a'} \land  v_r \subseteq \varread(\mathit{a'.\func}) \land v_w \subseteq \varwrite(\mathit{a'.\func})\}$
\end{proposition}
An interaction  $a \in \gamma$ is selected by a global pointcut $\tuple{p, v_r, v_w}$ if and only if  $a$ involves all the ports in $p$, and its update function reads from the variables in $v_r$ and writes to the variables in $v_w$.
The proposition states that during the execution, the interaction $a$ is matched (i.e., $a \vDash \mathit{gpc})$) iff the interaction $a$ is syntactically selected (i.e., $a \in \pcmatchg(\composite, \mathit{gpc})$).
\begin{example}[Interactions matched by a pointcut] \label{ex:global:match}
Figure \ref{fig:global-aspect-match} shows the interactions obtained by matching four pointcuts:
	\begin{enumerate}
		\item $\tuple{\setof{pa_1, pb_1}, \emptyset, \emptyset}$ selects all interactions including $\setof{pa_1,pb_1}$ in their ports, i.e., it selects only $a_0$ (e.g. $\pcmatchg(\composite, \tuple{\setof{pa_1, pb_1}, \emptyset, \emptyset}) = a_0$) as it is the only interaction involving both ports.
		\item $\tuple{\setof{pb_2}, \emptyset, \emptyset}$ selects all interactions including $\setof{ pb_2 }$ in their ports, i.e., it selects interactions $a_1$ and  $a_3$, since they both involve $pb_2$.
		\item $\tuple{\setof{pb_2}, \setof{x_b}, \emptyset}$ selects interactions including $\setof{ pb_2 }$ and which computation reads variable $x_b$ associated with $pb_2$.
The pointcut only selects $a_1$.
		\item $\tuple{\setof{pd_1}, \setof{x_d}, \setof{x_d}}$ selects interactions that include $\setof{ pd_1 }$ and which computation read and write the variable $x_d$ associated with $pd_1$ (to receive data).
The pointcut only selects $a_1$.
	\end{enumerate}
\end{example}

%
\subsection{Global Advice and Global Aspect}
\label{sec:global:advices}
%
A global advice defines the possible actions allowed on a global joinpoint $\tuple{q, a, q'}$.
These actions are restricted to two update functions $F_b$ and $F_a$ executed respectively before and after the interaction function $\mathit{a.\func}$.
Moreover, a global advice can modify only the ports that it matches, as interactions could include other ports.
Normally an update function of the interaction can modify the variables of all ports.
The non-matching ports and their variables are hidden from the advice as per application of Demeter's law~\cite{Lieberherr89}, as such we assume as little as possible on the interactions, and promote loose-coupling.
In the rest of the section, we consider a global pointcut $\mathit{pc} = \tuple{\setof{p_1, \ldots, p_n}, v_r, v_w}$, an advice is restricted to the set of ports $\setof{p_1, \ldots, p_n}$ and their variables, and a set of extra variables $V$ (inter-type variables).
\begin{definition}[Global advice]
\label{def:global-advice}
Given a set of ports $p \subseteq \bigcup_{B \in \atoms}\portsof{B}$ and a set of inter-type variables $V$, $X_{\mathrm{adv}} = V\ \cup\ \bigcup_{p_i \in p} (\varsof{p_i})$ is the set of \emph{advice variables}.
A global advice is a pair of update functions $\tuple{F_b, F_a}$ such that:
\begin{itemize}
    \item $(\varread(F_b) \cup \varwrite(F_b)) \subseteq X_{\mathrm{adv}}$, and
    \item $(\varread(F_a) \cup \varwrite(F_a)) \subseteq X_{\mathrm{adv}}$.
\end{itemize}
The global advice bound to $p$ and $V$ is noted $\mathrm{gadv}(p, V)$.
\end{definition}
Functions $F_b$ and $F_a$ are referred to as the \emph{before and after advice functions}, respectively.
For simplicity and clarity, we assume the update functions $F_b$ and $F_a$ to be uniquely determined and not empty.
To ensure this, one can add code markers at the start and end of each $F_b$ and $F_a$ which have no effect and are not present in the original system.
The variables that they read and write (captured with $\varread$ and $\varwrite$, respectively) should be variables of the advice.

We bind an advice to a pointcut expression with a global aspect.
The advice then applies to every joinpoint that the pointcut matches.
\begin{definition}[Global aspect]
\label{def:global-aspect}
A global aspect is a tuple $\tuple{\composite, V, \mathit{gpc}, \mathrm{gadv}(p,V)}$ s.t.:
	\begin{itemize}
		\item $\composite$ is a composite component;
		\item $V$ is the set of inter-type variables associated with the aspect;
		\item $\mathit{gpc} = \tuple{p, v_r, v_w}$ is the global pointcut (as per \defref{def:gpc});
		\item $\mathrm{gadv}(p, V)$ is the global advice (as per \defref{def:global-advice}).
	\end{itemize}
\end{definition}
A global aspect therefore associates a pointcut to an advice.
It ensures that the ports referred to the pointcut are the same for the advice, and that the advice has access to the variables of all ports in $p$ and $V$.
\begin{example}[Global advice and global aspect]
The global aspect
\[
	\tuple{\composite, \setof{v_0}, \tuple{\setof{pd_1}, \setof{x_d}, \setof{x_d}}, \tuple{\ffunc{v_0 := x_d},\ffunc{x_d := v_0}}}
\]
defines the inter-type variable $v_0$.
It also defines the pointcut to match the interactions that include port $pd_1$ and  which update function reads and writes to $x_d$.
The before and after update functions associated with the advice are respectively $\ffunc{v_0 := x_d}$ and $\ffunc{x_d := v_0}$; saving the value of $x_d$ in $v_0$ before the update function executes and then setting it back afterwards.
The pointcut matches $a_1$ as illustrated in \rfig{fig:global-aspect-match} and  specifies that  $a_1$ must execute the following sequence of instructions: $\ffunc{(v_0 := x_d), (x_d := x_d + x_b), (x_d := v_0)}$.
An advice function in this case can access only $\setof{v_0} \cup \varsof{pd_1}$.
The advice functions have no access to $x_b$, as it is not related to port $pd_1$ but $pb_2$.
This aspect disallows all interactions that read and write to $x_d$ to modify its value.
\end{example}
%
\subsection{Global Weaving}
\label{sec:global-weave}
%
Using the global aspect, the weaving procedure instruments the BIP model.
The procedure ensures that during the execution, the resulting BIP system will execute the advice whenever a joinpoint is reached.
Recall that interactions are stateless (i.e., they have no variables of their own), but they rely on data transfer from ports.
Variables can be defined only in atomic components.
Therefore, the weaving procedure must create an extra atomic component (so called inter-type component) that contains the variables of the aspect along with necessary ports and interactions.
The weaving operation is concerned only with syntactically modifying the BIP model.
For this purpose, we separate the two notions of matching to find the locations to modify from the instrumentation itself.
We therefore define first the transformation procedure and then its application with matching.

The transformation procedure uses the following input parameters:
\begin{itemize}
	\item a BIP composite component $\composite$ (the BIP model);
	\item a set of interactions $\subgamma$ resulting from matching with $\pcmatchg$ (Proposition~\ref{prop:global-match});
	\item a set of extra variables (i.e., the inter-type variables);
	\item the two functions $F_b$ and $F_a$ of the advice.
\end{itemize}
Accordingly, we create a new BIP composite component where the update function of each $a \in \subgamma$ is preceded by $F_b$ and followed by $F_a$.
In the following, we describe the weaving of a global aspect which requires weaving of the inter-type component and weaving of the advice.
\paragraph{Generating an inter-type component}
We first define the inter-type component.
\begin{definition}[Inter-type component]
\label{sec:global:intertype}
The inter-type component associated to the set of inter-type variables $V$ is defined as $B_V = \tuple{\setof{p_V}, \setof{\ell_0}, \setof{\tuple{\ell_0, p_V, \btrue, \ffunc{}, \ell_0}},  V}$ where $p_V = \tuple{p_V, V}$.
\end{definition}
$B_V$ contains $V$ as its variables, one port $p_V = \tuple{p_V, V}$ with all the variables attached to it, and one control location with a transition labeled with $p_V$ and guarded with the expression $\btrue$.
This ensures that the port will not stop any other interaction from executing once connected to it.
The inter-type component is added to the set of atomic components $\atoms$ of the BIP model.
\begin{example}[Adding an inter-type component to a model]
	\rfigb{fig:global-aspect-weave} depicts \\ $\pi(\gamma(\atoms \cup \setof{B_V}))$ where $V = \setof{v_0, v_1}$ and $\composite = \pi(\gamma(\atoms))$. A new atomic component $B_V$ is created. $B_V$ has the variables $v_0$ and $v_1$ and has its port $p_V$ always enabled.
	The variables in $V$ are attached to $p_V$.
\end{example}
\paragraph{Weaving the advice}
Once the inter-type component is added to the model, the advice is woven by connecting the existing interactions to it.
\begin{definition}[Global weave]
\label{def:global-weave}
Given a composite component $\composite = \pi(\gamma(\atoms))$, a set of interactions $\subgamma$, the set of inter-type variables $V$, and a global advice  $\mathit{adv} = \tuple{F_b, F_a}$, the global weave is defined as $\composite' = \fweaveg(\composite, \subgamma, V, adv)$ where $\composite' = \pi'(\gamma'(\atoms \cup \setof{B_V}))$ is the new composite component; with:
\begin{itemize}
	\item $B_V = \tuple{\setof{p_V}, \setof{\ell_0}, \setof{\tuple{\ell_0, p_V, \btrue, \ffunc{}, \ell_0}},  V}$ is the inter-type component identified by $V$ (as per \defref{sec:global:intertype});
	\item $\gamma'$ is defined as $\setof{\tmap(a) \mid a \in \gamma}$ with:\\
	$
	\tmap(a) : \gamma \rightarrow \gamma'  = \twopartdef
		{\tuple{\mathit{a.ports} \cup \setof{p_{V}}, \ffunc{F_b, \mathit{a.func}, F_a}, \mathit{a.guard}}}
		{\mbox{if } a \in \subgamma,}
		{a} {\mbox{otherwise.}}$
	\item $\pi' = \setof{\tuple{\tmap(a), \tmap(a')} \mid \tuple{a,a'} \in \pi}$ is the updated priority model.

\end{itemize}
Weaving a global aspect $\aspectglobal =$  $\tuple{\composite, V, \mathit{gpc, \tuple{F_b, F_a}}}$ into a composite component $\composite$ is noted $\composite \weaveg \aspectglobal$ and yields a new composite component
	 $\composite' = \fweaveg(\composite, \pcmatchg(\composite, gpc), V, \tuple{F_b, F_a})$.
\end{definition}
The inter-type component $B_V$ is added to the model.
The interactions that require instrumentation (i.e., those in $\gamma \cap \subgamma$) are extended with the port $p_V$ so as to have access to the inter-type variables and their computation function is prepended with $F_b$ and $F_a$.
The interactions not matched (i.e., those in $\gamma \setminus \subgamma$) are unmodified and copied.

The priority model ($\pi'$) is only extended to the changed interactions ($\tmap(a)$), but otherwise unmodified.
\begin{example}
	\rfigb{fig:global-aspect-weave} illustrates the weaving on the set of interactions $\setof{a_1}$ with the set of inter-type variables $V = \setof{v_0, v_1}$ of the advice $\tuple{F_b, F_a}$.
	A new atomic component $B_V$ is created that has two local variables $v_0$ and $v_1$ and has its port $p_V$ always enabled.
	The variables in $\fvar{V}$ are attached to $p_V$.
\begin{itemize}
	\item Interaction $\fvar{a_1}$ is connected to $p_V$ so as to allow access to $\fvar{V}$ on which $\fvar{F_b}$ and $\fvar{F_a}$ can operate.
	\item Update function $\fvar{F_b}$ is prepended to $\mathit{a_1.\func}$ so as to execute before and $\fvar{F_a}$ is appended to $\mathit{a_1.\func}$ so as to execute after.
	\item Since $p_V$ is always enabled, interaction $\fvar{a_1}$ will be enabled whenever $\fvar{pb_2}$ and $\fvar{pd_1}$ are both enabled and $\fvar{g_1}$ holds. The addition of $\fvar{p_V}$ does not affect enablement.
	\item Once $\fvar{a_1}$ is executed, $\fvar{F_b}$ and $\fvar{F_a}$ can modify the inter-type variables in $\fvar{B_V}$ by modifying $\varsof{p_{V}}$.
\end{itemize}
\end{example}
\paragraph{Correctness of weaving}
Consider $\events'$ (resp. $\events$) to be the set of executed interactions in the output (resp. initial) BIP system $\tuple{\composite', q'_0}$ (resp. $\tuple{\composite, q_0}$), where $\composite' = \composite \weaveg \tuple{\composite, V, gpc, \tuple{F_b, F_a}}$.
We begin by defining function $\erem_\vglobal : \events' \times \codes \times \codes \rightarrow \events \cup \setof{\epsilon}$.
Function $\erem_\vglobal$ removes the global advice from an interaction in $\events'$ and constructs a similar interaction in $\events$ or the empty interaction $\epsilon$ if it does not match the advice.
\[
	\erem_\vglobal(a, F_b, F_a) = \npartsdef{
		a' & \mbox{ if } \exists F : \mathit{a.\func} = \ffunc{F_b, F, F_a},\\
		\epsilon & \mbox{ otherwise},
	}
\]
with: $a' = \tuple{\mathit{a.ports} \setminus \setof{p_V}, F, \mathit{a.\guard}}$, where $\mathit{a.\func} = \ffunc{F_b, F, F_a}$.

The following proposition expresses the correct application of the advice on the joinpoints selected by a pointcut expression.
\begin{proposition}[Weaving correctness]
\label{prop:global-apply}
~\\
$\forall a \in \events', \exists F : \mathit{a.\func} = \ffunc{F_b, F, F_a} \mbox{ iff } (e' \neq \epsilon \land  e' \vDash \mathit{gpc}) \mbox{, with }
 e' = \erem_\vglobal(a, F_b, F_a)$.
\end{proposition}
We say that the update function of an interaction satisfies an advice weaving if its update function ($\mathit{a.func}$) starts with $F_b$ and ends with $F_a$ (i.e., the before and after update functions).
Proposition~\ref{prop:global-apply} states that any interaction in the execution satisfies an advice weaving iff one can construct an event $e'$ without the advice with $F_b$ and $F_a$ ($e' = \erem_\vglobal(a, F_b, F_a)$) which matches $\mathit{gpc}$ ($e' \vDash \mathit{gpc}$) in the initial system.
Since an advice can add extra behavior such as reading and writing to variables, it can cause the event to match more joinpoints, therefore we remove the extra update functions before matching with $\mathit{gpc}$.
\begin{example}[Correctness of global weave]
	We use the example with components $\code{Ping}$ and $\code{Pong}$, and its traces presented in \exref{ex:traces}.
	The global trace consists of the single interaction $a = \tuple{\setof{\code{send_1}, \code{recv_2}}, \ffunc{recv_2.p_2 := send_1.p_1},\btrue}$.
	Consider the global aspect
	$
		\mathrm{GA} = \tuple{\composite, \emptyset, \tuple{\setof{\code{send_1}}, \emptyset, \emptyset}, \tuple{F_\mathrm{b}, F_\mathrm{a}}}.
	$
	$\mathrm{GA}$ executes the advice when an interaction with port $\code{send_1}$ executes.
	The global trace of the system after weaving consists of the interaction \[
		a' = \tuple{\setof{\code{send_1}, \code{recv_2}}, \ffunc{F_\mathrm{b} \cdot recv_2.p_2 := send_1.p_1 \cdot F_\mathrm{a}},\btrue}.
	\]
	This is correct w.r.t Proposition~\ref{prop:global-apply}: we have $\erem_\vglobal(a', F_\mathrm{b}, F_\mathrm{a}) = a$, and $a \vDash \tuple{\setof{\code{send_1}}, \emptyset, \emptyset}$ holds.
	We note that the proposition ensures that $F_\mathrm{b}$ and $F_\mathrm{a}$ cannot occur in any interaction execution that does not match the pointcut.
	Furthermore, it is sufficient to directly edit $a$ in the model so that the edit appears during the execution (Proposition~\ref{prop:global-match}).
\end{example}
%

%
\section{Local Aspects} \label{sec:local}
After discussing concerns that arise at the global view and following the approach discussed in \secref{sec:aop_cbs:overview}, we now address the concerns arising in the local view.
An atomic component has control locations, variables,  and transitions labeled with ports, guards and computation functions.
For the local view, a component $B_k$ ``sees'' a sequence of local events (\defref{def:trmap}).
A local event consists of a source state containing the location and variable valuations, a transition, and a new state.
We define local joinpoints as local events, as they can be associated with the local execution of a component.
We are now concerned with the matching of these joinpoints.
We consider port execution and enablement, guard evaluation, access and modification of the state of the LTS (i.e., current location and valuation of local variables).
%
\subsection{Local Pointcuts}
\label{sec:lpc}
%
A local pointcut expression selects local joinpoints.
Considering the multitude of joinpoints, we use a grammar for specifying the pointcut expression, which makes it more readable and easier to combine.
Pointcut expressions are defined using the grammar in Listing~\ref{lst:grammar-match}, where $\ell \in \locsof{B_k}, x \in \varsof{B_k}, p \in \portsof{B_k}$.

\getcode[grammar-match]{grammar-match.g}{Local Pointcut Expression}
		{1}{style=grammar, escapechar=^}

\begin{definition}[Matching of a local joinpoint with a local pointcut expression]
\label{def:ljp-match}
A local joinpoint $e$  matches the pointcut expression $lpc$, noted $\evalLocal{e}{lpc}$, iff  match $lpc$ with\\
\[
\begin{array}{ll}
          \mid \phi \quad \mathrm{and} \quad \phi' & \rightarrow \evalLocal{e}{\phi} \land \evalLocal{e}{\phi'} \\
          \mid \pcloc(\ell)  &\rightarrow (e.l = \ell)\\
          \mid \pcguard(x)   &\rightarrow (\exists t \in \transof{B_k} : t.\source = e.l \land x \in \varguard(t)) \\
          \mid \pcfunc(x)    &\rightarrow (x \in \varread(\mathit{e.\tau.\func}))\\
          \mid \pcwrite(x)   &\rightarrow (x \in \varwrite(\mathit{e.\tau.\func})) \\
          \mid \pcportenabled(p) & \rightarrow (\exists q' : \tuple{e.q, p, q'} \in \rightarrow_{B_k})\\
          \mid \pcportexec(p) & \rightarrow (e.\tau.\port = p)
\end{array}
\]
\end{definition}
A component is considered at a location $\ell$ (predicate $\pcloc(\ell)$) if the component produces an event containing $\ell$ as the starting location of the transition of a local event.
\begin{remark}[Location]
While we can consider a component to be $\pcloc(\ell)$ when it produces an event containing $\ell$ as the ending location, we note that $\pcloc(\ell)$ includes evaluation of the guards and enabled ports.
Therefore, we restrict $\pcloc(\ell)$ to select only events that start with $\pcloc(\ell)$ as they capture better that reasoning since that evaluation is done to decide the transition to use and the next location.
As a convention, we assume that the component is at the location at the start of an event until the update function of the transition begins executing.
\end{remark}
The evaluation of a variable $x$ in the guard is checked on all outgoing transitions from a location regardless of the transition execution.
Therefore, for $\pcguard(x)$, we check if any transition has $x$ in its guard expression and originates from the location  $e.l$.
In contrast, an update function executes only when the transition is executed so we only examine the update function of $e.\tau$, and check if a variable $x$ is read ($\pcfunc(x)$) or modified ($\pcwrite(x)$).
For port enablement ($\pcportenabled(p)$), we check if there exists in the semantics of the component at least one transition labeled with the port $p$ from the event start state to any other state.
While for port execution ($\pcportexec(p)$), we only compare against $e.\tau$ as we are interested in the port that will execute.
\begin{remark}[Simplification]
Note that, from the semantics of BIP (\defref{semantics-composite}, p.~\pageref{semantics-composite}), whenever $\pcportexec(p)$ holds, then $ \pcportenabled(p)$ holds.
Therefore, it is possible to simplify a local pointcut expression by replacing, for the same $p$, `$\pcportexec(p) \mbox{ and } \pcportenabled(p)$' by $\pcportexec(p)$.
\end{remark}
%
\subsection{Matching Pointcuts}
\label{sec:local-match}
%
Unlike global pointcuts which map directly to interactions, local pointcuts require more complex instrumentation (i.e., location, transitions, guards, update functions).
Therefore, we group the syntactic elements and define which elements precede or succeed them.
This allows us to relate syntactic elements (in the BIP model) to a notion of before and after during the execution of the BIP system.\footnote{Similarly, in a regular program, adding extra code before a function call at the source-code level, results in executing the code before the function call during the execution of the program. However, we note that this is not as straightforward for the semantics of BIP.}
We use two levels of granularity.
The fine (resp. coarse) granularity focuses on a given transition (resp. a group of transitions grouped by source location).
We refer to a group of transitions grouped by source location $\ell$ as the location block $\ell$.
Transitions are grouped into blocks since according to the BIP semantics (\defref{def:semantics-atom}), to move from a location to another, it is necessary to evaluate all guards of all transitions originating from that location.
The guard evaluation determines which ports are enabled.
Each transition will then provide a jump to a location block.
At the coarse level, the elements that precede a block $\ell$ are the blocks that have transitions linking to $\ell$, while the elements that succeed $\ell$ are all blocks to which $\ell$ is a predecessor.
At the fine level, the elements that syntactically precede the transition are all transitions that lead to its block.
The elements that succeed the transition are all transitions in the block to which it refers to.
Given a set of transitions $M \subseteq \transof{B}$, we define:
\begin{itemize}
\item $\WOrigin(M) = \setof{\tau.\source \mid \tau \in M}$
to be the set of the origin/source locations of $M$ (blocks that contain the transitions in $M$);
\item $\WDest(M) = \setof{\tau.\dest \mid \tau \in M}$ to be the set of the destination locations of $M$
(blocks to which the transitions lead);
\item $\WSiblings(M) = \setof{\tau \in \transof{B} \mid  \tau.\source \in \WOrigin(M)}$ to be the set of transitions of $B$ that have their source locations within the origin locations of $M$ (selecting all transitions of the same block);
\item $\WGetP(M) = \setof{\tau \in \transof{B} \mid \tau.\dest \in \WOrigin(M)}$ to be the set of transitions of $B$ that have their destination locations within the origin locations of $M$ (transitions that lead to the block).
\end{itemize}

\begin{figure}[t]
	\centering
	\begin{subfigure}[b]{0.4\textwidth}
		\scalebox{0.8}{ \begin{tikzpicture}[bip]
 	\tikzset{
		setm/.style={line width=0.8mm},
		siblings/.style ={densely dotted, draw=black, line width=0.3mm},	
		origin/.style ={fill=black!50, text=white},	
		dest/.style ={pattern=north east lines, pattern color=black!30},	
		odest/.style ={preaction={origin}, dest, pattern color=black!80, text=white},	
		prevm/.style ={dashed}	
	}
    \component{c1}{atomic}{}{

      \node[place, origin, double] (c1l0) {$\ell_0$};
      \node[place, dest] (c1l1) [right=of c1l0] {$\ell_1$}; 
      \node[place, origin] (c1l2) [below=of c1l0] {$\ell_2$};
      \node[place ] (c1l3) [below=of c1l2] {$\ell_3$};
      \node[place, dest] (c1l4) [right=of c1l3] {$\ell_4$};

	\tconnect[t5]{siblings, setm}
		    {c1l0} {c1l1}
		    {0.1cm}{above}
		   {$g_5$}
		   {$p_2$}
		   {$f_5$}

	\tconnect[t0]{prevm}
		    {c1l0} {c1l2}
		    {0.5cm}{left}
		   {$g_0$}
		   {$p_1$}
		   {$f_0$}

	\tconnect[t1]{normal}
		    {c1l1} {c1l2}
		    {1cm}{right}
		   {$g_1$}
		   {$p_1$}
		   {$f_1$}

	\tconnect[t2]{siblings}
		    {c1l2} {c1l3}
		    {0.5cm}{left}
		   {$g_2$}
		   {$p_1$}
		   {$f_2$}
	
	\tconnect[t3]{siblings,setm}
		{c1l2}{c1l4}
		{1cm}{right}
		{$g_3$}
		{$p_2$}
		{$f_3$}
	\tconnect[t4]{siblings, loop, dashdotted, min distance=10mm,in=30,out=-30,looseness=2}
		{c1l2}{c1l2}
		{0.1cm}{right}
		{$g_4$}
		{$p_2$}
		{$f_4$}
	\input{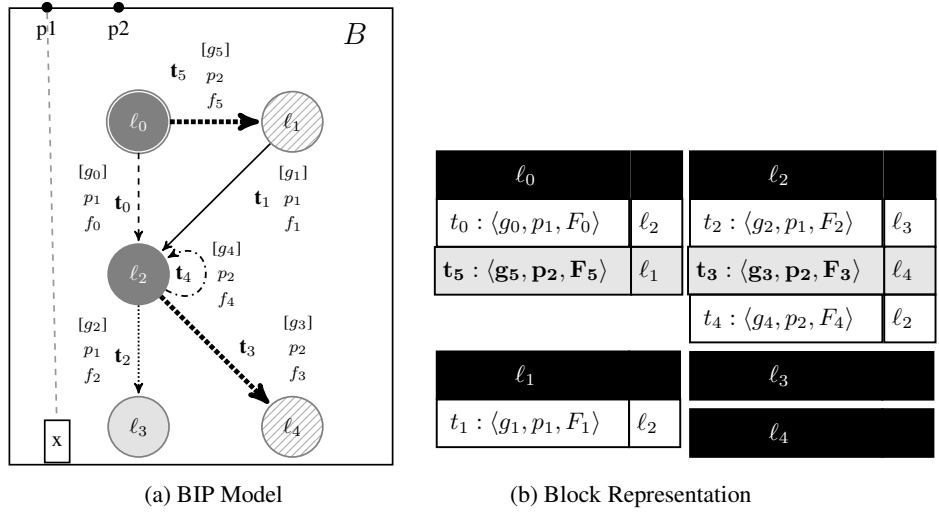}
  };

    \node[export] (c1p1) at ($(c1.north west)!.1!(c1.north east)$) [label=below:p1]    {};
    \node[export] (c1p2) [right=of c1p1] [label=below:p2] {};

    \node at ($(c1.north west)!.9!(c1.north east)$) [label=below:{\Large $B$}]    {};

	\node[var] (x) at ( $(c1.south west) + (0.8,0.4)$) []  {x};
	\draw[portvar] (c1p1) -- (x);
  \end{tikzpicture}
  }
        \caption{BIP Model}
        \label{fig:syntax-1}
    \end{subfigure}
	\begin{subfigure}[b]{0.4\textwidth}
		\scalebox{0.9}{\begin{tikzpicture}[bip]
\tikzset{
	cell/.style={rectangle,draw=black, align=center, anchor=center},
	space/.style={matrix of nodes,row sep=-\pgflinewidth,column sep=-\pgflinewidth}
}
\tikzset{
	bipblock/.style={space,%
		anchor=north west,%
		nodes={cell, font=\ttfamily, minimum height=2.0em, minimum width=8em},%
		row 1/.style={nodes={fill=black, text=white}},%
		column 2/.style={nodes={minimum width=1.2em}},%
	},%
}
\newcommand{\hsp}{{~\hspace{0.85em}}}
\matrix(l0) [bipblock] {
	$\ell_0$ &  \hsp \\
	$t_0: \tuple{g_0, p_1, F_0}$ & $\ell_2$ \\
	\hsel$\mathbf{t_5: \tuple{g_5, p_2, F_5}}$ & \hsel$\ell_1$ \\
};

\matrix(l1) [bipblock, below=0.55 of l0] {
	$\ell_1$ & \hsp \\
	$t_1: \tuple{g_1, p_1, F_1}$ & $\ell_2$ \\
};

\matrix(l2) [bipblock, xshift=3.7cm] {
	$\ell_2$ &  \hsp \\
	$t_2: \tuple{g_2, p_1, F_2}$ & $\ell_3$ \\
    \hsel$\mathbf{t_3: \tuple{g_3, p_2, F_3}}$ & \hsel$\ell_4$ \\
	$t_4: \tuple{g_4, p_2, F_4}$ & $\ell_2$ \\
};

\matrix(l3) [bipblock, below=0cm of l2, yshift=0.15cm] {
	$\ell_3$ & \hsp\\
};

\matrix(l4) [bipblock, below=0cm of l3, yshift=0.15cm] {
	$\ell_4$ & \hsp\\
};

\end{tikzpicture}}
        \caption{Block Representation}
        \label{fig:syntax-2}
    \end{subfigure}
	\caption{Syntax Representations}
	\label{fig:syntax}
\end{figure}

\begin{example}[Syntactic representation]
Figure~\ref{fig:syntax} shows an atomic component in the two views.
We consider the set of transitions $M = \setof{t_3,t_5}$ which is shown in bold.
Figure~\ref{fig:syntax-1} illustrates $\WOrigin(M)$, $\WDest(M)$, $\WSiblings(M)$ and $\WGetP(M)$.
The origin set contains the locations from which the transitions in $M$ are outbound: $\setof{\ell_0, \ell_2}$.
The destination set contains the locations to which the transitions in $M$ lead to: $\setof{\ell_1,\ell_4}$, and are highlighted with a pattern.
The dotted transitions belong to $\WSiblings(M)$, they are all transitions originating from $\WOrigin(M)$  i.e., $t_0, t_2, t_3, t_4$ and  $t_5$.
They are the transitions in the same block as $M$.
The dashed transitions belong to $\WGetP(M)$, they are $t_0$ and $t_4$.
These transitions lead to $\WOrigin(M)$.
\end{example}
\begin{definition}[Local pointcut selection]
\label{def:lpc-match}
A local pointcut expression selects a subset of $\transof{B_k}$: \\
$\pcmatch(B_k, lpc)  = $  match $lpc$ with \\
\[
\begin{array}{ll}
          \mid \pcloc(\ell)  &\rightarrow \setof{\tau \in \transof{B_k} \mid \tau.\source = \ell}\\
          \mid \pcguard(x)   &\rightarrow  \WSiblings(\setof{\tau \in \transof{B_k} \mid x \in\varguard(\tau) }) \\
          \mid \pcfunc(x)    &\rightarrow \setof{\tau \in \transof{B_k} \mid x \in \varread(\tau)}\\
          \mid \pcwrite(x)   &\rightarrow \setof{\tau \in \transof{B_k} \mid x \in \varwrite(\tau)}\\
          \mid \pcportexec(p) & \rightarrow  \setof{\tau \in \transof{B_k} \mid \tau.\port  = p}\\
          \mid \pcportenabled(p) & \rightarrow \WSiblings(\setof{\tau \in \transof{B_k} \mid \tau.\port = p})\\
          \mid \phi \hbox{ and } \phi' & \rightarrow \pcmatch(B_k, \phi) \cap \pcmatch(B_k, \phi')
\end{array}
\]
\end{definition}
Predicate $\pcloc(\ell)$ matches the transitions in block $\ell$.
Guards are evaluated at the level of the block.
Therefore, to match $\pcguard(x)$, we first select the transitions that contain $x$ in their guard, and then select their blocks with $\WSiblings()$.
An update function, however, is at the level of transitions.
Henceforth, $\pcfunc(x)$ and $\pcwrite(x)$ select only the transitions which update function reads $x$ and modifies $x$, respectively.
For ports, the execution of a port is at the level of a transition, therefore $\pcportexec(p)$ selects all transitions that contain the port $p$.
However, the enablement of a port happens at the level of the block, since multiple ports can be enabled but only one executes, therefore $\pcportenabled(p)$ extends the selection to the block of the transitions that would normally be selected by $\pcportexec(p)$.
When combining matches, the result must select transitions that are affected by both pointcuts.
For this, the transitions from both matches are intersected to ensure that the result has transitions present in both.
Note that since $\cap$ is associative and commutative, the set of obtained transitions is insensitive to the match order.

\begin{figure}[t]
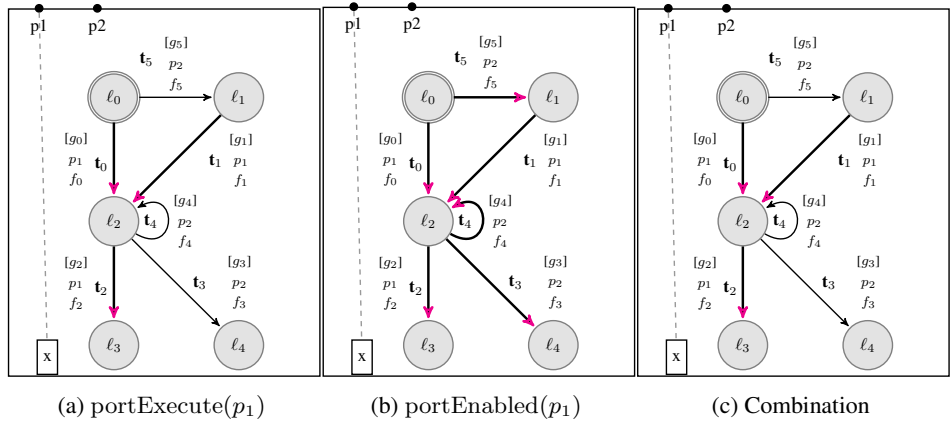

	\centering
	\begin{subfigure}[b]{0.3\textwidth}
		\scalebox{0.65}{  \begin{tikzpicture}[bip]
 
    \component{c1}{atomic}{}{

      \node[place,double] (c1l0) {$\ell_0$};
      \node[place] (c1l1) [right=of c1l0] {$\ell_1$}; 
      \node[place] (c1l2) [below=of c1l0] {$\ell_2$};
      \node[place] (c1l3) [below=of c1l2] {$\ell_3$};
      \node[place] (c1l4) [right=of c1l3] {$\ell_4$};

	\tconnect[t5]{normal}
		    {c1l0} {c1l1}
		    {0.1cm}{above}
		   {$g_5$}
		   {$p_2$}
		   {$f_5$}

	\tconnect[t0]{selected}
		    {c1l0} {c1l2}
		    {0.5cm}{left}
		   {$g_0$}
		   {$p_1$}
		   {$f_0$}

	\tconnect[t1]{selected}
		    {c1l1} {c1l2}
		    {1cm}{right}
		   {$g_1$}
		   {$p_1$}
		   {$f_1$}

	\tconnect[t2]{selected}
		    {c1l2} {c1l3}
		    {0.5cm}{left}
		   {$g_2$}
		   {$p_1$}
		   {$f_2$}
	
	\tconnect[t3]{normal}
		{c1l2}{c1l4}
		{1cm}{right}
		{$g_3$}
		{$p_2$}
		{$f_3$}
	\tconnect[t4]{normal, loop, min distance=10mm,in=30,out=-30,looseness=2}
		{c1l2}{c1l2}
		{0.1cm}{right}
		{$g_4$}
		{$p_2$}
		{$f_4$}
 	\input{tikz/toverlay}
	};

    \node[export] (c1p1) at ($(c1.north west)!.1!(c1.north east)$) [label=below:p1]    {};
    \node[export] (c1p2) [right=of c1p1] [label=below:p2] {};


	\node[var] (x) at ( $(c1.south west) + (0.8,0.4)$) []  {x};
	\draw[portvar] (c1p1) -- (x);
	
  \end{tikzpicture}
  }
        \caption{$\pcportexec(p_1)$}
        \label{fig:match-port-1}
    \end{subfigure}
	\begin{subfigure}[b]{0.3\textwidth}
		\scalebox{0.65}{  \begin{tikzpicture}[bip]
 
    \component{c1}{atomic}{}{

      \node[place,double] (c1l0) {$\ell_0$};
      \node[place] (c1l1) [right=of c1l0] {$\ell_1$}; 
      \node[place] (c1l2) [below=of c1l0] {$\ell_2$};
      \node[place] (c1l3) [below=of c1l2] {$\ell_3$};
      \node[place] (c1l4) [right=of c1l3] {$\ell_4$};

	\tconnect[t5]{selected}
		    {c1l0} {c1l1}
		    {0.1cm}{above}
		   {$g_5$}
		   {$p_2$}
		   {$f_5$}

	\tconnect[t0]{selected}
		    {c1l0} {c1l2}
		    {0.5cm}{left}
		   {$g_0$}
		   {$p_1$}
		   {$f_0$}

	\tconnect[t1]{selected}
		    {c1l1} {c1l2}
		    {1cm}{right}
		   {$g_1$}
		   {$p_1$}
		   {$f_1$}

	\tconnect[t2]{selected}
		    {c1l2} {c1l3}
		    {0.5cm}{left}
		   {$g_2$}
		   {$p_1$}
		   {$f_2$}
	
	\tconnect[t3]{selected}
		{c1l2}{c1l4}
		{1cm}{right}
		{$g_3$}
		{$p_2$}
		{$f_3$}
	\tconnect[t4]{selected, loop, min distance=10mm,in=30,out=-30,looseness=2}
		{c1l2}{c1l2}
		{0.1cm}{right}
		{$g_4$}
		{$p_2$}
		{$f_4$}
 	\input{tikz/toverlay}
	};

    \node[export] (c1p1) at ($(c1.north west)!.1!(c1.north east)$) [label=below:p1]    {};
    \node[export] (c1p2) [right=of c1p1] [label=below:p2] {};


	\node[var] (x) at ( $(c1.south west) + (0.8,0.4)$) []  {x};
	\draw[portvar] (c1p1) -- (x);
	
  \end{tikzpicture}
  }
        \caption{$\pcportenabled(p_1)$}
        \label{fig:match-port-2}
    \end{subfigure}
	\begin{subfigure}[b]{0.3\textwidth}
		\scalebox{0.65}{  \begin{tikzpicture}[bip]
 
    \component{c1}{atomic}{}{

      \node[place,double] (c1l0) {$\ell_0$};
      \node[place] (c1l1) [right=of c1l0] {$\ell_1$}; 
      \node[place] (c1l2) [below=of c1l0] {$\ell_2$};
      \node[place] (c1l3) [below=of c1l2] {$\ell_3$};
      \node[place] (c1l4) [right=of c1l3] {$\ell_4$};

	\tconnect[t5]{normal}
		    {c1l0} {c1l1}
		    {0.1cm}{above}
		   {$g_5$}
		   {$p_2$}
		   {$f_5$}

	\tconnect[t0]{selected}
		    {c1l0} {c1l2}
		    {0.5cm}{left}
		   {$g_0$}
		   {$p_1$}
		   {$f_0$}

	\tconnect[t1]{selected}
		    {c1l1} {c1l2}
		    {1cm}{right}
		   {$g_1$}
		   {$p_1$}
		   {$f_1$}

	\tconnect[t2]{selected}
		    {c1l2} {c1l3}
		    {0.5cm}{left}
		   {$g_2$}
		   {$p_1$}
		   {$f_2$}
	
	\tconnect[t3]{normal}
		{c1l2}{c1l4}
		{1cm}{right}
		{$g_3$}
		{$p_2$}
		{$f_3$}
	\tconnect[t4]{normal, loop, min distance=10mm,in=30,out=-30,looseness=2}
		{c1l2}{c1l2}
		{0.1cm}{right}
		{$g_4$}
		{$p_2$}
		{$f_4$}
 	\input{tikz/toverlay}
	};

    \node[export] (c1p1) at ($(c1.north west)!.1!(c1.north east)$) [label=below:p1]    {};
    \node[export] (c1p2) [right=of c1p1] [label=below:p2] {};


	\node[var] (x) at ( $(c1.south west) + (0.8,0.4)$) []  {x};
	\draw[portvar] (c1p1) -- (x);
	
  \end{tikzpicture}
  }
        \caption{Combination}
        \label{fig:match-port-3}
    \end{subfigure}

	\caption{Matching Ports}
	\label{fig:match-port}
\end{figure}

\begin{example}[Matching ports]
Figure~\ref{fig:match-port-1} shows (in red) the transitions matched with the pointcut $\pcportexec(p_1)$.
To match the execution of port $p_1$, all transitions labeled with $p_1$ are selected.
These transitions are executed only if port $p_1$ is executed.
Figure~\ref{fig:match-port-2} shows the transitions matched with $\pcportenabled(p_1)$.
Note that more transitions are selected since the enablement of $p_1$ may be followed by the execution of a port different than $p_1$ (e.g., $p_2$). Port $p_1$ is enabled at $\ell_2$ iff $g_2$  evaluates to $\btrue$.
Determining if $p_1$ is enabled requires pre-evaluating $g_2$.
Additionally, $p_2$ may be executed while $p_1$ is enabled at $\ell_2$. Then, the component may execute either $t_3$ or $t_4$. Therefore, the joinpoint must include $t_3$ and $t_4$.
However, if $g_2$ evaluates to $\bfalse$, $p_1$ is not enabled. Thus the joinpoint must not include $t_3$ and $t_4$.
To handle the evaluation of the guards at runtime and executing the advice properly, additional elements need to be instrumented, they are detailed in \secref{sec:weave-add}.
Figure~\ref{fig:match-port-3} shows the combination of the matches from the two, effectively showing $\pcportexec(p_1)$ since the execution implies enablement.
\end{example}
\begin{proposition} \label{prop:local-match}
 $\evalLocal{e}{lpc} \mbox{ iff } e.\tau \in \pcmatch(B_k, lpc)$ where $e$ is a local event, and we assume that $lpc$ does not contain $\pcportenabled(p)$.
\end{proposition}
This proposition states that a local event $e$ is a local joinpoint ($\evalLocal{e}{lpc}$) iff its transition $e.\tau$ is syntactically selected (i.e. $\tau \in \pcmatch(B_k, lpc)$).
Since state information is accessible only at runtime, determining enabled ports ($\pcportenabled(p)$) requires additional instrumentation.
%
\subsection{Local Advice}
\label{sec:local:advice}
%
The local advice defines the possible actions to be injected at a local joinpoint.
Similarly to a global advice (\defref{def:global-advice}), a local advice executes two functions before and after the local joinpoint.
Moreover, we introduce a constraint to restrict the variables accessible to the advice functions.
Advice variables consists of the variables of the atomic component and an extra set of inter-type variables $V$.
Furthermore, in order to increase the expressiveness of the local advice and the refinement of local behavior, a local advice may change the location of the atomic component depending on a specific guard.
However to ensure consistency, the location change happens once both functions of the advice have finished executing.
\begin{definition}[Local advice]
\label{def:local-advice}
A local advice $ladv(B_k, V)$ is a triple $\tuple{F_b, F_a, R}$. It has access to $X_{adv} = \varsof{B_k} \cup V$. It consists of:
\begin{enumerate}
	\item a before function $F_b$ such that $(\varread(F_b) \cup \varwrite({F_b})) \subseteq X_{adv}$;
	\item an after function $F_a$ such that $(\varread(F_a) \cup \varwrite({F_b})) \subseteq X_{adv}$;
	\item and a set $R$ of reset locations, defined as a set of tuples  $ \tuple{\ell, g}$ where
each tuple indicates that, after the end of the joinpoint, the component has to move to location $\ell$ if guard $g$ evaluates to $\btrue$.
\end{enumerate}
\end{definition}
A local aspect binds a local pointcut to a local advice and defines the inter-type variables.
\begin{definition}[Local aspect]
A local aspect is a tuple $\tuple{B_k, lpc, V, ladv(B_k , V)}$ where $B_k$ is an atomic component, $lpc$ is a pointcut expression, $V$ is the set of inter-type variables, $ladv(B_k , V)$ is the local advice to apply on the joinpoints.
\end{definition}
\begin{example}[Local aspect]
Let us consider the local aspect:
\[
\tuple{B_k, \pcloc(\ell_2), \setof{v_0}, \tuple{F_b, F_a, \setof{\tuple{\ell_1, x > 1}}}}
\]
It applies to component $B_k$.
The local pointcut $\pcloc(\ell_2)$ selects any joinpoint where $B_k$ is at location $\ell_2$.
The advice $\tuple{F_b, F_a, \setof{\tuple{\ell_1, x > 1}}}$ specifies that i) right before $B_k$ enters $\ell_2$, $F_b$ must execute, and ii) after $B_k$ exits $\ell_2$, $F_a$ must execute, and then iii) if $x > 1$ holds at runtime the component must move to location $\ell_1$.
\end{example}
%
\subsection{Local Weaving}
%
Similarly to global weaving (discussed in \secref{sec:global-weave}), the local weaving procedure instruments a BIP system around an atomic component.
It therefore weaves a local advice to local joinpoints selected by local pointcut expressions.
\paragraph{Edit frame.}
We recall that a local advice (see \defref{def:local-advice}) defines extra update functions to execute, and the possibility to change the location of the atomic component.
In order to match local joinpoints with local pointcuts, we need to instrument the atomic component while using the transition as the unit (\defref{def:lpc-match}).
Then, we need to define where the instrumentation occurs relative to the transition.
Using the syntactic representation in Section~\ref{sec:local-match}, we define \emph{edit points}.
Edit points provides hints to the weaving procedure to determine which elements need to be modified, so that during runtime, the advice executes appropriately.
We identify four edit points:
\begin{itemize}
	\item $\WCreate$ (Runtime) specifies that instrumentation is to detect the state at runtime, and it applies at the  block-level of the match (i.e., the location block of each transition in the match);
	\item $\WPreviousAfter$ (Previous End) specifies that the instrumentation applies at the level of the transitions that lead to the location blocks that include the match and specifically at the end of their update function;
	\item $\WCurrentBefore$ (Current Begin) specifies that the instrumentation applies at the level of the transitions in the match, at the beginning of their update function;
	\item $\WCurrentAfter$ (Current End) specifies that the instrumentation applies similarly to $\WCurrentBefore$ but instead, at the end of the update function.
\end{itemize}
The set of edit points is $\EP = \setof{\WPreviousAfter, \WCurrentBefore, \WCurrentAfter, \WCreate}$.
The edit point $\WCreate$ is a special hint that is associated with runtime information.
We use it to determine if a port is enabled as it requires the evaluation of the guards during runtime.
We use it to encode an ``if else'' statement, so that the guards are evaluated in a temporary state.
The temporary state is needed as advices have an update function for ``before'', and since we cannot wait for the evaluation to happen to execute the advice, we have to pre-evaluate the guards to determine if a port is enabled.

The \emph{edit frame} indicates the start and end edit points, it informally determines the start and end of the region necessary to match the pointcut, so that we can weave $F_b$ and $F_a$ respectively.
An edit frame is determined depending on the local pointcut expression as follows:
\begin{definition}[Edit frame]
\label{def:edit}
	The edit frame is a pair of edit points $\tuple{e_1, e_2}$ corresponding to the update functions of the local advice: $F_b$ and $F_a$ respectively. It is defined as:
$\pcedit(lpc)  = $  match $lpc$ with: \\
\[
\begin{array}{llll}
          \mid \pcloc(\ell)  	  &\rightarrow \tuple{\WPreviousAfter, \WCurrentBefore} &
          \mid \pcfunc(x)    	  &\rightarrow \tuple{\WCurrentBefore, \WCurrentAfter}\\
          \mid \pcguard(x)   	  &\rightarrow \tuple{\WPreviousAfter, \WCurrentBefore} &
          \mid \pcwrite(x)   	  &\rightarrow \tuple{\WCurrentBefore, \WCurrentAfter}\\
          \mid \pcportenabled(p) &\rightarrow \tuple{\WCreate, 	   \WCurrentBefore} &
          \mid \pcportexec(p) 	  &\rightarrow \tuple{\WCurrentBefore, \WCurrentAfter}\\
          \mid \phi \hbox{ and } \phi' & \rightarrow \tuple{\mathrm{max}(e_1,e_1'), \mathrm{max}(e_2,e_2')} &\\
\end{array}
\]
where $\tuple{e_1,e_2} = \pcedit(\phi)$ and  $\tuple{e_1', e_2'} = \pcedit(\phi')$, with the strict ordering $\WPreviousAfter \prec \WCurrentBefore \prec \WCurrentAfter \prec \WCreate$.
\end{definition}

An extra update function in the case of a location ($\pcloc(\ell)$) happens at the end of any transition that leads to the block $\ell$.
This ensures that it executes prior to the block $\ell$.
We consider the end of the block  whenever a transition is to execute, i.e., at the beginning of its update function.
This is done similarly for guards ($\pcguard(x)$) as guards are evaluated at the block level.
In the case of an update function $F$, any extra update function applies before $F$ at its beginning, and after $F$ at its end.
Therefore, for a variable read ($\pcfunc(x)$) or modification ($\pcwrite(x)$) any extra update function should happen before $F$.
We note that this is the finest level of granularity, we do not inspect the sequence of assignments in the update function, only its beginning and end.
Similarly, a port execution ($\pcportexec(p)$) is the execution of an update function of a transition so the edit points are identical.
Since state information is accessible only at runtime, determining enabled ports ($\pcportenabled(p)$) requires modification at the block level, mostly by inserting additional blocks that pre-evaluate the guards and do some extra computation.
Thus, we associate the edit frame $\tuple{\WCreate, \WCurrentBefore}$ to $\pcportenabled$ so that guards are pre-evaluated appropriately.

In the case of a combination of pointcuts, we must ensure that the frame can capture \emph{both pointcuts}.
Thus, we define a strict order on $\EP$: $\WPreviousAfter \prec \WCurrentBefore \prec \WCurrentAfter \prec \WCreate$.
The purpose of the order is to select the most relevant elements for weaving, considering our representation of the syntax of BIP (described in \secref{sec:local-match}).
It can be seen as the scope of elements that need to be instrumented.
Since $\WCreate$ requires runtime checks, it always requires the most modification, therefore it is the maximal element.
At the level of transitions, the order follows from the precedence: the start of an update function $(\WCurrentBefore)$ precedes its end $(\WCurrentAfter)$, and the entire transition is preceded by another transition that leads to its block ($\WPreviousAfter$).
At the level of the block, the transitions that lead to the block, precede it ($\WPreviousAfter$).
A combination of two frames must start when the elements overlap (i.e., the most delayed edit frame), and end once both end (i.e., at the most delayed one).
Also, the combination must include all elements that needs to be instrumented to match both frames.
To fulfill both of these conditions, we use $\mathrm{max}$, as it will include the hint to the maximal elements needed to be instrumented.
$\WCreate$ includes the instrumentation necessary at the entire block level for detecting $\pcportenabled$ at runtime.
Therefore, combining $\WCreate$ with any pointcut frame will still require runtime information.
Therefore, $\WCreate$ is defined as the maximum.
Combining port enabled with the other frames adds the port enablement condition on the existing condition (as per the semantics of $\pcportenabled$ in \defref{def:ljp-match}), therefore to capture the two cases: $\tuple{\WPreviousBefore, \WCurrentBefore}$ and $\tuple{\WCurrentBefore, \WCurrentAfter}$,  we discriminate using the second point.
When used alone, we consider $\pcportenabled$ to end similarly to $\pcloc$ and $\pcguard$, at the start of transition execution.

Note that, since $\mathrm{max}$ is associative and commutative the order does not matter.
Exhausting all possible combinations with $\mathrm{max}$, only the following frames are possible: $\tuple{\WPreviousAfter, \WCurrentBefore}$, $\tuple{\WCurrentBefore, \WCurrentAfter}$, $\tuple{\WCreate, \WCurrentBefore}$, and $\tuple{\WCreate, \WCurrentAfter}$.
\begin{example}[Edit frames]
We consider the pointcut expression: $\pcloc(\ell_1)$ and $\pcwrite(x)$.
These expressions have the frames $\tuple{\WPreviousAfter, \WCurrentBefore}$ and $\tuple{\WCurrentBefore, \WCurrentAfter}$, respectively.
This means that $\pcloc(\ell_1)$ starts right after the previous transitions leading to $\ell_1$ have finished executing their update function, while for $\pcwrite(x)$ it starts right after the transition executes (i.e., $\WCurrentBefore$, at the start of its update function).
Their combination must start when they both start, in this case $\WCurrentBefore$.
If we consider the start to be the earliest, then we can still pass $\WPreviousAfter$ but it is possible to execute  another transition that does not match $\pcwrite(x)$.
This is not consistent with the start of $\pcwrite(x)$.
They both end after (1) $\pcloc(\ell_1)$ is over and (2) $\pcwrite(x)$ is over, therefore they end when the transition ends ($\WCurrentAfter$).
\end{example}
\paragraph{Overview.}
Recall from \defref{def:local-advice} that both of the update functions $F_b$ and $F_a$ of an advice have read/write access to a component variables.
It is also possible that $F_b$ and $F_a$ belong to different transitions.
Therefore, it could be possible to update the variables in $F_b$ in a way that ensures $F_a$ is never executed.
The first requirement of the weaving procedure is to ensure that $F_a$ will always execute once $F_b$ has executed.
\begin{remark}[Deadlocks]
It is ensured that function $F_a$ executes after $F_b$ if the user's advice does not result in a deadlock state.
For example, the before computation may modify the state so as to have no outgoing enabled transition.
Hence, the after computation is not executed.
Since weaving transforms a BIP model, it is possible to use verification tools (such as DFinder~\cite{DFinder}) on the transformed model to check for deadlock freedom.
\end{remark}
In order to apply the requirement, the general weaving strategy uses one boolean variable per aspect $\aopvar{}$.
The variable \aopvar{} is set to $\bfalse$ right before reaching the joinpoint and set to $\btrue$ upon joinpoint entry, indicating a pointcut match.
It is then up to the weaving procedure to unset if necessary.
We use for brevity: $\aopset = \ffunc{\aopvar := \btrue}$ and $\aopclear = \ffunc{\aopvar := \bfalse}$.
In addition to $\aopvar$ we define the extra port $\aopport$.
The $\aopport$ port is associated with a singleton interaction with the highest priority in the system.
It is used to create high priority transitions in the component creating deterministic behavior.

In the following, we begin by weaving the reset location pairs, since it is independent of execution frames.
Later, we consider the weaving of the advice $adv = \tuple{F_b, F_a, \setof{}}$ on a set of transitions $M$ for each of the four pairs of frames:  $\tuple{\WPreviousAfter, \WCurrentBefore}$, $\tuple{\WCurrentBefore, \WCurrentAfter}$, $\tuple{\WCreate, \WCurrentBefore}$, $\tuple{\WCreate, \WCurrentAfter}$.
We fix $T'$ and $L'$ to be the set of transitions and locations of the new atomic component respectively.
A set of locations $\resetlocs$ contains the locations (blocks) on which the pointcut ends, so that it is possible to weave the reset locations.
An injective function $\tmap: \transof{B_k} \rightarrow 2^{T'}$ relates $\transof{B_k}$ to the new transitions by transforming them, creating extra transitions or copying them.
\paragraph{Weaving reset location pairs.}
\label{sec:weave-reset}
We consider a set $R$ of reset location pairs, and a set of locations $L_{\mathrm{R}}$.
The reset location transitions are defined as follows:
\[
	\weavereset(R, L_{\mathrm{R}}) = \bigcup_{\tuple{guard, dest} \in R}(\setof{\tuple{\ell, \aopport, \aopvar \land guard, \aopclear, dest} \mid \ell \in L_{\mathrm{R}}}).
\]
The transitions are guarded by $\aopvar$ so as to execute only if the pointcut matched.
They execute on $\aopport$ so as to have priority over other transitions at the location.
Once executed they invoke $\aopclear$ to indicate that the pointcut has ended.
Consequently it avoids a deadlock when the location resets to itself.
\begin{example}[Weaving reset location]
Figure \ref{fig:reset-locs} shows the weaving of reset locations, namely  $\weavereset(\setof{\tuple{\ell_4, g_a}, \tuple{\ell_1, g_b}}, \setof{\ell_1, \ell_2})$.
In total, four transitions are created as we have two locations and two reset pairs.
Transitions are highlighted differently for each destination location.
\end{example}
\getexamples[reset-locs]{weave/locs,weave/locs-1}
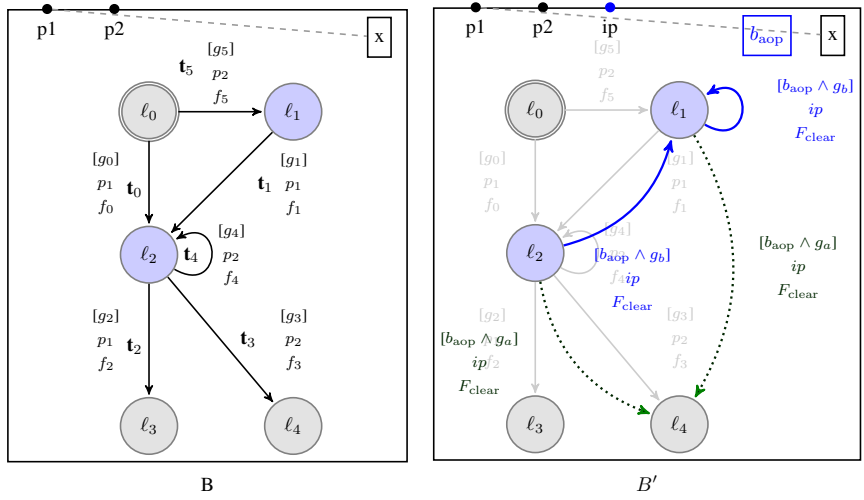
			{Weaving Reset Locations}{0.75}{2}

We now elaborate on the weaving of each edit frame by detailing the transitions that are modified, the extra locations created and specifying which locations are selected to apply the reset location transformation on.
\paragraph{Weaving $\tuple{\WCurrentBefore, \WCurrentAfter}$.}
The edit frame $\tuple{\WCurrentBefore, \WCurrentAfter}$ applies at the level of transitions in $\mathit{M}$.
Therefore, $F_b$ (resp. $F_a$) is simply added at the beginning (resp. end) of $F$, resulting in $\mathit{F'} = \ffunc{F_b, F, F_a}$.
All transitions leading to the transitions in $M$, namely $\WGetP(M)$ must invoke $\aopclear$ after finishing their computation.
Care should be taken in the case of loops as they are both in the match and lead to the match.
$\aopset$ is executed to indicate that the joinpoint matched followed by $F_a$.
No additional locations are necessary, $L' = \locsof{B_k}$.
$\resetlocs$ consists of the locations to which any transition in $M$ leads, they are then $\WDest(M)$.
We define $m$ as follows:
\[
	m(t = \tuple{\ell, p, g,F, \ell'}) = \npartsdef{
		\setof{\tuple{\ell, p, g, F \aopclear, \ell'}} & \mbox{ if } t \in  \WGetP(M) \setminus Mm \\
		\setof{\tuple{\ell, p, g, F_b F \aopset F_a , \ell'}} & \mbox { if } t \in  M, \\
		\setof{t} & \mbox{otherwise}.
	}
\]
The new set of transitions is then $T' = \bigcup_{t \in \transof{B_k}}{m(t)}$.
The whole procedure is defined as:
$\weaveframe_{\mathrm{cur}}(B_k, M, F_b, F_a) = \tuple{T', L', L_{\mathrm{R}}, m}$.
\paragraph{Weaving $\tuple{\WPreviousAfter, \WCurrentBefore}$.}
Both $\WPreviousAfter$ and $\WCurrentBefore$ apply at the level of transitions.
$\WPreviousAfter$ indicates that $F_b$ must be woven on the transitions leading to those in $M$ namely $\WGetP(M)$, at the end of their update function.
$\WCurrentBefore$ indicates that $F_a$ must be woven on the transitions in $M$, at the beginning of their update function.
In this case, loops require more instrumentation.
Let the set of loop transitions be $T_{\mathrm{L}} = \WGetP(M) \cap M$, if we have such transitions $T_{\mathrm{L}} \neq \emptyset$, we create the set of extra locations $L_{\mathrm{temp}} = \setof{\ell^\bot \mid \ell \in \WOrigin(T_{\mathrm{L}})}$.
For a given loop transition $\tuple{\ell, p ,g, F, \ell}$, we set the update function to $\ffunc{F_a, \aopset, F}$ and change the destination to $\ell^\bot$ and add a transition that executes $F_b$ on port $ip$ from $\ell^\bot$ to $\ell$.
Thus if a reset location is to happen, it would happen on $\ell^\bot$, without executing function $F_b$ again.
\[
\begin{array}{ll}
	\resetlocs  & = \npartsdef{
			\WDest(M)\setminus \WOrigin(T_{\mathrm{L}}) \cup L_{\mathrm{temp}}
			& \mbox{ if } T_{\mathrm{L}} \neq \emptyset \\
			\WDest(M) & \mbox{otherwise}
	}\\
 \quad
	L' & = \npartsdef{
			\locsof{B_k} \cup L_{\mathrm{temp}}
			& \mbox{ if } T_{\mathrm{L}} \neq \emptyset \\
			\locsof{B_k} & \mbox{otherwise}
	}
\end{array}
\]

\[
	m(t = \tuple{\ell, p, g,F, \ell'}) = \npartsdef{
		\setof{\tuple{\ell, p, g, F \aopclear F_b, \ell'}} & \mbox{ if } t \in  \WGetP(M) \setminus M \\
		\setof{\tuple{\ell, p, g, F_a \aopset F, \ell^\bot}} & \mbox{ if } t \in  M \cap \WGetP(M) \\
		\setof{\tuple{\ell, p, g, F_a \aopset F , \ell'}} & \mbox { if } t \in  M \setminus \WGetP(M) \\
		\setof{t} & \mbox{otherwise}
	}
\]
\[
	T' = \setof{m(t) \mid t \in \transof{B_k}} \cup
		 \setof{\tuple{\ell^\bot, \aopport, \btrue, \aopclear F_b, \ell}
				\mid \ell^\bot \in L_{\mathrm{temp}}}
\]
The whole procedure is defined as:
$\weaveframe_{\mathrm{prev}}(B_k, M, F_b, F_a) = \tuple{T', L', L_{\mathrm{R}}, m}$.

\paragraph{Weaving $\tuple{\WCreate, \WCurrentBefore}$.}
\label{sec:weave-add}
This frame requires block instrumentation to handle $\pcportenabled$.
$\WCreate$ indicates that additional computations needs to be handled. These computations refer to port enablement since only matching at least one port enablement can lead to $\WCreate$.
In the first step, we define the selected ports.
Selected ports are the ports  matched as part of the pointcut expression $\mathit{lpc}$ using $\mathrm{sp}(\mathit{lpc}) = $ match $\mathit{lpc}$ with
\[
\begin{array}{llllll}
	      | \pcloc(\ell)  	  &\rightarrow \emptyset &
          | \pcfunc(\mathit{x})    	  &\rightarrow \emptyset &
          | \pcguard(\mathit{x})   	  &\rightarrow \emptyset \\
          | \pcwrite(\mathit{x})   	  &\rightarrow \emptyset &
          | \pcportenabled(\mathit{p}) &\rightarrow \setof{\mathit{p}} &
          | \pcportexec(\mathit{p}) 	  &\rightarrow \emptyset\\
          | \phi \hbox{ and } \phi' & \rightarrow \mathrm{sp}(\phi) & \cup \, \mathrm{sp}(\phi')
	\end{array}
\]
First we begin by creating the guard expression for the enabled ports.
A port $p$ is enabled in a location $\ell$ if there exists at least one transition which guard evaluates to $\btrue$.
In the case of multiple ports, they must all be enabled.
\[
	\mkAddGuard(SP, \ell, M) = \bigwedge_{p \in SP} (\bigvee_{\tau \in M \land \tau.\source = \ell \land \tau.\port = p} (\tau.\guard)
)
\]

Second, since port enablement can be detected only when the guards are evaluated at the location, to execute function $F_b$ we need to pre-evaluate the guards before entering the location.
To match a port enablement at a location $\ell$, we create a temporary location $\ell^\bot$ that pre-evaluates the guard. By applying this to the entire match $M$, we have $L_{\mathrm{temp}} = \setof{\ell^\bot \mid \ell \in \WOrigin(M)}$.
We then connect each $\ell^\bot$ to $\ell$ with two transitions, the first executes $F_b$ and does $\aopset$, indicating the pointcut match, and the second does $\aopclear$.
Both these transitions execute on $\aopport$ so as to not be enabled or execute any existing port.
The added transitions are:
\[
 T_{\mathrm{cr}} =
\bigcup_{\ell^\bot}(\setof{
			\tuple{\ell^\bot, \aopport, g, \aopset F_b, \ell},
			\tuple{\ell^\bot, \aopport, \neg g, \aopclear, \ell}
			} \mbox{ with }  g =\mkAddGuard(\mathrm{sp}(lpc), \ell, M) )
\]

Third, since port enablement is determined dynamically, we simulate an \code{if/else} construct.
Outgoing transitions from $\ell$ are duplicated. This results in two versions.
The first checks for the joinpoint match (if $\aopvar$ holds) and applies $F_a$.
The second checks for $(\aopvar = \bfalse)$ and does not apply $F_a$ weaving in $\WCurrentBefore$.

Lastly, all the incoming transitions from $\ell$ are redirected from $\ell$ to $\ell^\bot$ so as to reach $\ell^\bot$ to pre-evaluate the guard before $\ell$.
Thus, we get the resulting new set of locations $L'$ wand transitions $T'$.
\[
\begin{array}{ll}
	L' = \locsof{B_k} \cup L_{\mathrm{temp}}
& \quad
	T' = \setof{m(t) \mid t \in \transof{B_k}} \cup T_{\mathrm{cr}}
\end{array}
\]
with $t = \tuple{\ell, p, g,F, \ell'} $ and
\[
	m(t) = \npartsdef{
		\setof{
			\tuple{\ell, p, \aopvar \land g, F_a F, \ell'},
			\tuple{\ell, p, \neg\aopvar \land g, F, \ell'}
		}
		& \mbox{ if } t \in M\\
		\setof{\tuple{\ell, p, g,F, \ell'^\bot}} & \mbox{ if } t \in \WGetP(M) \setminus M \\
		\setof{t} & \mbox{otherwise}
	}
\]
In the case of loops, a reset location should happen on the temporary locations.
\[
	\resetlocs = \WDest(M) \setminus \WOrigin(M) \cup L_{\mathrm{temp}}
\]
The whole procedure is defined as:
$\weaveframe_{\mathrm{runb}}(B_k, M, F_b, F_a, lpc) = \tuple{T', L', L_{\mathrm{r}}, m}$.
\begin{example}[Dynamic weave]
\rfigb{fig:weave-add} depicts the weaving of the advice $\tuple{F_b, F_a, \emptyset}$ with the pointcut $\mathit{lpc} = \pcloc(\ell_2)$ $\mathrm{and}$ $\pcportenabled(p_1)$ $\mathrm{and}$ $\pcportenabled(p_2)$.
The selected ports are $\mathrm{sp}(\mathit{lpc}) = \setof{p_1,p_2}$, the origin is determined, $\WOrigin(\setof{t_2,t_3,t_4}) = \setof{\ell_2}$.
The condition for the enabled ports is: $g_\mathrm{add} = \mkAddGuard(\setof{p_1,p_2},$ $ \ell_2, \setof{t_2,t_3,t_4}) = g_2 \land (g_3 \lor g_4)$.
Both ports $\setof{p_1,p_2}$ are enabled when $p_1$ is enabled ($g_2$ is $\btrue$) and $p_2$ is enabled ($g_3 \lor g_4$ is $\btrue$).
We create a set of temporary locations: $L_\mathrm{temp} = \setof{\ell_2^{\bot}}$.
Two transitions are created per location.
The first executes iff the ports are both enabled (pointcut matched), its update function is $F_b$ and $\aopset$.
The second executes iff one of the ports is not enabled (pointcut not matched), its update function does $\aopclear$.
Transitions with $\ell_2$ as destination are redirected to $\ell_2^\bot$.
Then, we copy over the originally outgoing transitions from $\ell_2$, creating two versions of them: one has $F_a$ and is guarded by $\aopvar$ and another executes normally and is guarded by $\neg\aopvar$.
\end{example}

\getexamples[weave-add]{weave/add,weave/add-3}
			{Weaving $\tuple{\WCreate, \WCurrentBefore}$}{0.62}{2}
\paragraph{Weaving $\tuple{\WCreate, \WCurrentAfter}$.}
The frame is woven similarly to $\tuple{\WCreate, \WCurrentBefore}$.
The $\WCreate$ edit point indicates that $\pcportenabled$ must hold, so the transformations described in the previous part are similar.
However, we modify the order of execution of $F_b$ and $F_a$.
Since $F_b$ executes on $\WCurrentBefore$, we do not execute it on the added transitions that detect port enabled.
\[
 T_{\mathrm{cr}} =
\bigcup_{\ell^\bot}(\setof{
			\tuple{\ell^\bot, \aopport, g, \aopset, \ell},
			\tuple{\ell^\bot, \aopport, \neg g, \aopclear, \ell}
			} \mbox{ with }  g =\mkAddGuard(\mathrm{sp}(lpc), \ell, M) ).
\]
Instead, $F_b$ needs to be added to the transition which matches the port enablement.
Therefore,  $F_b$ and $F_a$ are added to the duplicated transitions that match the guard $g$.
\[
\begin{array}{ll}
	L' = \locsof{B_k} \cup L_{\mathrm{temp}}
& \quad
	T' = \setof{m(t) \mid t \in \transof{B_k}} \cup T_{\mathrm{cr}}
\end{array}
\]
with $t = \tuple{\ell, p, g,F, \ell'} $ and
\[
	m(t) = \npartsdef{
		\setof{
			\tuple{\ell, p, \aopvar \land g, F_b F F_a, \ell'},
			\tuple{\ell, p, \neg\aopvar \land g, F, \ell'}
		}
		& \mbox{ if } t \in M,\\
		\setof{\tuple{\ell, p, g,F, \ell'^\bot}} & \mbox{ if } t \in \WGetP(M) \setminus M, \\
		\setof{t} & \mbox{otherwise}.
	}
\]
When considering reset location, loops and temporary locations must be adjusted similarily to weaving $\tuple{\WCreate, \WCurrentBefore}$:
\[
	\resetlocs = \WDest(M) \setminus \WOrigin(M) \cup L_{\mathrm{temp}}
\]
The whole procedure is defined as:
$\weaveframe_{\mathrm{runa}}(B_k, M, F_b, F_a, lpc) = \tuple{T', L', L_{\mathrm{r}}, m}$.
\paragraph{Weaving a local aspect.}
The local weave operation weaves an advice on set of transitions $M$ with the set of intertype variables $V$ and a local pointcut expression $lpc$.
\begin{definition}[Local Weave]
\label{def:local-weave}
	The local weave is defined as:
\[
	\tuple{\composite', \tmap} = \fweavel(\composite, B_k, V, lpc, M, ladv(B_k, V) = \tuple{F_b, F_a, R})
\]
where $\composite' = \pi(\gamma'((\atoms \setminus B_k) \cup B'_k))$ is the resulting composite component; with:
\begin{itemize}
	\item $B_k' = \tuple{P \cup \setof{ip}, L', X \cup V \cup \setof{\aopvar}, T' \cup T_{\mathrm{R}}}$ is the resulting atomic component;
	\item $\aopvar$ is the aspect boolean variable described earlier;
	\item $\tuple{T', L', L_{\mathrm{R}} , m} = \npartsdef{
	\weaveframe_{\mathrm{cur}}(B_k, M, F_b, F_a)
	& \mbox{ if } \pcedit(lpc) = \tuple{\WCurrentBefore, \WCurrentAfter}\\
	\weaveframe_{\mathrm{prev}}(B_k, M, F_b, F_a)
	& \mbox{ if } \pcedit(lpc) = \tuple{\WPreviousAfter, \WCurrentBefore}\\
	\weaveframe_{\mathrm{runb}}(B_k, M, F_b, F_a, lpc)
	& \mbox{ if } \pcedit(lpc) = \tuple{\WCreate, \WCurrentBefore}\\
	\weaveframe_{\mathrm{runa}}(B_k, M, F_b, F_a, lpc)
	& \mbox{ if } \pcedit(lpc) = \tuple{\WCreate, \WCurrentAfter}\\
}$
	\item $T_{\mathrm{R}} = \weavereset(R, L_{\mathrm{R}})$ is obtained from weaving reset locations;
	\item $\gamma' = \gamma \cup \setof{a_{\aopport}}$ where $a_{\aopport} = \tuple{\aopport, \btrue, \ffunc{}}$ is the high priority interaction for the aop port;
	\item $\pi'= \pi \cup \setof{\tuple{a, a_{\aopport}} \mid a \in \pi} $ is the new set of priorities.
\end{itemize}
Weaving a local aspect on a composite component $\composite = \pi(\gamma(\atoms))$ is defined as:
\[
\composite' = \composite \weave \tuple{B_k \in \atoms, lpc, V, ladv(B_k, V)}
\]
where: $\composite'$ is the result from the local weave: $\tuple{\composite', \tmap} = \fweavel(\composite, B_k, V, lpc, \pcmatch(B_k, lpc), $ $ ladv(B_k, V))$.
\end{definition}
The new atomic component $B'_k$ has one extra port ($\aopport$), and has $\varsof{B_k} \cup V \cup \setof{\aopvar}$ as the set of variables.
The edit frame is determined using operator $\pcedit$ (\defref{def:edit}) and transitions and locations are instrumented accordingly.
The obtained composite component $\composite'$ has one extra singleton interaction $a_{\aopport}$ associated with port $ip$.
Additionally, interaction $a_{\aopport}$ is given the highest priority w.r.t. predefined interactions.
\paragraph{Correctness of local weave.}
Informally, to verify the correct addition of the advice w.r.t joinpoint matching, we need to first verify that the before and after update functions were placed correctly when a joinpoint is matched during runtime.
This is similar to the correctness of global weaving expressed in Proposition~\ref{prop:global-apply}.
However, we note that the before update function ($F_b$) must, in some cases, execute in the local event preceding the match.
Secondly, in the presence of reset location, the local event succeeding the matched event should indicate that the local component is in the appropriate location.
As such, the order of the events is also checked in the case of local weaving.

We first need to make the distinction between the frames that weave $F_b$ on the previous transitions with those that do not.

We define predicate $\mathrm{early}(lpc) = \npartsdef{
	\bfalse & \mbox{ if } \pcedit(lpc) \in \setof{ \tuple{\WCurrentBefore, \WCurrentAfter}, \tuple{\WCreate, \WCurrentAfter}},\\
	\btrue & \mbox{ otherwise}.
}$

We consider $\events'_k$ (resp. $\events_k$) to be reachable events in $B'_k$ (resp. $B_k$).
We define $\erem_k : \events'_k \times \codes \times \codes \rightarrow \events_k \cup \setof{\epsilon}$, the function that removes the local advice from a local event.
\[
	\erem_k(\tuple{\tuple{l, v}, \tau, q}, F_b, F_a) = \npartsdef{
	\tuple{\tuple{l, v'}, \tau', q'}
	& \mbox{ if } l \in  \locsof{B_k}, \\
	\epsilon & \mbox{ otherwise}.
}
\]
The constructed event is the following:
\begin{itemize}
	\item $v'$ excludes the valuations of the inter-type $V$ from $v$ while the location is maintained.
	\item If the event has a location not found in $\locsof{B_k}$, then a similar event cannot be constructed. The resulting event is $\epsilon$.
	\item $\tau' = \tuple{l, \mathit{\tau.\port}, g, F, l'}$ is the similar transition where $g$ does not contain $\aopvar$, $F$ does not contain $\aopset, \aopclear, F_b, F_a$ and $l'$ is any location (since $q'$ is never matched against a joinpoint so it is not relevant).
\end{itemize}
By removing the advice, we ensure that $lpc$ must not match read/writes introduced by $F_b$ and $F_a$.

We use $\mathrm{before}(e_i, F, d)$ and $\mathrm{after}(e_i, F, d)$ where $e_i$ is a local event, $F$ is the before or after update function, and $d$ is the value of the predicate $\mathrm{early}$, to verify the correct application of the before and after update functions respectively.
\begin{align*}
	\mathrm{before}(e_i, F, d) & \mbox{ iff } (b \land i \neq 0 \land \exists F' : \mathit{e_{i-1}.\tau.\func} = \ffunc{F', F}) \\
		& \quad \quad \lor (\neg b \land \exists F' : \mathit{e_{i}.\tau.\func} = \ffunc{F, F'}) \\
	\mathrm{after}(e_i, F, d) & \mbox { iff } (b \land \exists F' : \mathit{e.\tau.\func} = \ffunc{F, F'}) \lor (\neg b \land \exists F' : \mathit{e.\tau.\func} = \ffunc{F', F})
\end{align*}
In the case where the predicate $\mathrm{early}(lpc)$ holds, the predicate $\mathrm{before}(e_i, F, d)$ checks if the function of the preceding event's update function ($\mathit{e_{i-1}.\tau.\func}$) ends with $F$, with the exception of the first event ($i \neq 0$), while the predicate $\mathrm{after}$ checks that the current event's function ($\mathit{e_i.\tau.\func}$) starts with $F$.
In the case where the predicate $\mathrm{early}(lpc)$ does not hold, the predicate $\mathrm{before}$ (resp. $\mathrm{after}$) ensures that the current event's update function ($e_{i}.\tau.\func$) starts (resp. ends) with $F$.

For the case of a reset location pair $r = \tuple{\mathit{guard}, \mathit{loc}}$ we check if its guard holds \btrue on the next event. If it does, we verify that the location $\mathit{e_{i+1}.l}'$ is the destination location in a reset location pair.
\[
	\mathrm{reset}(e_i, r) \mbox{ iff } (guard(\mathit{e_{i+1}.v}) \implies \mathit{e_{i+1}.l'} = \mathit{loc})
\]
We can now express the correct application as follows:
\begin{proposition}{Correctness of local advice weaving.}
\label{prop:local-apply}
Consider $T'$ to be the sequence of global events of the BIP System $\tuple{\composite', Q_0}$ where $\composite' = \composite \weave \tuple{B_k, \mathit{lpc}, V, ladv(B_k, V)}$ and $B'_k$ is the new atomic component with $F_b$ and $F_a$ as the advice before and after update functions. Let $T'_k = \trproject(T', B'_k) = (e_0 \cdot e_1 \cdot \hdots)$, $d = \mathrm{early}(lpc)$, we have $\forall e_i \in T'_k$, $e'_i = \erem_k(e'_i, F_b, F_a)$:
\begin{align}
	(e'_i \neq \epsilon \land \evalLocal{e'_i}{\mathit{lpc}}) & \mbox { iff } \mathrm{before}(e_i, F_b, d)
								   \land \mathrm{after}(e_i, F_a, d)\\
	(e'_i \neq \epsilon \land \evalLocal{e'_i}{\mathit{lpc}}) & \implies  (R \neq \emptyset \implies \exists r \in R : \mathrm{reset}(e_i, r))
\end{align}
\end{proposition}
We begin by constructing an event $e'_i = \erem_k(e_i, F_b, F_a)$ from $e_i$ by removing the local advice.
The proposition states that $e'_i$ is a joinpoint ($\evalLocal{e_i}{lpc}$) in the original system iff the update function associated with $e_i$ verifies the rules for before ($ \mathrm{before}(e_i, F_b, d)$) and after ($\mathrm{after}(e_i, F_a, d)$).
In the presence of reset location pairs (2),  if the event $e'_i$ is a joinpoint in the original system, the next event includes at least one reset location pair ($R \neq \emptyset \implies \exists r \in R : \mathrm{reset}(e_i, r)$).

%
\section{Weaving Strategies}\label{sec:containers}
An aspect is the single association of a pointcut expression to a joinpoint.
However, when weaving more than one aspect, specific problems and extra considerations arise.
This section identifies possible issues when weaving multiple aspects and presents two basic procedures to coordinate the weaving.

Furthermore, this section can be seen as a preliminary method to provide a grouping of aspects and basic strategies for weaving them.
It can be seen as the interface between the user and the transformations described earlier in the paper for the two views.
As such, while we present a basic overview of the strategies to combine and weave multiple aspects, we note that it is possible to use the transformations to build more complex ones that best suit the users' needs.
%
\subsection{Interference}
%
Recall that multiple concerns may happen at one joinpoint.
This can be seen as the \emph{tangling} phenomenon.
When a new concern is added to the joinpoint, it is possible to have existing concerns at the same joinpoint.
This situation is referred to as \emph{interference}.
Since not all concerns are independent, interference is an important issue to study.
\begin{example}[Interference]
Consider $\mathit{gpc} = \tuple{\setof{p_1}, \emptyset, \emptyset}$, $V = \setof{x}$, the two global aspects
	$\mathit{GA_1} = \tuple{\composite, V, \mathit{gpc}, \tuple{F_b = \ffunc{x := 3}, F_a}}$, and
	$\mathit{GA_2} = \tuple{\composite, V, \mathit{gpc}, \tuple{F'_b = \ffunc{x := 2}, F'_a}}$.
Both of these aspects' advices operate on the same inter-type variable $x$ and on the same matched joinpoints.
There are four possibilities for weaving the advices, depending on the order of $F_b, F'_b, F_a$ and $F'_a$.
In the case of $F_b$ and $F'_b$ if we have $\ffunc{F_b, F'_b}$ (resp. $\ffunc{F'_b, F_b}$) then $x$ will be $2$ (resp. $3$) at runtime.
\end{example}
Defining weaving strategies helps to deal with interference in a more predictable way.
To do so, we examine in the following: (1) a modular unit that groups aspects, and (2) the operations that weave multiple aspects.
%
\subsection{Containers}
%
Aspect containers encapsulate a group of aspects.
Local containers (resp. global containers) apply to local (resp. global) aspects.
Aspect containers seek to group interfering aspects and define extra restrictions so as to manage their weaving.
By doing so, we expose multiple aspects to the user as a coherent unit.
\begin{definition}[Global and local containers]
A global container is a tuple $\tuple{\tuple{\aspectglobal_1, \ldots, \aspectglobal_n},V'}$ such that any $\aspectglobal_j \in \setof{\aspectglobal_1, \ldots, \aspectglobal_n} $ has the inter-type $V'$.
A local container is a tuple $\tuple{\tuple{\aspectlocal_1, \ldots, \aspectlocal_m}, B, V}$ such that any $\aspectlocal_i \in  \setof{\aspectlocal_1, \ldots, \aspectlocal_m}$ is applied to an atomic component $B$ and has the inter-type $V$.
\end{definition}
Containers define an order on the aspects they encapsulate.
They permit the definition of a weaving order for aspects.
Moreover, containers ensure that aspects share the same inter-type variables.
Sharing allows the inter-type to be encapsulated in the container.
In the case of local containers, local aspects are required to operate on the same atomic component encouraging encapsulation.
The local aspects operating on different atomic components do not interfere and cannot share inter-type variables.
%
\subsection{Weaving Procedures}
%
\paragraph{The $\weaveserial$ procedure.}
Given a global or local container, one example of weaving procedure weaves aspects in the order they are contained.
The aspects are presented in a sequence  $\tuple{asp_1, \ldots, asp_n}$ where $asp_1, \ldots, asp_n$ are all either global or local aspects.
\begin{definition}[$\weaveserial$] \label{def:composition:wserial} Depending on the aspect type we have two operations:
\begin{itemize}
	\item The procedure $\weaveserial_\ell$ applied to a sequence of local aspects  $\tuple{asp_1, \ldots, asp_n}$ is: $\composite' = \weaveserial_\ell(\composite, \tuple{asp_1, \ldots, asp_n}) = ((((\composite \weave asp_1) \weave asp_2) \weave ..) \weave asp_n)$
	\item The procedure $\weaveserial_g$ applied to a sequence of global aspects  $\tuple{asp_1, \ldots, asp_m}$ is: $\composite' = \weaveserial_g(\composite, \tuple{asp_1, \ldots, asp_m}) = ((((\composite \weaveg asp_1) \weaveg asp_2) \weaveg ..) \weaveg asp_m)$
\end{itemize}
\end{definition}
The $\weaveserial$ procedure allows the pointcut of any aspect $asp_i$ ($i \in \setof{2, \ldots, n}$) to match changes introduced by all aspects $asp_j$ prior to it (for $j < i$).

While \defref{def:composition:wserial} presents a simple weaving strategy, we can consider alternative strategies.
Firstly, we can consider that the matching of the pointcut and the weaving are separate.
This is the case for $\weaveall$ presented in the next paragraph.
Secondly, we note that it is possible to modify the order of the aspects in the sequence.
For example, it is possible to order aspects by (1) the type of the pointcut's edit frame or (2) by whether or not their advice performs only read or also modifies the variables.
\paragraph{The $\weaveall$ procedure.}
The weaving procedure for local aspects introduces new locations and transitions in the cases of reset locations and port enabled.
The second compositional approach $\weaveall$ allows the pointcut of a local aspect $asp_i$ in the sequence to not match extra transitions added by $asp_j$ (for $j : j < i$).
While $\weaveserial$ can be seen as iteratively doing match and weave for each aspect, $\weaveall$ makes it is possible to match the joinpoints of the entire sequence of aspects before weaving them.
The procedure $\weaveall$ matches all the pointcuts of the sequence, then weaves the aspects according to their order based on their original match projected onto the new component.
For a local aspect $\aspectlocal = \tuple{B, \mathit{pc}, V, F_b, F_a, R}$ we denote $B, pc, V, \tuple{F_b, F_a, R}$ by $\mathit{\aspectlocal.B}$, $\mathit{\aspectlocal.pc}$, $\mathit{\aspectlocal.V}$ and $\mathit{\aspectlocal.adv}$, respectively.

If we consider two local aspects $asp_1$, $asp_2$ and their local pointcut matches $M_1$ and $M_2$ respectively.
Weaving $asp_1$ on its match $M_1$, using $\fweavel(\composite_0, asp_1.B, asp_1.V, $ $asp_1.pc,$ $M_1$, $ asp_1.adv)$, results in $\tuple{\composite_1, \tmap_1}$.
The weaving of $asp_2$ needs to apply on $\composite_1$ and not on $\composite_0$ on which $asp_1.pc$ was matched.
Therefore its original match $M_2$ needs to apply to transitions in $\composite_1$ as the local weave could change the transitions.
To do so we use the transformation function $m_1$ (see \defref{def:local-weave}) to get the transitions in $\composite_1$ from $\composite_0$.
Therefore the new set of matches in $\composite_1$ given $M_2$ is determined using $\mathrm{newM}(M_2, \tmap_1)$.
\[
	\mathrm{newM}(M, \tmap) = \setof{\tmap(t) \mid t \in M}
\]
Since $\tmap$ applies only to one weave, we generalize it to the $k^{th}$ weave by successive application over the $k-1$ weaves using $\tuple{\tmap_1 \ldots, \tmap_{k - 1}}$.
\[
	\mathrm{follow}(M, \tuple{\tmap_1, \ldots, \tmap_{k-1}}) =  \twopartdef{\mathrm{newM}(\mathrm{newM}(\mathrm{newM}(M, \tmap_1), \ldots), \tmap_{k-1})} {k - 1 > 1}
								{M} {\text{otherwise}}
\]
\begin{definition}[$\weaveall$]
The $\weaveall$ for a sequence of $n$ local aspects on a composite component $\composite_0$ is defined recursively as $\tuple{\composite_n, \tmap_n} = \weaveall(\composite_0, \tuple{asp_1, \ldots, asp_n})$ such that $\forall i \in \setof{1, \ldots, n}$:
\begin{align*}
	\tuple{\composite_i, \tmap_i} &= \fweavel(\composite_{i - 1}, follow(M_i, \tuple{\tmap_1, \ldots, \tmap_{i-1}}), \mathit{asp_i.V}, \mathit{asp_i.adv})\\
	M_i &= \pcmatch(\composite_0, asp_i.pc)
\end{align*}
\end{definition}
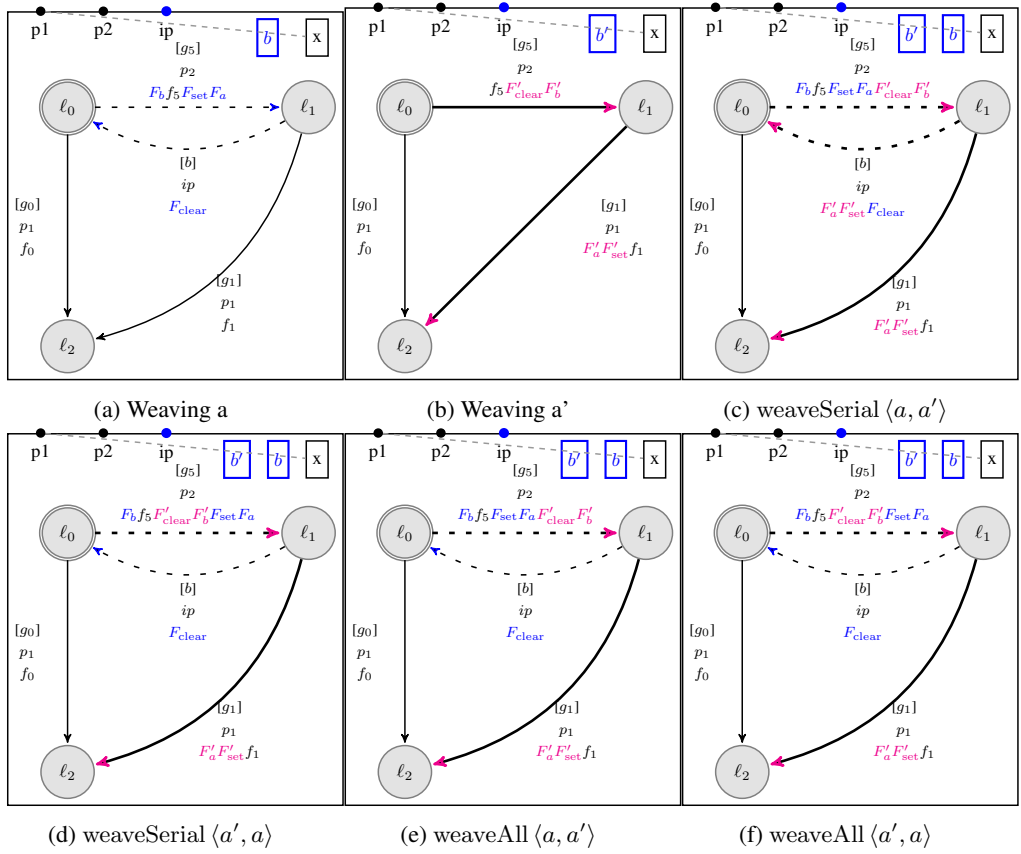
\begin{figure}[t]
	\centering
	\begin{subfigure}[b]{0.3\textwidth}%
		\scalebox{0.7}{  \begin{tikzpicture}[bip]
 
    \component[explode]{c1}{atomic}{}{

      \node[place,double] (c1l0) {$\ell_0$};
      \node[place] (c1l1) [right=of c1l0] {$\ell_1$}; 
      \node[place] (c1l2) [below=of c1l0] {$\ell_2$};

	\tconnect[t5]{previous}
		    {c1l0} {c1l1}
		    {0.1cm}{above}
		   {$g_5$}
		   {$p_2$}
		   {$\chigh{F_b} f_5 \chigh{\aopset F_a}$}

	\tconnect[t0]{normal}
		    {c1l0} {c1l2}
		    {0.5cm}{left}
		   {$g_0$}
		   {$p_1$}
		   {$f_0$}

	\tconnect[t1]{normal, bend left}
		    {c1l1} {c1l2}
		    {0.1cm}{below}
		   {$g_1$}
		   {$p_1$}
		   {$f_1$}
	\tconnect[r1]{previous, bend left}
			{c1l1}{c1l0}
			{0.1cm}{below}
			{$b$}
			{$ip$}
			{\chigh{\aopclear}}

};

    \node[export] (c1p1) at ($(c1.north west)!.1!(c1.north east)$) [label=below:p1]    {};
    \node[export] (c1p2) [right=of c1p1] [label=below:p2] {};
    \node[export, high] (c1ip) [right=of c1p2] [label=below:ip] {};
   
	\node[var] (x) at ( $(c1.north east) + (-0.5, -0.5)$) []  {x};
	\node[var, high] (b2)  [left=of x, shift={(0.5,0)}]  {$b$};	
	\draw[portvar] (c1p1) -- (x);
	
  \end{tikzpicture}
  }%
		\caption{Weaving a}%
		\label{fig:inter-a}%
	\end{subfigure}%
	\hspace{0.3cm}
	\begin{subfigure}[b]{0.3\textwidth}%
		\scalebox{0.7}{  \begin{tikzpicture}[bip]
 
    \component[explode]{c1}{atomic, explode}{}{

      \node[place,double] (c1l0) {$\ell_0$};
      \node[place] (c1l1) [right=of c1l0] {$\ell_1$}; 
      \node[place] (c1l2) [below=of c1l0] {$\ell_2$};

	\tconnect[t5]{selected}
		    {c1l0} {c1l1}
		    {0.1cm}{above}
		   {$g_5$}
		   {$p_2$}
		   {$f_5 \cselect{\aopclear' F'_b} $}

	\tconnect[t0]{normal}
		    {c1l0} {c1l2}
		    {0.5cm}{left}
		   {$g_0$}
		   {$p_1$}
		   {$f_0$}

	\tconnect[t1]{selected}
		    {c1l1} {c1l2}
		    {1cm}{right}
		   {$g_1$}
		   {$p_1$}
		   {$\cselect{F'_a \aopset'} f_1$}

};

    \node[export] (c1p1) at ($(c1.north west)!.1!(c1.north east)$) [label=below:p1]    {};
    \node[export] (c1p2) [right=of c1p1] [label=below:p2] {};
    \node[export, high] (c1ip) [right=of c1p2] [label=below:ip] {};
   
	\node[var] (x) at ( $(c1.north east) + (-0.5, -0.5)$) []  {x};
	\node[var, high] (b2)  [left=of x, shift={(0.5,0)}]  {$b'$};	
	\draw[portvar] (c1p1) -- (x);
	
  \end{tikzpicture}
  }%
		\caption{Weaving a'}%
		\label{fig:inter-b}%
	\end{subfigure}%
	\hspace{0.3cm}
	\begin{subfigure}[b]{0.3\textwidth}%
		\scalebox{0.7}{  \begin{tikzpicture}[bip]
 
    \component[explode]{c1}{atomic}{}{

      \node[place,double] (c1l0) {$\ell_0$};
      \node[place] (c1l1) [right=of c1l0] {$\ell_1$}; 
      \node[place] (c1l2) [below=of c1l0] {$\ell_2$};

	\tconnect[t5]{previous, selected}
		    {c1l0} {c1l1}
		    {0.1cm}{above}
		   {$g_5$}
		   {$p_2$}
		   {$\chigh{F_b} f_5 \chigh{\aopset F_a} \cselect{\aopclear' F'_b}$}

	\tconnect[t0]{normal}
		    {c1l0} {c1l2}
		    {0.5cm}{left}
		   {$g_0$}
		   {$p_1$}
		   {$f_0$}

	\tconnect[t1]{selected, bend left}
		    {c1l1} {c1l2}
		    {0.1cm}{below}
		   {$g_1$}
		   {$p_1$}
		   {$\cselect{F_a' \aopset'} f_1$}
	\tconnect[r1]{previous, selected, bend left}
			{c1l1}{c1l0}
			{0.1cm}{below}
			{$b$}
			{$ip$}
			{$\cselect{F_a' \aopset'} \chigh{\aopclear}$}

};

    \node[export] (c1p1) at ($(c1.north west)!.1!(c1.north east)$) [label=below:p1]    {};
    \node[export] (c1p2) [right=of c1p1] [label=below:p2] {};
    \node[export, high] (c1ip) [right=of c1p2] [label=below:ip] {};

	\node[var] (x) at ( $(c1.north east) + (-0.5, -0.5)$) []  {x};
	\node[var, high] (b2)  [left=of x, shift={(0.7,0)}]  {$b$};	
	\node[var, high] (b1)  [left=of b2, shift={(0.7,0)}]  {$b'$};	
	
	\draw[portvar] (c1p1) -- (x);
	
  \end{tikzpicture}
  }%
		\caption{$\weaveserial \tuple{a, a'}$}%
		\label{fig:inter-c}%
	\end{subfigure}%

	\begin{subfigure}[b]{0.3\textwidth}%
		\scalebox{0.7}{  \begin{tikzpicture}[bip]
 
    \component[explode]{c1}{atomic}{}{

      \node[place,double] (c1l0) {$\ell_0$};
      \node[place] (c1l1) [right=of c1l0] {$\ell_1$}; 
      \node[place] (c1l2) [below=of c1l0] {$\ell_2$};

	\tconnect[t5]{previous, selected}
		    {c1l0} {c1l1}
		    {0.1cm}{above}
		   {$g_5$}
		   {$p_2$}
		   {$\chigh{F_b} f_5 \cselect{\aopclear' F'_b} \chigh{\aopset F_a}$}

	\tconnect[t0]{normal}
		    {c1l0} {c1l2}
		    {0.5cm}{left}
		   {$g_0$}
		   {$p_1$}
		   {$f_0$}

	\tconnect[t1]{selected, bend left}
		    {c1l1} {c1l2}
		    {0.1cm}{below}
		   {$g_1$}
		   {$p_1$}
		   {$\cselect{F_a' \aopset'} f_1$}
	\tconnect[r1]{previous, bend left}
			{c1l1}{c1l0}
			{0.1cm}{below}
			{$b$}
			{$ip$}
			{$\chigh{\aopclear}$}

};

    \node[export] (c1p1) at ($(c1.north west)!.1!(c1.north east)$) [label=below:p1]    {};
    \node[export] (c1p2) [right=of c1p1] [label=below:p2] {};
    \node[export, high] (c1ip) [right=of c1p2] [label=below:ip] {};

	\node[var] (x) at ( $(c1.north east) + (-0.5, -0.5)$) []  {x};
	\node[var, high] (b2)  [left=of x, shift={(0.7,0)}]  {$b$};	
	\node[var, high] (b1)  [left=of b2, shift={(0.7,0)}]  {$b'$};	
	\draw[portvar] (c1p1) -- (x);
	
  \end{tikzpicture}
  }%
		\caption{$\weaveserial \tuple{a', a}$}%
		\label{fig:inter-d}%
	\end{subfigure}%
	\hspace{0.3cm}
	\begin{subfigure}[b]{0.3\textwidth}%
		\scalebox{0.7}{  \begin{tikzpicture}[bip]
 
    \component[explode]{c1}{atomic}{}{

      \node[place,double] (c1l0) {$\ell_0$};
      \node[place] (c1l1) [right=of c1l0] {$\ell_1$}; 
      \node[place] (c1l2) [below=of c1l0] {$\ell_2$};

	\tconnect[t5]{previous, selected}
		    {c1l0} {c1l1}
		    {0.1cm}{above}
		   {$g_5$}
		   {$p_2$}
		   {$\chigh{F_b} f_5 \chigh{\aopset F_a} \cselect{\aopclear' F'_b}$}

	\tconnect[t0]{normal}
		    {c1l0} {c1l2}
		    {0.5cm}{left}
		   {$g_0$}
		   {$p_1$}
		   {$f_0$}

	\tconnect[t1]{selected, bend left}
		    {c1l1} {c1l2}
		    {0.1cm}{below}
		   {$g_1$}
		   {$p_1$}
		   {$\cselect{F_a' \aopset'} f_1$}
	\tconnect[r1]{previous, bend left}
			{c1l1}{c1l0}
			{0.1cm}{below}
			{$b$}
			{$ip$}
			{$\chigh{\aopclear}$}

};

    \node[export] (c1p1) at ($(c1.north west)!.1!(c1.north east)$) [label=below:p1]    {};
    \node[export] (c1p2) [right=of c1p1] [label=below:p2] {};
    \node[export, high] (c1ip) [right=of c1p2] [label=below:ip] {};

	\node[var] (x) at ( $(c1.north east) + (-0.5, -0.5)$) []  {x};
	\node[var, high] (b2)  [left=of x, shift={(0.7,0)}]  {$b$};	
	\node[var, high] (b1)  [left=of b2, shift={(0.7,0)}]  {$b'$};	
	
	\draw[portvar] (c1p1) -- (x);
	
  \end{tikzpicture}
  }%
		\caption{$\weaveall \tuple{a, a'}$}%
		\label{fig:inter-e}%
	\end{subfigure}%
	\hspace{0.3cm}
	\begin{subfigure}[b]{0.3\textwidth}%
		\scalebox{0.7}{  \begin{tikzpicture}[bip]
 
    \component[explode]{c1}{atomic}{}{

      \node[place,double] (c1l0) {$\ell_0$};
      \node[place] (c1l1) [right=of c1l0] {$\ell_1$}; 
      \node[place] (c1l2) [below=of c1l0] {$\ell_2$};

	\tconnect[t5]{previous, selected}
		    {c1l0} {c1l1}
		    {0.1cm}{above}
		   {$g_5$}
		   {$p_2$}
		   {$\chigh{F_b} f_5 \cselect{\aopclear' F'_b} \chigh{\aopset F_a}$}

	\tconnect[t0]{normal}
		    {c1l0} {c1l2}
		    {0.5cm}{left}
		   {$g_0$}
		   {$p_1$}
		   {$f_0$}

	\tconnect[t1]{selected, bend left}
		    {c1l1} {c1l2}
		    {0.1cm}{below}
		   {$g_1$}
		   {$p_1$}
		   {$\cselect{F_a' \aopset'} f_1$}
	\tconnect[r1]{previous, bend left}
			{c1l1}{c1l0}
			{0.1cm}{below}
			{$b$}
			{$ip$}
			{$\chigh{\aopclear}$}

};

    \node[export] (c1p1) at ($(c1.north west)!.1!(c1.north east)$) [label=below:p1]    {};
    \node[export] (c1p2) [right=of c1p1] [label=below:p2] {};
    \node[export, high] (c1ip) [right=of c1p2] [label=below:ip] {};
   
	\node[var] (x) at ( $(c1.north east) + (-0.5, -0.5)$) []  {x};
	\node[var, high] (b2)  [left=of x, shift={(0.7,0)}]  {$b$};	
	\node[var, high] (b1)  [left=of b2, shift={(0.7,0)}]  {$b'$};	
	\draw[portvar] (c1p1) -- (x);
	
  \end{tikzpicture}
  }%
		\caption{$\weaveall \tuple{a', a}$}%
		\label{fig:inter-F}%
	\end{subfigure}%
	\caption{Weave Procedures}%
	\label{fig:inter}%
\end{figure}

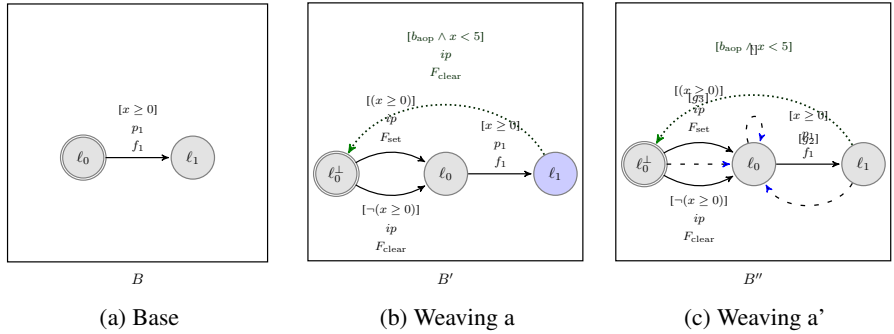
\begin{figure}[t]%
	\centering
	\begin{subfigure}[b]{0.3\textwidth}%
	\centering
		\scalebox{0.57}{  \begin{tikzpicture}[bip]
 
    \component{c1}{minimum size=6cm}{}{

      \node[place,double] (c1l0) {$\ell_0$};
      \node[place] (c1l1) [right=of c1l0] {$\ell_1$}; 

	\tconnect[t0]{normal}
		    {c1l0} {c1l1}
		    {0.1cm}{above}
		   {$x \geq 0$}
		   {$p_1$}
		   {$f_1$}

 	}; 

    \node at ($(c1.south west)!.5!(c1.south east)$)  [label=below:$\bipAtomicExample$]    {};

  \end{tikzpicture}
  }
		\caption{Base} %
		\label{fig:inter-locs-1} %
	\end{subfigure}%
	\begin{subfigure}[b]{0.3\textwidth}%
	\centering
		\scalebox{0.57}{  \begin{tikzpicture}[bip]
 
    \component{c1}{minimum size=6cm}{}{

      \node[place,double] (c1l0) {$\ell_0^\bot$};
      \node[place] (c1l1) [right=of c1l0] {$\ell_0$}; 
      \node[place, fill=\colorhigh!20] (c1l2) [right=of c1l1] {$\ell_1$}; 

	\tconnect[t0]{normal}
		    {c1l1} {c1l2}
		    {0.1cm}{above}
		   {$x \geq 0$}
		   {$p_1$}
		   {$f_1$}

	\tconnect[t01]{normal, bend left}
		{c1l0}{c1l1}
		{0.2cm}{above}
		{$(x\geq0)$}
		{$ip$}
		{$\aopset$}

	\tconnect[t02]{normal, bend right}
		{c1l0}{c1l1}
		{0.2cm}{below}
		{$\neg(x\geq0)$}
		{$ip$}
		{$\aopclear$}

	\tconnect[loc1]{created, bend right=60}
		{c1l2}{c1l0}
		{0.6cm}{above}
		{$\aopvar \land x < 5$}
		{$ip$}
		{$\aopclear$}
 	}; 

    \node at ($(c1.south west)!.5!(c1.south east)$)  [label=below:$\bipAtomicExample'$]    {};

  \end{tikzpicture}
  }
		\caption{Weaving a} %
		\label{fig:inter-locs-2} %
	\end{subfigure}%
	\begin{subfigure}[b]{0.3\textwidth}%
	\centering
		\scalebox{0.57}{  \begin{tikzpicture}[bip]
 
    \component{c1}{minimum size=6cm}{}{

      \node[place,double] (c1l0) {$\ell_0^\bot$};
      \node[place] (c1l1) [right=of c1l0] {$\ell_0$}; 
      \node[place] (c1l2) [right=of c1l1] {$\ell_1$}; 

	\tconnect[t0]{normal, sample}
		    {c1l1} {c1l2}
		    {0.1cm}{above}
		   {$x \geq 0$}
		   {$p_1$}
		   {$f_1$}

	\tconnect[t01]{normal, sample,  bend left}
		{c1l0}{c1l1}
		{0.2cm}{above}
		{$(x\geq0)$}
		{$ip$}
		{$\aopset$}

	\tconnect[t02]{normal, sample,  bend right}
		{c1l0}{c1l1}
		{0.2cm}{below}
		{$\neg(x\geq0)$}
		{$ip$}
		{$\aopclear$}

	\tconnect[loc1]{created, sample, bend right=60}
		{c1l2}{c1l0}
		{0.2cm}{above}
		{$\aopvar \land x < 5$}
		{}
		{}
 	\tconnect[loc2]{previous, sample, bend left=60}
		{c1l2}{c1l1}
		{0.6cm}{above}
		{$g_2$}
		{}
		{}
 	\tconnect[loc2]{previous, sample, left=60}
		{c1l0}{c1l1}
		{0.6cm}{above}
		{$g_3$}
		{}
		{}
 		
	\tconnect[loc3]{previous, loop above, sample}
		{c1l1}{c1l1}
		{0.6cm}{above}{}{}{}
}; 

    \node at ($(c1.south west)!.5!(c1.south east)$)  [label=below:$\bipAtomicExample''$]    {};

  \end{tikzpicture}
  }
		\caption{Weaving a'} %
		\label{fig:inter-locs-3} %
	\end{subfigure}%

	\caption{Reset Location Interference} %
	\label{fig:inter-locs} %
\end{figure}

\begin{example}
\rfigb{fig:inter} shows the different results obtained by weaving two local aspects $a$ and $a'$ with the two proposed procedures.
The pointcuts are $\mathit{a.pc} = (\pcloc(\ell_0)$ $\mathrm{and}$ $\pcportexec(p_2))$ and $\mathit{a'.pc} = \pcloc(\ell_1)$.
The corresponding advices are $\mathit{a.adv} =$ \\ $\tuple{F_b, F_a, \setof{\tuple{\ell_0, \btrue}}}$ and $\mathit{a'.adv} =$ $\tuple{F'_b, F'_a, \setof{}}$.
\begin{itemize}
	\item
\rfigb{fig:inter-a} and \rfig{fig:inter-b} show the weaving of each aspect individually.
Let $\mathit{M}$ (resp. $\mathit{M'}$) be the match of $\mathit{a.pc}$ (resp. $\mathit{a'.pc}$) and $\mathit{t_5}$ the transition guarded by $\mathit{g_5}$, $\mathit{t_5}$ is both in $M$ and $\WGetP(M')$.
In this case, the advices from $\mathit{a}$ and $\mathit{a'}$ overlap when woven on $\mathit{t_5}$.
	\item
\rfigb{fig:inter-c} and \rfig{fig:inter-d} illustrate the $\weaveserial_\ell$ operation.
	\begin{enumerate}
		\item
\rfigb{fig:inter-c} presents the serial weave of $\mathit{a}$ followed by $\mathit{a'}$. The weave of $\mathit{a}$ results in $\tuple{\composite_1, \tmap_1}$. $\composite_1$ is shown in \rfig{fig:inter-a}.
Upon weaving $\mathit{a'}$ its pointcut will match the reset location as it is outbound from $\mathit{\ell_1}$ and will therefore prepend $\mathit{F'_a}$ to it.
The pointcut will also match $\tmap_1(\mathit{t_5})$ as it is inbound, therefore it appends $\ffunc{F'_b, \aopclear'}$ to its existing function ($\ffunc{F_b, f_5, \aopset, F_a}$) on which $\mathit{a}$ was already woven.
		\item
\rfigb{fig:inter-d} presents the serial weave on $\mathit{a'}$ followed by $\mathit{a}$. The weave of $a'$ results in $\tuple{\composite'_1, g'_1}$. The component is shown in \rfig{fig:inter-b}.
Upon weaving $\mathit{a}$ the pointcut will match $\tmap'_1(\mathit{t_5})$. Its function $\ffunc{f_5, F'_b, \aopclear'}$ is then appended with $\ffunc{F_a, \aopset}$.
Additionally, the reset location is then added guarded by $b \land \btrue$.
	\end{enumerate}
	\item
\rfigb{fig:inter-F} illustrates the $\weaveall$ operation and how it differs from $\weaveserial$.
The pointcut of $\mathit{a'}$ will always match against the original component, matching always the transition guarded by $\mathit{t_5}$.
The $\weaveall$ procedure projects the match.
The new match result will be $\mathit{t_5}$ if $\mathit{a'}$ is woven first or $\tmap_1(\mathit{t_5})$ if $\mathit{a}$ is woven first.
The joinpoint will therefore not contain the reset location in both cases.
The order of the aspects still defines the order of the advices woven on overlapping transitions.
\rfig{fig:inter-e} displays the different advice order.
\end{itemize}
\end{example}
\paragraph{Reset location.}
In both cases, whenever multiple reset locations are woven on the same location, they are all woven with an extra port appended to the component (the \aopport{} port).
Since at every weave a singleton interaction has the highest priority, the last local aspect in the sequence has the highest priority on their port.
Thus, if multiple reset locations are found on one location (and all their guards evaluate to $\btrue{}$), the reset location that was woven last will always execute.
In the case of $\weaveserial{}$, it is possible to match the transitions that originate in the intermediary location, and thus can include different reset locations than expected.
Since aspects are assumed to be woven on a new system at every iteration, it is possible to have reset locations on intermediary states created by $\pcportenabled$.
This could cause undesirable side effects, such as weaving on different locations or transitions (as shown in the following example).
By defining more elaborate weaving strategies, one can better manage the expectations of the user and define properties that are preserved or affected by the interference.
\begin{example}[Reset location interference]
\rfigb{fig:inter-locs} shows interference when weaving two local aspects $\mathit{a}$ and $\mathit{a'}$ on a base component $\mathit{B}$ (shown in \rfig{fig:inter-locs-1}).
The pointcuts expressions for $\mathit{a}$ and $\mathit{a'}$ are respectively : $\pcportenabled(p_1)$ and $\pcguard(x)$.
Both $\mathit{a}$ and $\fvar{a'}$ use the same reset location set: $\setof{\tuple{\ell_0, x < 5}}$.
In this case, we match the transition and create the additional location $\ell_0^\bot$ as shown in \rfig{fig:inter-locs-2}.
However, when matching $\pcguard(x)$ on $\fvar{B'}$, all transitions (including created ones) match since they contain $\fvar{x}$ in their guards.
In this case the destination locations are $\ell_0^\bot, \ell_0$ and $\ell_1$.
Thus, three transitions to $\fvar{\ell_0}$ are then created (one from each destination location) when weaving $\fvar{a'}$ as shown in \rfig{fig:inter-locs-3}.
In this case we note that the transitions are not created on $\fvar{\ell_0^\bot}$ and execute before the reset location of $\fvar{a}$.
A side effect of the new reset location transitions makes it possible to completely skip evaluating the $\pcportenabled$ pointcut while moving from $\fvar{\ell_0^\bot}$ to $\fvar{\ell_0}$.
\end{example}
\begin{remark}[Strategies and properties]
While we explored specific procedures to weave aspects, and possible issues that are generated from weaving multiple aspects, we stress that for the current work we consider each weave to be on a fresh component.
Thus, in our view, the designer, when weaving multiple aspects, is actually looking at the changes in the system from each single weave and determining what should be woven next.
\end{remark}

%
\section{\aopbip{}: Aspect-Oriented Programming for BIP Systems}
\label{sec:evaluation}
We present \aopbip{}, a prototype tool that implements our approach.
To test our approach we consider first a network protocol then show the applicability of our approach to monitor CBSs.
We begin by describing the network protocol using BIP.
Then, we identify cross-cutting concerns and describe them using the \aopbip{} language.
Finally, we instrument the network model to include these concerns using the \aopbip{} tool.
We illustrate the weaving of each concern by looking at scattering and tangling.
Scattering is measured by counting the elements (affected transitions or interactions in the model) which contain the concern, while tangling is measured by counting the number of elements on which multiple concerns overlap.
The scattering and tangling of a concern serve as an indicator of the complexity to implement a concern manually without using \aopbip{} tool.

\subsection{Tool Overview}
\label{sec:evaluation:tool}
\aopbip{} is a proof-of-concept, aspect-oriented extension to BIP written in Java ($\sim$~4300 LOC).
\paragraph{Using the \aopbip{} tool.}
The command-line front end of \aopbip{} takes as input:
\begin{itemize}
	\item A \code{.bip} file that represents a BIP system written in the BIP language~\cite{BIPTools};
	\item The name of the weaving procedure to apply when weaving the aspects on the BIP model;
	\item A list of \code{.abip} files that describe the aspects.
\end{itemize}
\aopbip{} produces the BIP model and the aspect containers by parsing the \code{.bip} file and \code{.abip} files, respectively.
It selects a weaving procedure to compose the aspects \emph{per container} as described in \secref{sec:containers}.
\aopbip{} then  weaves the containers onto the BIP model resulting in an output BIP model.
\paragraph{Overview of the \aopbip{} language.}
The BIP language is described in full in~\cite{BIPTools}.
We use only a subset of the language to illustrate the concepts of this paper.
The BIP language is typed.
Components, ports and data are associated with types.
In addition to a description of the BIP system, \code{.bip} files also contain a module declaration to encapsulate the system and a \code{header} section.
The header contains arbitrary \code{C} code that is used during code generation.

The \aopbip{} language follows the same ideas presented in this paper with a few extensions.\footnote{The full grammar can be found in~\ref{app:grammar}.}
At the file level an \aopbip{} file contains a \code{header} followed by multiple containers.
The \code{header} provides additional \code{C} code to merge with the \code{.bip} header.
This is useful to define extra functions or include extra libraries.
Aspects are grouped into containers.
A container is defined by declaring the \code{Aspect} keyword followed by its identifier\footnote{We chose to use \code{Aspect} instead of \code{Container} as an implementation decision to have a similar notion than that of AspectJ, since AspectJ defines multiple pairs of pointcut and advice to be an aspect.}.
If the container defines local aspects, then it must specify the atomic component it targets right after its identifier.
The inter-type variables are not included in the individual aspects, but at the container level.
We select an atomic component or an interaction by its \emph{identifier} in the system.
Therefore, local aspects apply to a specific instance of the atomic type and global aspects apply to a specific instance of the connector type.
The global pointcut syntax includes a port specification: \code{portspec}.
The port specification is used to alias a port identifier.
For example, using the specification  \code{p:c1.stop} allows the port \code{stop} in component \code{c1} to be referenced as \code{p} in the rest of the pointcut or the advice, if \code{c1.stop} has a variable \code{x}, it can be referenced as \code{p.x}.
This is provided merely as syntactic sugar to simplify referring to the port variables in the read and modified variables in the pointcut  and the advice's update functions.
\subsection{The \code{Network} Example}
\label{sec:evaluation:case}

	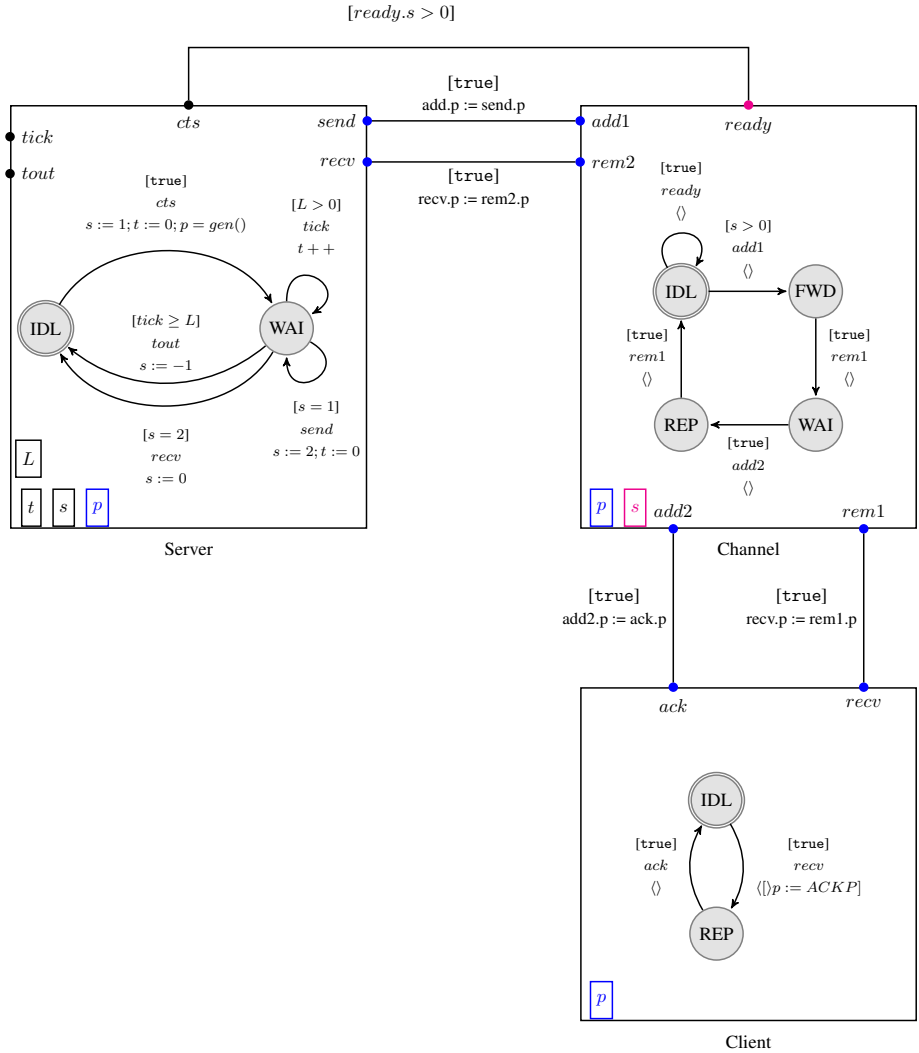
\begin{figure}[t!] %
		\centering %
		\scalebox{0.7}{\begin{tikzpicture}[bip]
\tikzset{
  myloop/.style={loop, looseness=2, min distance=10mm},
  casestudy/.style={minimum height=8cm},
	v1/.style  = {color=blue},
	v2/.style  = {color=magenta},
}
  \componentb[explode]{S}{atomic, casestudy}{}{

  \node[place,double] (S_IDLE) {IDL};
  \node[place] (S_WAIT) [right=of S_IDLE] {WAI}; 
	
  \tconnect[S_CTS]{normal, bend left=60}
    {S_IDLE}{S_WAIT}
    {0.3cm}{above}
    {$\btrue$}
    {$cts$}
    {$s := 1; t := 0; p = gen()$}

  \tconnect[S_SEND]{normal,myloop,in=-90,out=-30}
    {S_WAIT}{S_WAIT}
    {0.4cm}{below}
    {$s = 1$}
    {$send$}
    {$s := 2; t := 0$}	

  \tconnect[S_TICK]{normal,myloop,in=30,out=90}
    {S_WAIT}{S_WAIT}
    {0.4cm}{above}
    {$L > 0$}
    {$tick$}
    {$t++$}

  \tconnect[S_RECV]{normal,bend left=60}
    {S_WAIT}{S_IDLE}
    {0.4cm}{below}
    {$s = 2$}
    {$recv$}
    {$s := 0$}	

  \tconnect[S_TOUT]{normal, bend left=40}
    {S_WAIT}{S_IDLE}
    {0.2cm}{above}
    {$tick \geq L$}
    {$tout$}
    {$s := -1$}

	}{Server};

	\node[export] (S_cts) at (S.north) [label=below:$cts$] {};
	\node[export, v1, yshift=-0.3cm] (S_send) at (S.north east) [label=left:$send$] {};
	\node[export, v1, below=0.6cm of S_send] (S_recv) [label=left:$recv$] {};
	\node[export, yshift=-0.6cm] (S_tick) at (S.north west) [label=right:$tick$] {};
	\node[export, yshift=-1.3cm] (S_tout) at (S.north west) [label=right:$tout$] {};

  \node[var] (S_t) at ( $(S.south west) + (0.4,0.4)$) []  {$t$};
  \node[var] (S_l) [above right= 0.2cm and -0.5cm of S_t]  {$L$};
  \node[var] (S_s) [right=0.2cm of S_t]  {$s$};
  \node[var,v1] (S_p) [right=0.2cm of S_s]  {$p$};
 
  \componentb{CH}{atomic, right=4cm of S, casestudy}{}{
    
  \node[place,double] (CH_IDL) {IDL};
  \node[place, right=of CH_IDL] (CH_FWD) {FWD};
  \node[place, below=of CH_FWD] (CH_WAI) {WAI};
  \node[place, below=of CH_IDL] (CH_REP) {REP};
 
   \tconnect[CH_ADD1]{normal}{CH_IDL}{CH_FWD}{0.2cm}{above}{$s > 0$}{$add1$}{\ffunc} 
   \tconnect[CH_REM1]{normal}{CH_FWD}{CH_WAI}{0.2cm}{right}{$\btrue$}{$rem1$}{\ffunc} 
   \tconnect[CH_ADD2]{normal}{CH_WAI}{CH_REP}{0.2cm}{below}{$\btrue$}{$add2$}{\ffunc} 
   \tconnect[CH_REM2]{normal}{CH_REP}{CH_IDL}{0.2cm}{left }{$\btrue$}{$rem1$}{\ffunc} 
   \tconnect[CH_REDY]{normal, ,myloop, in=60, out=120}{CH_IDL}{CH_IDL}{0.2cm}{above }{$\btrue$}{$ready$}{\ffunc} 

  }{Channel};
	\node[export,v2] (CH_ready) at (CH.north) [label=below:$ready$] {};
	\node[export,v1,yshift=-0.3cm] (CH_add1) at (CH.north west) [label=right:$add1$] {};
	\node[export,v1,below=0.6cm of CH_add1] (CH_rem2) [label=right:$rem2$] {};
	\node[export,v1,xshift=-1cm] (CH_rem1) at (CH.south east) [label=above:$rem1$] {};
	\node[export,v1,left=3.4cm of CH_rem1] (CH_add2) [label=above:$add2$] {};

  \node[var,v1] (CH_p) at ( $(CH.south west) + (0.4,0.4)$) []  {$p$};
  \node[var,v2] (CH_s) [right=0.2cm of CH_p]  {$s$};

  \componentb{C}{atomic, below=3.0cm of CH}{}{
      \node[place,double] (C_IDL) {IDL};
      \node[place, below=of C_IDL] (C_REP) {REP};
      \tconnect[C_recv]{normal,bend left}{C_IDL}{C_REP}{0.2cm}{right}{$\btrue$}{$recv$}{$\ffunc[p := ACKP]$} 
      \tconnect[C_ack] {normal,bend left}{C_REP}{C_IDL}{0.2cm}{left }{$\btrue$}{$ack$}{\ffunc} 
  }{Client};
	\node[export,v1,xshift=-1cm] (C_recv) at (C.north east) [label=below:$recv$] {};
	\node[export,v1,left=3.4cm of C_recv] (C_ack) [label=below:$ack$] {};
  \node[var,v1] (C_p) at ( $(C.south west) + (0.4,0.4)$) []  {$p$};

  \draw[-] (S_cts.north)
			      -- ([shift={(0, 1cm)}]$(S_cts.north)$)node {}
			      -- ([shift={(4cm, 1cm)}]$(S_cts.north)$)node [above] {\ifunc{$ready.s > 0$}{}}
			      -- ([shift={(0, 1cm)}]$(CH_ready.north)$)node {}
			      -- (CH_ready.north);
  \draw[-] (S_send) 
            -- ($(S_send.east)!.5!(CH_add1.west)$) node [above] {\ifunc{$\btrue$}{add.p := send.p}} 
            -- (CH_add1);
  \draw[-] (S_recv) 
            -- ($(S_recv.east)!.5!(CH_rem2.west)$) node [below] {\ifunc{$\btrue$}{recv.p := rem2.p}} 
            -- (CH_rem2);

 \tconnector{CH_add2.south}{C_ack.north} {0,0}{left}{\ifunc{$\btrue$}{add2.p := ack.p}}{i1}
 \tconnector{CH_rem1.south}{C_recv.north}{0,0}{left}{\ifunc{$\btrue$}{recv.p := rem1.p}}{i2}

\end{tikzpicture}} %
		\caption{The \code{Network} Component} %
		\label{fig:case} %
	\end{figure} %

A network protocol is used to illustrate the handling of crosscutting concerns in BIP and is shown in \rfig{fig:case}.
The network protocol is an augmented version of the one presented in~\cite{Borzoo}.
The \code{Network} composite component consists  of a \code{Server}, a \code{Client} and a \code{Channel}.
The double circles denote the start locations for each component.
The \code{Server} waits for the \emph{clear-to-send} signal on its \code{cts} port.
This indicates that a channel is available.
It then generates a packet and sends it to the channel.
The channel forwards the packet to the client which acknowledges it.
The channel will then send the acknowledgement back to the server.

\sideaspect[firstline=21]{faults}{0.42}{0.4cm}{Authentication aspects and resulting execution. The aspects extends the behavior by verifying a hash of the messages and forwarding or blocking the message.}{linebackgroundcolor={\hl{1}\hl{2}\hl{3}\hl{11}\hl{12}\hl{13}}}
\paragraph{Crosscutting concerns.}
The network protocol is augmented by refining its specification.
The correctness of our transformation ensures that the crosscutting concerns are rigorously handled.
First, logging is introduced by capturing port executions locally in all components.
Second, security is added in the form of authentication.
A signature (hash) is added to the packet and checked.
To accomplish the above we introduce two local aspects.
The various aspects along with the output are displayed in \rfig{lst:aspect-faults}.
The first intercepts the \code{Server}'s \code{cts} port execution and adds the signature once the server is ready to send, by modifying the packet stored in the local variable $p$.
The second intercepts the \code{Channel}'s \code{add1} port execution; this port executes when a packet from the \code{Server} is sent.
The advice verifies the signature (using \code{check(p)}) and stores the result (logical 0 or 1) in an inter-type variable \code{clear}.
The advice also adds a reset location to $\code{IDL}$ if the verification failed (\code{clear == 0}), preventing the \code{Channel} from forwarding the packet to the \code{Client}.
The \code{Carol} aspect is added to modify the packet in transit to display a failed authentication.
The global pointcut expression matches: (1) ports $\code{server.send}$ and $\code{channel.add1}$, and (2) variable $\code{server.send.r}$.
The advice in the \code{Carol} aspect changes the value of $\code{channel.add1.r}$ after the execution of any interaction that matches the pointcut.
Normally, the system executes the packet transfer by reading \code{a.r} and modifying \code{b.r}.
The advice function instead will override \code{b.r} by generating a fake packet from \code{a.r} using  $\code{pfake(a.r)}$.
The output displays a successful and an unsuccessful attempt.
The packet is represented by a number and the signature is the last digit in the number.
We notice that the first aspect added \code{|6} to the packet \code{886} and the second aspect removed it when the channel forwarded it (\code{Channel1.rem1}).
\code{Carol} replaces the first packet \code{886} with \code{386}, which both have the same signature (\code{6}).
The verification succeeds in this case unlike in the second try, when \code{763} is replaced by \code{736} since the signature of \code{736} is \code{6} but not \code{3}.
Third, congestion avoidance is added by computing the round-trip time of the message and then waiting before sending further messages.
Fourth, basic fault tolerance is introduced in the form of a failsafe mechanism. The system deadlocks and then terminates safely, after the server fails to receive a certain number of acknowledgments.
\paragraph{Coverage of concerns.}
The coverage of the concerns is shown in \rtable{table:case}.
Column Transitions reports the number transitions that have been modified including the number of added transitions for reset locations.
Column Interactions reports the number of modified interactions.
Column OT (resp. OI) reports the number of transitions (resp. interactions) that overlap with other concerns.
Interfering concerns are reported in column OC.
Concerns are indicated by label (1-4).
We illustrate concerns that target multiple areas in the system.
Without using \aopbip{}, implementing these concerns would require one to edit a significant part of the system.
For instance, in the case of logging, the code must be inserted in 10 transitions, of which half overlap with other concerns.

\begin{table}[t]
\centering
\renewcommand{\arraystretch}{1.2}
\caption{Crosscutting concerns in \code{Network}}
\label{table:case}
\begin{tabular}{ r l c c c c l }
  \toprule
  \textbf{\#}& \textbf{Concern} & \textbf{Transitions} & \textbf{Interactions} & \textbf{~OT~} & \textbf{~OI~} & \textbf{~OC~}\\
  \midrule
  1 & Logging    & 10 & 0 & 5 & 0 & 2,3\\
  \rowcolor[gray]{0.9}
  2 & Authentication  & 2  & 1 & 2 & 1 & 1,3.4 \\
  3 & Congestion   & 5  & 0 & 4 & 0 & 1,2 \\
  \rowcolor[gray]{0.9}
  4 & Fault Tolerance  & 0  & 3 & 0 & 1 & 2 \\
  \midrule
  \multicolumn{2}{c}{\code{Network}}  & 12 & 5 & \multicolumn{3}{c}{}\\
  \bottomrule
\end{tabular}
\end{table}

\subsection{Applicability to Runtime Verification of CBSs}
\label{sec:evaluation:rv}
\paragraph{Overview}
Runtime verification (RV) is a lightweight verification technique used to verify whether a run of a specific system verifies a specific property~\cite{RuntimeTutorial,BartocciFFR18,series/lncs/10457}.
It consists in extracting a sequence of events, which is then fed to a monitor that verifies it against a specification.
RV frameworks for CBSs, and particularly for BIP systems (RV-BIP~\cite{FalconeJNBB15} and RVMT-BIP~\cite{NazarpourFBB17}) have been already developed.
They define specific transformations to instrument components and insert monitors as components in the new system (RV-BIP for sequential systems and RVMT-BIP for multi-threaded systems).
However, since runtime verification is a crosscutting concern, it is possible to instrument a system with aspects (both global and local) to generate necessary events for monitoring.
At the global level, it is possible to monitor interactions by intercepting their ports and variable accesses.
Thus, by describing global pointcuts, we can generate events that are global, and synthesize global aspects that implement monitors.
Since we allow for inter-type declarations at the global level, a monitor state can be stored in the inter-type component.
The component can then be used to describe a specification for a monitor.
At the local level, it is possible to monitor the component state by using local pointcuts.
Thus we can generate events that are local to the component.
Using local aspects, we can then describe local monitors that are embedded in the component to check for local events.
Certain properties however require information from multiple local monitors, thus it is impossible to handle the synchronization with our current approach.
Directly writing monitors as aspects is not handled for these types of properties.
However, it is possible for each local monitor to print out an event, and a separate monitoring mechanism to verify the entirety offline.
While we do not tackle the automatic synthesis of monitors from a specification, we show next how \aopbip{} can be used to write manual monitors for specific properties.
\paragraph{Dala robot}
A robotic application is used as an example in~\cite{FalconeJNBB15}.
The Dala robot~\cite{DALA} is a large and realistic interactive system which consists of a set of modules.
Each module is a set of services that corresponds to different tasks and a set of \emph{posters} that are used to exchange data with other modules.
A simplified simulation of the modules consists of various services classified as \emph{readers} and \emph{writers} accessing data of a \emph{poster} component and a global clock component \code{clock}.
Reader services read the data in the poster, and writer services modify it.
A writer simulates a large data transfer consisting of two phases.
In the first phase, the \code{writer}  writes a task id to the poster using an interaction that exchanges the task id between the two ports: \code{writer.writev}, and \code{poster.writev}.
We assume that the task ids represent an abstract workload assigned to the various services and components of the robot.
Upon completion, the interaction consisting of the ports \code{writer.finishWrite} and \code{poster.finishWrite} is executed.
We use the example to monitor the following properties: mutual exclusion between \emph{writers}, data freshness and ordering of tasks.
The monitors are presented in \rcode{lst:rv-mons}.
\paragraph{Monitoring mutual exclusion}
The first property focuses on \emph{writers}.
The writing process consists of two steps: \code{write} and \code{finishWrite}.
The \code{writev} port is triggered when a writer starts writing to the poster service, and upon completion \code{finishWrite} is executed.
The property checks that no two writers are writing at the same time.
To monitor the property, we create a local aspect on the \code{poster} component, and use an extra inter-type variable $c$ initialized to zero.
Upon execution of the port \code{writev}, we check if $c$ is non-zero and increment it by one, and upon execution of the port \code{finishWrite} we decrement $c$ by one.
Thus, if two executions of the port \code{writev} happen, the monitor can verify the property ($c > 0$).
\paragraph{Monitoring freshness}
The second property focuses on \emph{readers}, when a \emph{reader} reads data posted by a \emph{writer}, the freshness property checks for constraints on timestamps.
The goal is to make sure that the data being read is up-to-date, i.e. the data has been read at most after a fixed amount of ticks.
To do so, the property compares the timestamp stored in the \emph{poster} against the current timestamp stored in the \emph{clock} component, whenever a \emph{reader} reads data.
Thus, our monitor considers the interaction consisting of the ports \code{reader.read}, \code{poster.read} and \code{clock.getTime}.
The monitor intercepts the port variables to get the timestamp, then computes the difference and verifies whether it is below a threshold of 2 ticks.
\paragraph{Monitoring ordering}
The third property concerns the interaction between the \emph{writers} and the \emph{poster}; when a task id is set, the ordering property expresses constraints on the order of tasks.
In this example, we check if a task with a larger id executes before one with a smaller id.
To do so, we use an inter-type variable \code{lastTask} and initialize it to zero, indicating that the next expected task is 1.
We intercept all the interactions that involve the port \code{poster.writev} and check for those that write to the port variable (which contains the task id).
In this case, the global advice is executed after the call, so as to catch the value written to the port's variable.
Upon writing to the variable we check if the new task is in the correct order by comparing it to \code{lastTask + 1} and then store the new task id as the \code{lastTask}.

\getcode[rv-mons]{monitors.abip}{Monitors}{}{style=bip,language=aopbip,tabsize=2,basicstyle=\footnotesize\ttfamily}
%

\section{Related Work} \label{sec:related}
\paragraph{Modularizing AOP.}
Multiple approaches have sought to (i) improve the applicability of AOP, and (ii) enhance its modularity.
These approaches include, but are not limited to Ptolemy~\cite{Ptolemy}, XPIs~\cite{XPI}, AspectJML~\cite{AJML}, and using substitution~\cite{Molderez2015}.
They mainly focus on defining contracts and improving matching and advice applications in a modular way, and provide a better decoupling mechanism for the aspects of the systems.
These approaches provide us with a perspective on how to better integrate, in a component-based manner, aspects in CBSs.
However, they were mainly conceived to improve on the initial AOP methods that target non-CBS systems (i.e., AspectJ~\cite{AspectJ}).
They perceive the system as arbitrary function calls or message passing between objects that is not constrained.
As such, they do not inherently target component-based semantics, such as synchronization or data transfers between various modules.
\paragraph{AOP for CBSs.}
Pessemier et al.~\cite{PessemierSDC08} present a framework to deal with crosscutting concerns in CBSs.
It is a symmetric approach, i.e. it uses the same language of the system to describe AOP concepts.
It presents aspects as components containing the advice and additional interfaces, and are therefore integrated homogeneously within the system.
Interaction with the advice is done through these added interfaces.
The implementation of a concern is found in what is called an \emph{aspect component}.
Aspect components are regular components augmented with extra interfaces.
They contain the advices necessary to implement a crosscutting concern.
Interaction with the advice happens through additional interfaces called \emph{advice interfaces}.
Moreover, aspects components expose regular interfaces. Thus, they can be seen as regular components.
Joinpoints are a combination of interfaces of different components.
Thus, components are seen as black boxes.
The interception model is based on composition filters~\cite{AksitBV92} extended from objects to components.
The model is mapped onto Fractal, a modular and extensible component model~\cite{Fractal}.
This approach has several advantages.
First it explicitly models dependencies between aspects and components, and allows for their composition at an architectural level.
Second, it allows the aspects to be manipulated and reconfigured at runtime.
Third, it clearly defines the relationships (1) between aspects and other aspects, and (2) between aspects and the components they modify.
This approach, however, does not consider the semantics of interactions.
It targets arbitrary interface signatures, so the implementation itself must explicitly address the synchronization amongst the different components and data transfer.
Notions of \emph{before}, \emph{after} differ from just executing a function.
In the simplest case, a BIP interaction requires ports to be all enabled, therefore \emph{before} and \emph{after} execute upon synchronization of all involved components.

Similarly, other works such as~\cite{Duclos02} and \cite{Lieberherr03} integrate AOP into CBSs as well.
These approaches are, however, asymmetric (i.e., they use an external language to represent AOP concepts) and subsumed by~\cite{PessemierSDC08}.
Duclos's approach~\cite{Duclos02} defines two languages to integrate aspects.
Lieberherr's approach~\cite{Lieberherr03} defines aspects as part of the modules they apply to, and compares the expressiveness of the approach with both AspectJ and HyperJ~\cite{HyperJ}.

SAFRAN~\cite{SAFRAN} differs from the above approaches by using AOP in the Fractal component model to define adaptation policies.
\paragraph{Formalization of aspects.}
The aforementioned approaches for CBSs do not rely on formal models.
Work to formalize aspects in programs has been undertaken by~\cite{KatzCategories}.
The approach specifies different categories of aspects and how they affect various classes of properties (safety, liveness).
Aspects are then assigned to a category by syntactic analysis.
The work has been extended by Djoko et al.~\cite{Djoko} by expanding the categories and defining languages of aspects.
The languages of aspects ensure by construction that aspects written with them fit a specific category.
Additional tools for verification and analysis of aspects and their interference have been developed and are presented in~\cite{KatzTools}.
The approach focuses on object-oriented systems, and not component-based systems, but it also provides theoretical results on property preservation which our approach does not study.
By formalizing aspects in the context of CBS semantics, our approach paves the way to extend these works to CBSs.

Larissa~\cite{Larissa} is a language for handling crosscutting concerns in reactive systems modeled as the composition of Mealy automata.
The matching is done by assigning monitor programs that look for a specific execution trace.
Joinpoints are then associated with the input history.
Advices consist of two types: \code{toInit} and \code{recovery}.
The \code{toInit} advice places the program back in its original state.
The \code{recovery} advice consists of restoring the program to the last recovery state it was in.
Since it is impossible to play the input backwards for recovery, a set of global recovery points is determined.
A recovery state is determined by a monitor: the recovery program. The recovery states are associated with specific execution traces and are matched similarity to joinpoints.
Compared to our approach, Larissa supports joinpoints based on the input history.
It can also be seen as symmetric since aspects are introduced in the synchronous language used.
However, the underlying model is conceived for reactive systems, and not CBSs, it does not have a clear distinction between communication and components, and thus does not distinguish between aspects related to components and communications.
The communication model is based on simple input/output matching.
Moreover, advices are not expressive and only consider reset/restore the state of the system.
Formalizing aspects in the BRIC component model has been undertaken by~\cite{BRIC}.
BRIC formalizes the behavior of components and their interactions using the Communicating Sequential Processes (CSP) language.
Unlike our approach, BRIC is symmetric: aspects, pointcuts, and advices are described in CSP, and woven using CSP operators.
Additionally, BRIC targets interactions and not the components themselves.
It regards components as black boxes.
Similarly to BIP, CSP benefits from compositional verification of the properties and has a well-defined semantics.
CSP uses denotational semantics as opposed to BIP which uses operational semantics.
Verification on the resulting woven system is possible in both approaches.
However, BIP has a strong expressive synchronization primitive~\cite{BliudzeS08} which is more expressive than CSP~\cite{Hoare85}.
This allows more concerns to be formalized.
%

\section{Conclusions} \label{sec:conclusion}
\subsection{Summary}
This paper deals with crosscutting concerns in CBS using the AOP paradigm.
It targets the two stages in the construction of CBSs by defining local and global aspects to refine components and their compositions, respectively.
Local joinpoints capture concerns found in local components.
Local pointcuts select joinpoints based on location, guards, variables in update functions and ports.
They are associated with advices to add extra functionalities (e.g., computation or location change).
Global joinpoints capture the interactions between components through their interfaces without having information about the internal representation of components.
Global pointcuts select global joinpoints based on interactions, by considering ports and their respective data transfer operations.
They are associated with global advices, to add extra functionality to the interaction model (e.g., data transfer, storing global information, etc.).
Furthermore, to remedy interference and to increase expressiveness, we present a way to compose multiple aspects.
Global and local joinpoints are mapped to BIP semantics and pointcut matching and advice weaving are implemented using model-to-model transformation on BIP models.
We implement the proposed method in the \aopbip{} tool-chain.
\aopbip{} consists of a language to describe both local and global aspects and provides an implementation of matching, weaving and composition.
We study the automatic integration of various crosscutting concerns (logging, security, performance, fault handling) on a given input BIP system.
Furthermore, we focus on a particular crosscutting concern, namely runtime verification and use \aopbip{} to monitor three properties on a robotic application designed as a CBS.

\subsection{Future Work}
Future work comprises four directions.
The first consists in capturing more joinpoints and extending the possible behavior of advices.
This would facilitate, in particular, the support for runtime verification~\cite{RuntimeTutorial,BartocciFFR18} and runtime enforcement~\cite{Falcone10,FalconeFM12}, possibly in a timed setting~\cite{ColomboPS09,BauerLS11,PinisettyFJMRN14,FalconeJMP16} where the elapsing of physical time influences the behavior of monitors.
Possible new joinpoints include variables in interaction guard, specific values of variables.
Advices can be extended to modify guards on matching transitions and interactions.
The second consists in applying CBS methods to define advices and aspect composition.
This would help integrating AOP in BIP symmetrically, where aspects are implemented as components and interactions within the existing system.
Moreover, this would allow to enable or disable aspects in the system, and specify more complex advices (i.e., advices as components instead of update functions and extra transitions).
The third consists in elaborating new ways to compose aspects by finding new criteria to order them.
Aspects can be re-ordered in a container based on their pointcut expressions, by grouping those that affect the same transitions or interactions, and whether or not they modify the existing variables (read/write aspects).
Additionally, the language could be extended to allow the explicit definition of precedence rules.
The fourth consists in implementing model-to-model transformations using Domain Specific Languages inspired by ATL~\cite{JouaultABK08} targeting the BIP model and comparing their expressiveness with our approach.
\section*{Acknowledgment}
The authors acknowledge the support of the ICT COST (European Cooperation in Science and Technology) Action IC1402 Runtime Verification beyond Monitoring (ARVI) and the University Research Board (URB) at American University of Beirut.

\section*{References}
\bibliographystyle{elsarticle-harv}
\bibliography{biblio-nolinks}
\appendix
\section{Proofs} \label{app:proofs}
The common assumption for the proofs that include the advice $F_b$ and $F_a$ functions, is that $F_b$ and $F_a$ can be uniquely determined and are not empty.
To ensure this, one can add code markers at the start and end of each $F_b$ and $F_a$ which have no effect and are not present in the original system.

\begin{proof}[of Proposition~\ref{prop:global-match}]
We consider a global event $\tuple{q, a, q'} \in \events$ and a global pointcut expression $\tuple{p, v_r, v_w}$.
The proof follows from the definition of $\pcmatchg(\composite, gpc)$ which selects all interactions matching the criteria that should be matched by $(E_i = \tuple{ q , a, q '} \vDash \mathit{gpc})$.
\begin{align*}
	(a \vDash \tuple{p, v_r, v_w})
		&  \mbox{ iff } p \subseteq \portsof{a}\\
					&\quad	\land v_r \subseteq \varread(a.\func)\\
					&\quad 	\land v_w \subseteq \varwrite(a.\func)
		&& \hbox{(\defref{def:gpc-match})}\\
		& \mbox { iff } a \in \pcmatchg(\gamma, \mathit{gpc})
		&& \hbox{(Def. of  $\pcmatchg$)}\\
\end{align*}
\end{proof}
\begin{proof}[of Proposition~\ref{prop:global-apply}]
Given a global advice $\tuple{F_b, F_a}$.
We consider an executed interaction $a \in \events'$, and $\erem_\vglobal(a, F_b, F_a) = a'$ the constructed interaction without the advice.
\begin{align*}
	\exists F :  a.\func = \tuple{F_b, F, F_a}
		& \mbox{ iff } \exists a' \in \gamma : \tmap(a') = a \land a' \in  \subgamma)
		&& \hbox{(\defref{def:global-weave}, \tmap())}\\
		& \mbox{ iff } a' \in  \pcmatchg(\composite, \mathit{gpc})
		&& \hbox{(\defref{def:global-weave})}\\
		& \mbox{ iff } (\tuple{q'_s, a', q'_e} \vDash \mathit{gpc})
		&& \hbox{(Proposition~\ref{prop:global-match})}\\
		& \mbox{ iff } \erem_\vglobal(a, F_b, F_a) \vDash \mathit{gpc}
\end{align*}
An executed interaction's update function $\mathit{a.func}$ starts with $F_b$ and ends with $F_a$, according to the definition of $\tmap()$ in \defref{def:global-weave} iff it is the result of weaving the advice on it from an interaction $a'$ ($\exists a' \in \gamma : \tmap(a') = a \land a' \in \subgamma$).
The interaction $a'$ is in $\subgamma$ iff it was selected by the local pointcut expression ($a' \in  \pcmatchg(\composite, \mathit{gpc})$).
According to Proposition~\ref{prop:global-match}, any interaction $a' \in  \pcmatchg(\composite, \mathit{gpc})$ is a joinpoint, specifically $a' = \erem_\vglobal(a, F_b, F_a)$.

\end{proof}

\begin{proof}[of Proposition~\ref{prop:local-match}]
Let $\mathit{e} = \tuple{\tuple{l,v}, \tau, \tuple{l',v'}}$, we assume that $\fvar{lpc}$ does not contain $\pcportenabled(p)$.
The proof follows by induction on the structure of $\fvar{lpc}$.
\[
	\evalLocal{e}{\fvar{lpc}} \mbox{ iff } \tau \in \pcmatch(B_k, \fvar{lpc})
\]
Base cases:
	\begin{align*}
		\evalLocal{e}{\pcloc(\ell)} & \mbox{ iff } l = \ell
		&& \hbox{(\defref{def:ljp-match})}\\
		& \mbox{ iff } \tau.\source = \ell\\
		& \mbox{ iff } \tau \in \setof{t \in \transof{B_k} \mid t.\source = \ell}\\
		& \mbox{ iff } \tau \in \pcmatch(B_k, \pcloc(\ell))
		&& \hbox{(\defref{def:lpc-match})}\\
	\evalLocal{e}{\pcguard(x)} & \mbox{ iff }  (\exists t \in \transof{B_k} : t.\source = l = \tau.\source \\ & \land x \in \varguard(t))
		&& \hbox{(\defref{def:ljp-match})}\\
		& \mbox{ iff } \tau \in \WSiblings(t)
		&& \hbox{(Def. $\WSiblings$)}\\
		& \mbox{ iff } \tau \in \pcmatch(B_k, \pcguard(x))
		&& \hbox{(\defref{def:lpc-match})}\\
	\evalLocal{e}{\pcfunc(x)} & \mbox{ iff }  (x \in \varread(\tau.\func))
		&& \hbox{(\defref{def:ljp-match})}\\
		& \mbox{ iff } \tau \in \setof{t \in \transof{B_k} \mid x \in \varread(t.\func)}\\
		& \mbox{ iff } \tau \in \pcmatch(B_k, \pcfunc(x))
		&& \hbox{(\defref{def:lpc-match})}\\
	\evalLocal{e}{\pcportexec(p)} & \mbox{ iff }  (\tau.\port = p))
		&& \hbox{(\defref{def:ljp-match})}\\
		& \mbox{ iff } \tau \in \setof{t \in \transof{B_k} \mid t.\port = p}\\
		& \mbox{ iff } \tau \in \pcmatch(B_k, \pcportexec(x))
		&& \hbox{(\defref{def:lpc-match})}
\intertext{\small $\evalLocal{e}{\pcwrite(x)} \mbox{ iff } \tau \in \pcmatch(B_k, \pcwrite(x))$ is shown by replacing in the above: $\pcfunc(x)$ with $\pcwrite(x)$.}
\shortintertext{Inductive step:}
	\evalLocal{e}{\phi \mbox{ and } \phi'}
	&\mbox{ iff } \evalLocal{e}{\phi} \land \evalLocal{e}{\phi'}\\
	&\mbox{ iff } \tau \in \pcmatch(B_k, \phi) \land \tau \in \pcmatch(B_k, \phi')
	&&\hbox{(Hypothesis)}\\
	&\mbox{ iff } \tau \in \pcmatch(B_k, \phi) \cap \pcmatch(B_k, \phi')
\end{align*}
\end{proof}

\begin{proof}[of Proposition~\ref{prop:local-apply}]
We consider $F_b$ and $F_a$ to be the advice before and after update functions, and the local event $\fvar{e_i} \in T'_k$ and $\fvar{e'} = \erem_k(e_i, F_b, F_a)$ to be the constructed event without the advice.
The proof is split into two parts: the first proves the correct application of an advice's update functions, the second proves the correct application of reset locations.
\begin{proofpart}
\[
(e'_i \neq \epsilon \land \evalLocal{e'_i}{\fvar{lpc}}) \mbox { iff } \mathrm{before}(e_i, F_b, d) \land \mathrm{after}(e_i, F_a, d)
\]
We distinguish two cases based on $\mathrm{early}(\fvar{lpc})$.
\paragraph{Case 1: $\mathrm{early}(\fvar{lpc}) = \bfalse$}
This case includes the edit frames $\tuple{\WCurrentBefore, \WCurrentAfter}, \tuple{\WCreate, \WCurrentAfter}$.

\noindent
$\mathrm{before}(e_i, F_b, \bfalse) \land \mathrm{after}(e_i, F_a, \bfalse)$ simplifies to $\exists \fvar{F} : e_i.\tau.\func = \ffunc{F_b, F, F_a}$.
\paragraph{Case 1.1}
We consider that $\pcportenabled$ is not included in $\fvar{lpc}$, then  the advice is woven on the update function $F$, $e_i.\tau = \tuple{\ell, p, g, F_b F \aopset F_a, \ell'}$ iff it is the result of a matching transition $\exists t : t = \tuple{\ell, p, g, F, \ell'} \in M$ according to $\weaveframe_{\mathrm{cur}}$ in \defref{def:local-weave} with $e_i.\tau \in \tmap(t)$.
We have $t = \erem_k(e_i, F_b,F_a).\tau$ and $t \in \pcmatch(B_k, \fvar{lpc})$ iff $\evalLocal{\erem(e_i,F_b,F_a)}{\fvar{lpc}}$ from Proposition~\ref{prop:local-match}.
\paragraph{Case 1.2}
We consider that $\pcportenabled$ is included in $\fvar{lpc}$, the set of enabled ports in the expression is $\fvar{P} = \mathrm{sp}(\fvar{lpc})$.
Using the definition of $\weaveframe_{\mathrm{runa}}$ in \defref{def:local-weave}, the advice is woven on the transition $e_i.\tau = \tuple{\ell, p, \aopvar \land g, \ffunc{F_b, F, F_a}, \ell'}$ iff $\exists t : e_i.\tau \in \tmap(t)$ with $t = \tuple{\ell, p, g, F, \ell'} = \erem_k(e_i, F_b, F_a).\tau$.
We decompose $\evalLocal{\erem_k(e_i, F_b, F_a)}{\fvar{lpc}}$ into two conditions $\evalLocal{\erem_k(e_i,F_b, F_a)}{c_1} \land \evalLocal{\erem_k(e_i, F_b, F_a)}{c_2}$, where $c_1$ contains the conjunction of all the $\pcportenabled$ pointcuts and $c_2$ contains all the rest (as per \defref{def:ljp-match}).
\begin{enumerate}
	\item $e'_i.\tau \in \pcmatch(B_k, \fvar{lpc})$ iff $\evalLocal{e'_i}{c_2}$ by applying Proposition~\ref{prop:local-match} to the syntactic part.
	\item $e_i.\tau$ executed iff $\aopvar$ is $\btrue$. $\aopvar$ is $\btrue$ iff $\mkAddGuard(P, \ell, M)$ since the instrumentation guarantees that the only transitions that set $\aopvar$ are guarded by $\mkAddGuard$.
$\fvar{M}$ contains all the transitions outbound from $\ell$ since $\pcportenabled$ uses $\WSiblings$ (\defref{def:lpc-match}).
However, by the definition of $\mkAddGuard$, $\mkAddGuard$ iff $\forall p_j \in P, \exists \tuple{\erem_k(e_i, F_b, F_a).l : p_j, g_j, f_j, \ell_j}$ with $g_j(\erem_k(e_i, F_b, F_a).v) = \btrue$.
Therefore, $\mkAddGuard$ iff $\evalLocal{e'_i}{c_1}$.
\end{enumerate}
Therefore, we have:
\[
	\mathrm{before}(e_i, F_b, \bfalse) \land  \mathrm{after}(e_i, F_a, \bfalse) \mbox{ iff } e'_i \neq \epsilon \land \evalLocal{e'_i}{\fvar{lpc}}
\]
\paragraph{Case 2: $\mathrm{early}(\fvar{lpc}) = \btrue$}
This case includes the edit frames $\tuple{\WPreviousAfter, \WCurrentBefore}, \tuple{\WCreate, \WCurrentBefore}$.

\noindent
$\mathrm{before}(e_i, F_b, \btrue) \land \mathrm{after}(e_i, F_a, \btrue)$ simplifies to $\exists F, F': e_{i-1}.\tau.\func = \ffunc{F, F_b} \land e_i.\tau.\func = \ffunc{F_a, F}$.
The proof is similar to Case 1, we assume $i > 0$ for simplicity and decompose into two cases: $\fvar{lpc}$ does not contain (resp. contains) $\pcportenabled$ and use the definition of $\weaveframe_{\mathrm{prev}}$ and $\weaveframe_{\mathrm{runb}}$ from \defref{def:local-weave} respectively.
 $e_{i}.\tau$ corresponds to a $t \in M$.
Since we consider the event before $e_{i_1}.\tau$ we note that both frames weave on all transitions that lead to $M$.
In the case of $\weaveframe_{\mathrm{prev}}$, $F_b$ is woven at the end of all transitions in $\WPrevious(M) \setminus M$ and in the case of the loop from the created locations to the locations from which $M$ is outbound therefore including all possible previous transitions.
In the case of $\weaveframe_{\mathrm{runb}}$, $F_b$ is woven at the created transition guarded by $\mkAddGuard$ (which is $\btrue$).
\end{proofpart}

\begin{proofpart}
\[
	(e'_i \neq \epsilon \land \evalLocal{e'_i}{\fvar{lpc}}) \implies  (R \neq \emptyset \implies \exists r \in R : \mathrm{reset}(e_i, r)
\]
Since it is possible to constructs $e'_i$ and $e'_i$ is a joinpoint, we decompose $\evalLocal{e'_i}{\fvar{lpc}} = \evalLocal{e'_i}{c_1} \land \evalLocal{e'_i}{c_2}$ into a conjunction of two conditions (similarly to Case 1.2).
$c_1$ is a conjunction of all $\pcportenabled$ pointcuts and $c_2$ all the rest.
\begin{enumerate}
	\item Since $\evalLocal{e'_i}{c_1}$, we have $\forall p_j \in \mathrm{sp}(\fvar{lpc}) : \exists \tau' = \tuple{e'_i.l, p_j, g_{\tau'}, f_{\tau'},\ell'_{\tau'}}$ by \defref{def:ljp-match} with $e'_i.\tau \in \WSiblings(\tau')$ (definition of $\WSiblings$).
Therefore, $e'_i.\tau \in \pcmatch(B_k, c_1)$
	\item From Proposition~\ref{prop:local-match}, $\evalLocal{e'_i}{c_2}$ iff $e'_i.\tau \in \pcmatch(B_k, c_2)$.
\end{enumerate}
We have from the above
\[
e'_i.\tau \in \pcmatch(B_k, c_1) \land e'_i.\tau \in \pcmatch(B_k, c_2) \text{ iff } e'.\tau \in (\pcmatch(B_k,c_1) \cap \pcmatch(B_k, c_2)).
\]
Therefore  $\evalLocal{e'_i}{\fvar{lpc}}$ iff $e'_i.\tau \in \pcmatch(B_k, \fvar{lpc})$ (i.e., $e'_i.\tau \in M$ and $e_i.\tau \in \tmap(e'_i.\tau)$).
The local weave guarantees that since $e'_i.\tau \in M$ then $\aopset$ has been called either on $e_i.\tau$ in the case where $\pcportenabled$ is not present, or $e_{i-1}.\tau$ ($\aopset$ is in the update function of the created transition precedes $e_i.\tau$ guarded by $\mkAddGuard$).
The local weaving (\defref{def:local-weave}) ensures for all cases of edit frames, that the transitions will lead to a location on which reset location is woven (by using $\WDest(M)$ and adding the appropriate created locations).
Therefore, it ensures that the next location after $e_i.\tau$ executes, contains all reset location pairs if $R \neq 0$.
Using the definition of $\weavereset$ (\defref{def:local-weave}), for any $r_j = \tuple{g_j, \mathit{dest}_j} \in R$, we have a transition $\mathit{reset}_j = \tuple{e_{i+1}.l, \aopport, \aopvar \land g_j, f_j, \mathit{dest}_j}$.
The port $\aopport$ corresponds to an interaction with the highest priority, therefore if $\aopport$ is enabled, i.e., $\aopvar \land g_j(e_{i+1}.v)$ then it will execute with higher priority than other transitions on the location.
Since $\aopvar = \btrue$, if $g_i(e_{i+1})$ then the component will execute the reset location transition and move to $\mathit{dest}_j$ (i.e., $e_{i+1}.l' = dest_j$).
\end{proofpart}
\end{proof}

\section{\aopbip{} Language} \label{app:grammar}
The \aopbip{} language (overviewed in \secref{sec:evaluation:tool}) follows the ideas presented in this paper.
Because of the differences between the global and the local view, the grammar distinguishes global from local pointcut expressions and advices.
Furthermore, the identifiers specified in the AOP model generated from an \code{.abip} file are expected to reference identifiers in the BIP model generated from the \code{.bip} file.
The \code{validators} modules (both local and global) in the tool are responsible for verifying that the identifiers match.

\getcode[grammar-contain]{grammar-pc.g}{The $\aopbip$ Language Grammar}
		{0.75}{style=grammar, basicstyle=\normalsize\ttfamily, float=h}

\end{document}